\documentclass[preprint]{JHEP3} 

\JHEPspecialurl{http://jhep.sissa.it/JOURNAL/JHEP3.tar.gz}

\usepackage{epsfig,multicol}

\def\beq{\begin{equation}}

\def\eeq{\end{equation}}
\def\eq#1{{Eq.~(\ref{#1})}}

\newcommand{\bas}{\bar{\alpha}_S}

\newcommand{\ket}[1]{|#1\rangle}

\newcommand{\Tr}{{\rm Tr}}

\newcommand{\bea}{\begin{eqnarray}}

\newcommand{\eea}{\end{eqnarray}}

\parskip=\itemsep               

\newcommand{\beqar}[1]{\begin{eqnarray}\label{#1}}

\newcommand{\eeqar}{\end{eqnarray}}

\def\thefootnote{\fnsymbol{footnote}}

\title{
{\Large  \bf Odderon and seven Pomerons: QCD Reggeon field theory from JIMWLK evolution}}
\author{A. Kovner and M. Lublinsky \\
Physics Department, University of Connecticut, \\ 2152 Hillside
Road, Storrs, CT 06269-3046, USA}

\abstract{ We reinterpret the JIMWLK/KLWMIJ evolution equation as the QCD Reggeon
field theory (RFT). The basic "quantum Reggeon field" in this theory is
the unitary matrix $R$ which represents the single gluon scattering
matrix. We discuss the peculiarities of the Hilbert space on which the RFT
Hamiltonian acts. We develop a perturbative expansion in the RFT
framework, and find several eigenstates of the zeroth order Hamiltonian.
The zeroth order of this perturbation preserves the number of $s$ -
channel gluons. The eigenstates have a natural interpretation in terms of
the $t$ - channel exchanges. Studying the single $s$ - channel gluon sector we find the
eigenstates which include the reggeized gluon and five other colored Reggeons.
In the two ($s$ - channel) gluon sector we study only singlet color
exchanges. We find five charge conjugation even states. The bound state of
two reggeized gluons is the standard BFKL Pomeron. The intercepts of the
other Pomerons in the large $N$ limit are $1+\omega_P=1+2\omega$ where $1+\omega$ is the intercept of the BFKL Pomeron,
but their coupling in perturbation theory is suppressed by at least
$1/N^2$ relative to the double BFKL Pomeron exchange. For the
$[27,27]$ Pomeron we find $\omega_{[27,27]}=2\omega+O(1/N)>2\omega$. We also
find three charge conjugation odd exchanges, one of which is the unit
intercept Bartels-Lipatov-Vacca Odderon, while another one has an interecept greater than unity. We explain in what sense our calculation goes beyond the standard BFKL/BKP calculation.
We make additional comments and discuss open questions in our approach.
}

\begin{document}



\def\thefootnote{\arabic{footnote}}

\section{Introduction}

The theoretical approach to high energy scattering in the pre-QCD era was established by V. Gribov and is known as Gribov`s Reggeon Calculus
\cite{Gribov}. The idea was further developed and formulated  as an effective Reggeon field theory (RFT)
in Refs. \cite{Amati,Baker}. Since the advent of QCD more than three decades ago great effort has been made to reformulate these ideas in the framework of the QCD Lagrangian and thus to obtain RFT from first principles as a bona fide high energy limit of the theory of strong interactions.
Despite significant progress this goal has not been achieved yet. Although various elements of RFT in QCD have been available for some time, a coherent formulation of RFT is still not at hand.

The study of high energy limit in QCD began with the derivation of the Balitsky-Fadin-Kuraev-Lipatov (BFKL) Pomeron \cite{BFKL}, which was obtained as
an infinite sum of perturbative diagrams computed in the leading logarithmic approximation (LLA). The key element of the BFKL theory
is the concept of gluon reggeization, which is proven both in leading order (LO) \cite{LipatovRgluon,FS}
and next to leading order (NLO) \cite{Fadin}.
In particular, the BFKL Pomeron  is understood as a "bound state" of two reggeized gluons in the $t$-channel.

The BFKL Pomeron has an intercept greater than unity and leads to scattering amplitudes that grow as a power of energy, hence violating
the $s$-channel unitarity.
It was shortly realized that in order to restore unitarity, which undoubtedly has to be satisfied by the full QCD amplitudes, one needs to consider
$t$-channel states with more than just two reggeized gluons. This was put forward by Bartels \cite{ELLA} and is
today known as generalized leading log approximation (GLLA). The important milestone result was a derivation of the BKP
equation, which governs (perturbative) high energy evolution of an amplitude due to an exchange of an arbitrary but fixed number of reggeized gluons
\cite{BKP}.  It was further realized that the $s$-channel unitarity cannot be achieved without introducing
transitions between states with different number of reggeized gluons \cite{GLR}.
This is known as extended GLLA (EGLLA) \cite{3P,3P1}.
The vigorous program of solving the BKP equations in the large $N_c$ limit has been pursued in recent years by Lipatov and Korchemsky  with collaborators \cite{LipatovInteg},\cite{FK},\cite{Korchemsky},\cite{devega}.

The ideas put forward in \cite{3P} have been under intense investigation during the last decade or so \cite{BLW,Lotter,BV,BE,ES,BBV,BLV,Braun1,Braun2}. The transition vertex between 2 and 4 reggeized gluons was derived in \cite{3P,3P1,BV} and $2 \rightarrow 6$ in \cite{BE}. At present,
this approach therefore gives elements of an effective theory in terms of  $t$-channel gluon states and transition vertices.
These elements have been put together in \cite{Braun2} by Braun into an
 effective theory of BFKL Pomerons interacting via triple pomeron vertex of \cite{3P,3P1}.
This model is meant to describe nucleus-nucleus collisions at high energies at LLA and large $N_c$. In principle it contains Pomeron loops
\cite{braunlast}, even though they were not included in the semiclassical analysis of \cite{Braun2}.

A different approach to high energy scattering in QCD has been developed in the last 10 years or so based on the ideas of gluon saturation. The seminal paper of Gribov, Levin and Ryskin \cite{GLR} put forward the idea that nonlinear effects in QCD evolution lead to saturation of gluonic density at high enough energy. This gluon saturation than should restore the $s$-channel unitarity which is badly violated by the BFKL Pomeron. The GLR equation - the nonlinear evolution equation for gluon density in the double logarithmic approximation was derived
in \cite{MUQI}.
The gluon saturation ideas have been further developed in a series of papers by Mueller\cite{Mueller}, who also introduced the notion of
QCD dipoles as a convenient basis for the discussion of high energy processes, and has related it to BFKL Pomeron. The dipole version of
 the triple Pomeron vertex was obtained in Ref. \cite{Reggeon,NP}.

McLerran and Venugopalan recast the problem of saturation into that of studying nonlinearities of the classical Yang-Mills field theory directly in the path integral approach\cite{mv}. This provided an impetus for the formulation of the nonlinear QCD evolution approach to  high energy scattering\cite{balitsky},\cite{Kovchegov},\cite{JIMWLK},\cite{cgc}.
This line of research lead to two equivalent formulations of the nonlinear QCD evolution.
One is formulated in terms of infinite hierarchy of evolution equations for multiparton scattering amplitudes\cite{balitsky} - the so called Balitsky hierarchy; while the other one is based on a functional evolution equation for the probability distribution functional that determines the probabilities of various field configurations in the hadronic wave function - the so called JIMWLK equation\cite{JIMWLK},\cite{cgc}. The relation between the two formulations is roughly the same as between the infinite set of Dyson-Schwinger equations and the second quantized Hamiltonian in a quantum field theory. In this paper we find it more
convenient to use the JIMWLK formulation, and will refer to the functional evolution kernel as the JIMWLK Hamiltonian.

In the last year or so further progress has been made triggered by the realization that the JIMWLK equation does not include  Pomeron loops
\cite{IM,shoshi1,IMM,IT}.
 It was realized that the JIMWLK Hamiltonian must be extended in such a way that makes it self dual under the Dense-Dilute Duality transformation \cite{something,kl4}. Although the complete selfdual kernel has not yet been derived, the dual of the JIMWLK Hamiltonian (which we will refer to as KLWMIJ) was shown to describe the perturbative evolution of a dilute
hadronic wave function \cite{kl}. The efforts to consistently include Pomeron loops into the evolution are ongoing and some progress in this respect has been made \cite{IM,
balitsky1,kozlov,MSW,IT,LL3,kl1,kl4,kl5,SMITH,genya,Balitsky05,zakopane,braunlast}.

We note that there also has been earlier work on the subject of the high energy evolution which is similar in spirit. In particular \cite{Verlinde} considered an effective action for high energy QCD with Wilson lines as effective degrees of freedom.
The very same degrees of freedom appear in more recent studies of Refs. \cite{kl,kl1,SMITH,Balitsky05}. Yet another approach is due to Lipatov
and collaborators \cite{LipatovFT}, who derived an effective action with both real and reggeized gluons as effective degrees of freedom. This action  respects
the unitarity of full QCD, but its complexity has so far precluded any progress in understanding its physical consequences.

Obviously establishing the direct relation between
the JIMWLK formalism, which is firmly based in
 perturbative QCD and the Gribov Reggeon field theory ideas should be useful in unifying the two approaches and bringing closer the two "wings" of the high energy QCD community. Some attempts in this direction have been made starting with the classic paper of Gribov, Levin and Ryskin \cite{GLR}, where the Pomeron fan diagrams were analyzed. Most of the attempts in recent years have concentrated on the mean field approximation to the dipole
(large $N_c$) limit of high energy evolution \cite{Mueller,Reggeon,NP,Braun1,BLV}, which is technically a much simpler problem. These approaches converge to the Balitsky-Kovchegov (BK) equation \cite{balitsky,Kovchegov}
which sums fan-type Pomeron  diagrams \cite{GLR} in LLA  and describes the nonlinear corrections to the
BFKL Pomeron\cite{BLV} and the Odderon
\cite{KSW}.  Beyond the large $N_c$ limit it is known that the leading perturbative order of the JIMWLK evolution reproduces the BKP hierarchy
\cite{JIMWLK} and the perturbative Odderon solution \cite{oderon}.

Still the relations studied so far are somewhat indirect and do not provide a complete picture for translating between the JIMWLK and the Reggeon languages.
 The aim of the present paper is precisely to fill in this gap. We demonstrate that the JIMWLK/KLWMIJ Hamiltonian is nothing but the proper quantum field theoretical formulation of the Reggeon field theory. Our approach in the present paper will be based on the KLWMIJ form of the evolution, but as will become clear later on using JIMWLK in this context is entirely equivalent.
We note, that we do not address here the problem of Pomeron loops in the JIMWLK framework, and therefore the Reggeon field theory we deal with is not the complete one. However the mapping between the two languages is independent of the exact form of the  Hamiltonian and should also encompass the complete and as yet unknown self dual generalization.

This paper is structured as follows. In Section 2 we briefly review the general setup of calculating scattering amplitudes in the high energy eikonal approximation, the JIMWLK/KLWMIJ evolution kernel and discuss the structure of the Hilbert space on which the KLWMIJ Hamiltonian acts.

In Section 3 as a warmup to the main part of the paper we discuss the dipole limit of the JIMWLK evolution\cite{kl1,kl4}. We show that when the dipole model Hamiltonian is treated as a quantum field theory, one can set up a formal perturbative expansion around its quadratic part. The quadratic part preserves the dipole number, and so perturbative solution can be given in a sector with fixed number of dipoles. Solving the quadratic part itself one finds eigenfunctions which correspond to multi - BFKL Pomeron states, with the eigenvalues given by the multiples of the BFKL trajectory. The process of the diagonalization can be viewed (with some disclaimers) as finding such $s$ - channel dipole impact factors which couple exclusively to the $n$-Pomeron states in the $t$-channel.

Section 4 is the main part of this paper. Our aim here is to study the 2+1 dimensional quantum field theory defined by the full KLWMIJ Hamiltonian and to relate its perturbative spectrum with the results of the perturbative BFKL/BKP approach. This is of course much more complicated than for the dipole model, as the basic degrees of freedom now carry color indices. In fact the symmetry of this Hamiltonian is $SU(N)\otimes SU(N)\otimes Z_2\otimes Z_2$. One discrete $Z_2$ factor corresponds to the charge conjugation symmetry of QCD ($C$-parity). The second $Z_2$ appears due to the symmetry between $s$ and $u$ channel scattering amplitudes in the eikonal approximation. It naturally corresponds with the Reggeon signature extensively discussed in the literature. The $SU(N)\otimes SU(N)$ symmetry also has a natural interpretation as an independent  rotation of the incoming and outgoing color indices of the gluon scattering matrix. In the perturbative regime, where the gluon scattering matrix is close to unity, this symmetry is spontaneously broken down to diagonal $SU_V(N)$. The disordered regime on the other hand corresponds to the black body limit, since in physical terms it means that the color of the outgoing state after scattering is completely uncorrelated with the color of the incoming one.
Like in the dipole model we develop perturbation theory around a "perturbative vacuum state", that is the state in which the single gluon scattering matrix is close to unity. We develop an approximation which we will refer to as the partonic approximation to the KLWMIJ Hamiltonian, which preserves the number of $s$ channel gluons throughout the evolution. We split the KLWMIJ Hamiltonian into the perturbative part and the interaction part. The perturbative part  can be solved separately in every sector with fixed number of $s$ - channel gluons. However its structure is much more complicated than in the dipole model, since it is not quadratic but rather quartic in the gluon creation and annihilation operators.

We solve the one gluon sector of the theory and find states which directly correspond to reggeized gluon as well as other nonsinglet Reggeons. In our setup the color quantum number of the state "exchanged in the $t$ channel" is easily identified as the representation of the eigenstate with respect to the unbroken $SU_V(N)$ symmetry.
We find the octet $d$-Reggeon which is degenerate with the reggeized gluon as well as Reggeons corresponding to other irreducible representations which can be obtained from multiplying two adjoints $8\otimes 8\,=\,1\,+\,8_A\,+\,8_S\,+\,10\,+\,\overline{10}\,+\,27\,+\,{\cal R}_7$
\footnote{We follow in this paper the notations and nomenclature of \cite{NSZ}. Thus the different representations of the $SU(N)$ group are labeled by the dimensionality of their counterparts in $SU(3)$, except for ${\cal R}_7$
which does not exist in the $SU(3)$ case.}. The reggeized gluon wave function and trajectory are the same as in the BFKL calculation. The non octet Reggeons have all the same spatial wave function and their trajectories are simple factors (proportional to the second Casimir of the representation) multiplying the trajectory of the reggeized gluon. Notably the singlet Reggeon exchange leads to energy independent contribution to cross section.

We then go on to solve the two particle sector, but limit our solution to singlet exchanges in the $t$-channel. The structure we find can be easily understood in term of "bound states" of the Reggeons we found in the single $s$ - channel gluon sector.
We find states which are even and odd under the charge conjugation and the $Z_2$ signature symmetry. The $C$-parity and signature even sector contains the bound state of two reggeized gluons - the standard BFKL Pomeron with the usual wave function and trajectory. The other $C$-parity and signature even states include the bound state of two octet $d$-Reggeons (which is degenerate with the BFKL Pomeron), as well as bound states of  two
$27$ , two ${\cal R}_7$ Reggeons and a (color)symmetric bound state of $10$ and $\overline{10}$ Reggeons. All of those in the large $N_c$ limit have an intercept $1+\omega_P=1+2\omega$ where $1+\omega$ is the intercept of the BFKL Pomeron. The $1/ N_c$ correction to  the intercept of ${\cal R}_7$ Reggeon bound state is negative. The $1/N_c$ correction to the intercept of the $27$ Reggeon bound state is positive, making it the dominant exchange. We show that the $[8_S,8_S],\ [27,27]$ and $[{\cal R}_7,{\cal R}_7]$ Pomerons contain at least four gluons in the $t$-channel, while the
$[10+\overline 10,10+\overline 10]$ Pomeron contains at least six $t$ - channel gluons, making its coupling subleading in the dilute target limit.

In the $C$-parity odd sector we find three odderon states (for a recent review on Odderon see ref. \cite{Ewerzod}).
 One of them is signature odd and corresponds to the
bound state of the reggeized gluon and the d-Reggeon antisymmetric in color space and coordinate space separately. We show that the wave function of this state contains at least  three gluons in the $t$ - channel and it is therefore identified with the Bartels-Lipatov-Vacca Odderon (BLV)
 \cite{BLVod}.  Another Odderon state is signature even. Its interpretation is that of the antisymmetric bound state of the $10$ and $\overline {10}$ Reggeons. Its wave function contains at least six $t$-channel gluons.
Correspondingly its trajectory is twice that of the BLV Odderon. The intercept of both of these states is unity.
The third state we find is quite peculiar and interesting. It has a negative signature and can be thought of as a bound state of the reggeized gluon and the d-Reggeon symmetric in color space and coordinate space separately. Its intercept is greater than one and equal to that of the BFKL Pomeron. In general this state has at least three $t$-channel gluons and thus its coupling is not suppressed relative to the BLV Odderon. We show however that it
 decouples from a quark dipole.

In Section 5 we discuss the relation of our calculation to the BFKL/BKP setup. The BKP approximation arises as the limit of the partonic approximation to the KLWMIJ Hamiltonian when each gluon emitted in the process of the evolution is allowed to interact with the target only once. The partonic approximation is more general since it allows multiple scatterings of the emitted gluons. We also discuss the way in which the "composite states" of many Reggeons contribute to the unitarization of the scattering amplitudes and the limitations of this picture.


We conclude in Section 6 with discussion of our results and open questions.
Three Appendices contain details of our calculations as well as a summary of
useful properties of projectors for $SU(N)$ group.

\section{High energy scattering: the general setup}

The process we are interested in is the scattering of a highly energetic left moving projectile consisting of gluons on a hadronic target. We are working in the light cone gauge natural to the projectile wave function, $A^-=0$.
In this gauge the high energy scattering matrix of a single  gluon projectile at
 transverse position $x$ on the target is given by
the eikonal factor  \footnote{In our convention
the variable $x^-$ is rescaled to run from 0 to 1.}
\begin{eqnarray}\label{S}
S(x;0,x^-)\,\,=\,\,{\cal P}\,\exp\{i\int_{0}^{x^-} dy^-\,T^a\,\alpha^a_T(x,y^-)\}\,;
\,\,\,\,\,\,\,\,\,\,\,\,\,\,\,\,\,\,\,S(x)\,\equiv\,S(x;0,1)\,.
\end{eqnarray}
where $T^a_{bc}=if^{abc}$ is the generator of the $SU(N)$ group in the adjoint representation.

The field $\alpha_T$ is the large $A^+$ component created by the target color charge density. It
obeys the classical equation of motion  and is determined by the color charge density of the target $\rho_T(x)$ via\cite{first,JIMWLK,cgc}
\beq\label{alpha}
\alpha^a_T(x,x^-)T^a\,\,=g^2\,\,{1\over \partial^2}(x-y)\,
\left\{S^\dagger(y;0,x^-)\,\,\rho^{a}_T(y,x^-)\,T^a\,\,S(y;0,x^-)\right\}
\eeq

For a composite projectile which has some distribution of gluons in its wave function
 the eikonal factor can be written in the form analogous to
$S(x)$, see \cite{something}
\begin{equation}
\Sigma^P[\alpha_T]\,\,=\,\,\int D\rho_P\,\,W^P[\rho_P]\,\,
\,\exp\left\{i\int_{0}^{1} dy^-\int d^2x\,\rho_P^a(x,y^-)\,\alpha_T^a(x,y^-)\right\}\label{s}
\end{equation}
with $x_i$ - the transverse coordinate.  The quantity $\rho_P(x_i)$ is the color charge density in the projectile wave function at a given transverse position, while $W^P[\rho]$ can be thought of as the weight functional
which determines the probability density to find a given configuration of color charges in the projectile. For a single gluon $\rho^a(x_i)=T^a\delta^2(x_i-x^0_i)$, and eq.(\ref{s}) reduces to eq.(\ref{S}).

The total $S$-matrix of the scattering process at a given rapidity $Y$ is given by
\begin{equation}
{\cal S}(Y)\,=\,\int\, D\alpha_T^{a}\,\, W^T_{Y_0}[\alpha_T(x,x^-)]\,\,\Sigma^P_{Y-Y_0}[\alpha_T(x,x^-)]\,.
\label{ss}
\end{equation}
In \eq{ss} we have restored the rapidity variable and have chosen the frame where the target has rapidity $Y_0$ while the projectile carries the rest of the total rapidity $Y$.
Lorentz invariance  requires ${\cal S}$ to be independent of $Y_0$.

The high energy evolution is generally given by the following expression:
\begin{equation}
\frac{d}{d\,Y}\,{\cal S}\,=\,\int\, D\alpha_T^{a}\,\, W^T_{Y_0}[\alpha_T(x,x^-)]\,\,\,
\chi\left[\alpha_T,\frac{\delta}{\delta\,\alpha_T}\right]\,\,\,
\Sigma^P_{Y-Y_0}[\alpha_T(x,x^-)]\,.
\label{hee}
\end{equation}
Here $\chi$ stands for a generic Hermitian kernel of high energy evolution, which can be viewed as acting
either to the right or as Hermitian conjugated to the left:
\begin{equation}
{\partial\over\partial Y}\,\Sigma^P\,\,=\,\,\chi\left[\alpha_T,\,{\delta\over\delta\alpha_T}\right]\,\,\Sigma^P[\alpha_T]\,;
\ \ \ \ \ \ \ \ \
{\partial\over\partial Y}\,W^T\,\,=\,\,
\chi\left[\alpha_T,\,{\delta\over\delta\alpha_T}\right]\,\,W^T[\alpha_T]\,.
\label{dsigma}
\end{equation}
As was shown in \cite{something} in order for the total $S$-matrix to be Lorentz invariant and symmetric between the projectile and the target, the evolution kernel $\chi$ must be self dual. That is it has to be invariant under the Dense-Dilute Duality transformation
\begin{equation}\label{duality}
\alpha^a(x,x^-)\rightarrow i{\delta\over\delta\rho^a(x,x^-)},\ \ \ \ \ \ \ \, {\delta\over\delta\alpha^a(x,x^-)}\rightarrow -i\rho^a(x,x^-)
\end{equation}
However if one considers the situation where the target is large and the projectile is small, the symmetry between the target and the projectile is irrelevant. In the limit when the color charge density of the target is parametrically large ($\rho^a=O(1/\alpha_s)$) the kernel is given by
 the JIMWLK expression \cite{JIMWLK,cgc}
\begin{eqnarray}\label{JIMWLK}
&&\chi^{JIMWLK}\,=\,\hat K_{x,y,z} \,\left\{  \,2\,tr\left[{\delta\over \delta S^\dagger_x}\,T^a\,S_x\right]\,
S^{ab}_z\,tr\left[S_y\,T^b\,
{\delta\over \delta S^\dagger_y}\right] \right. \nonumber \\ &&\\
&-&\left.
tr\left[ {\delta\over \delta S^\dagger_x}\,T^a\,S_x\right]\,
tr\left[{\delta\over \delta S^\dagger_y}\,T^a\,S_y\right]\,-\,
 tr\left[S_x\,T^a\, {\delta\over \delta S^\dagger_x}\right]\,
tr\left[S_y\,T^a\,{\delta\over \delta S^\dagger_y}\right]
\right\}\,. \nonumber
\end{eqnarray}

\beq\label{kernel}
 K_{x,y,z}\,\equiv\,\frac{\alpha_s}{2\pi^2}\, {(z-x)_i(z-y)_i
\over (z-x)^2(z-y)^2}\,;\ \ \ \ \ \ \ \ \ \ \ \ \ \ \ \ \
\hat K_{x,y,z} \,f\,\equiv\,\int_{x,y,z}\,K_{x,y,z}\,f(x,y,z)
\eeq
The functional derivatives in this expression do not act on $S_z$ in the kernel.

Referring back to eq.(\ref{s}) we see that $\Sigma^P[\alpha]$ is the functional Fourier transform of $W^P[\rho]$. It thus follows that if $\Sigma^P$ evolves according to eq.(\ref{dsigma})with the JIMWLK kernel, $W^P[\rho]$ must evolve with the dual kernel, which we call KLWMIJ
\begin{equation}
{\partial\over\partial Y}\,W^P\,\,=\,\,
\chi^{KLWMIJ}\left[\rho,\,{\delta\over\delta\rho}\right]\,\,W^P[\rho]\,.
\label{dsigma1}
\end{equation}
with \cite{kl}
\begin{eqnarray}\label{KL}
&&\chi^{KLWMIJ}\,=\,
\hat K_{x,y,z} \,\left\{ \,2\,tr\left[{\delta\over \delta R^\dagger_x}\,T^a\,R_x\right]\,R^{ab}_z\,
tr\left[R_y\,T^b\,
{\delta\over \delta R^\dagger_y}\right] \right. \nonumber \\ &&\\
&&-\left.
tr\left[ {\delta\over \delta R^\dagger_x}\,T^a\,R_x\right]\,
tr\left[{\delta\over \delta R^\dagger_y}\,T^a\,R_y\right]\,-\,
 tr\left[R_x\,T^a\, {\delta\over \delta R^\dagger_x}\right]\,
tr\left[R_y\,T^a\,{\delta\over \delta R^\dagger_y}\right]
\right\}\,. \nonumber
\end{eqnarray}
where the "dual Wilson line" $R$ is defined as
\begin{equation}
R(z)^{ab}=
\left[{\cal P}\exp{\int_{0}^{1} d z^-T^c{\delta\over\delta\rho^c(z,\,z^-)}}\right]^{ab}
\label{rr}
\end{equation}
Just like in \eq{JIMWLK} the functional derivatives with respect to $R$  do not act
on $R(z)$. This can be rewritten in the normal ordered form, that is with all derivatives acting only on external factors.

\begin{eqnarray}\label{KL1}
\chi^{KLWMIJ}&=&-\,
\hat K_{x,y,z} \,:\left\{ - \,2\,tr\left[{\delta\over \delta R^\dagger_x}\,T^a\,R_x\right]\,R^{ab}_z\,
tr\left[R_y\,T^b\,
{\delta\over \delta R^\dagger_y}\right] \right. \nonumber \\ &&\nonumber \\
&+&\left.
tr\left[ {\delta\over \delta R^\dagger_x}\,T^a\,R_x\right]\,
tr\left[{\delta\over \delta R^\dagger_y}\,T^a\,R_y\right]\,+\,
 tr\left[R_x\,T^a\, {\delta\over \delta R^\dagger_x}\right]\,
tr\left[R_y\,T^a\,{\delta\over \delta R^\dagger_y}\right]
\right\}\nonumber \\ && \nonumber \\
&-&\,\hat K_{x,x,z} \,\left\{ - \,2\,R^{ab}_z\,
tr\left[T^a\,R_x\,T^b\,
{\delta\over \delta R^\dagger_x}\right] \,+\,
2\,N\,tr\left[ {\delta\over \delta R^\dagger_x}\,R_x\right]
\right\}:
\end{eqnarray}
We will find it convenient to rewrite the KLWMIJ kernel in the following form
\begin{eqnarray}\label{KL2}
&&\chi^{KLWMIJ}\,=\,-\,
\hat K_{x,y,z} \,:\left\{ - \,2\,tr\left[{\delta\over \delta R^\dagger_x}\,T^a\,(R_x\,-\,R_z)\right]\,R^{ab}_z\,
tr\left[(R_y\,-\,R_z)\,T^b\,
{\delta\over \delta R^\dagger_y}\right] \right. \nonumber \\ &&\nonumber \\
&&+\left.
tr\left[ {\delta\over \delta R^\dagger_x}T^a(R_x-R_z)\right]
tr\left[{\delta\over \delta R^\dagger_y}T^a(R_y-R_z)\right]\,+\,
 tr\left[(R_x-R_z)T^a {\delta\over \delta R^\dagger_x}\right]
tr\left[(R_y-R_z)T^a{\delta\over \delta R^\dagger_y}\right]
\right\}\nonumber \\ && \nonumber \\
&&-\,\hat K_{x,x,z} \,\left\{ - \,2\,R^{ab}_z\,
tr\left[T^a\,(R_x\,-\,R_z)\,T^b\,
{\delta\over \delta R^\dagger_x}\right] \,+\,
2\,N\,tr\left[ {\delta\over \delta R^\dagger_x}\,(R_x\,-\,R_z)\right]
\right\}:
\end{eqnarray}
To arrive at this form we used explicitly the fact that $R$ is a unitary matrix in the adjoint representation. This form of the kernel is particularly convenient, since each term in eq.(\ref{KL2}) separately is ultraviolet finite. The complete KLWMIJ kernel is ultraviolet finite, that is apparent ultraviolet divergencies at the point $x=y,\ \ z=x$ cancel between different terms in eq.(\ref{KL1}). In the form eq.(\ref{KL2}) this finiteness is explicit due to the factors $R_z-R_x$ etc.

Note that the KLWMIJ kernel possesses the global $SU_L(N)\otimes SU_R(N)$ symmetry. The symmetry transformations rotate
the left and right indices of the matrix $R$ independently. Furthermore there is a discrete $Z_2$ symmetry
$R^{ab} \rightarrow R^{ba}$. This transformation interchanges the color states of the incoming and outgoing
gluons. Thus this symmetry reflects
the $s-u$ crossing invariance of the QCD scattering amplitudes. The quantum number (charge) associated with the $Z_2$ symmetry
goes under the name of {\it signature}. The fact that the KLWMIJ Hamiltonian preserves this symmetry is nothing
else but the Gribov`s signature conservation rule. Below we will see the signature conservation rule at work.
The kernel is also charge conjugation invariant. We will discuss the $Z_2$ charge conjugation transformation in detail in Section 4.

An important point which we have not mentioned so far, is that $W[\rho]$ can not be an arbitrary functional.
This issue has been discussed in detail in \cite{kl,kl5} and more recently in \cite{Hatta,Fukushima}.
The point is that the longitudinal coordinate $x^-$ is introduced in our formalism as an "ordering" coordinate. To begin with the transverse charge density $\hat\rho^a(x)$ depends only on transverse coordinates. However $\hat\rho^a(x)$ are quantum operators and therefore do not commute between themselves. To emulate calculation of correlation functions of a product of such operators in the projectile wave function
we have introduced in \cite{kl} an additional variable $x^-$ whose values simply track the order of appearance of the operator factors of
$\hat\rho^a(x)$ in the product. The consistency requirement on $W[\rho]$ is then such that classical correlation functions calculated with
$W$ as the weight functional must reproduce the relations between the correlators of $\hat\rho^a(x)$ which follow from the
$SU(N)$ algebra satisfied by $\hat\rho$'s. We have shown in \cite{kl5} that the functionals of the following form satisfy all
 the required relations
\begin{equation}\label{hilbert}
W[\rho]\,=\,\Sigma[R]\,\delta[\rho(x,x^-)]
\end{equation}
with an arbitrary functional $\Sigma$ which depends on the "dual Wilson line" $R$ only.
Eq.(\ref{hilbert}) is a simple restriction on the Fourier transform of $W$. Inverting it we find
\begin{equation}\label{hilbert1}
\int d\rho\, \exp\left\{i\int dx^-d^2x\,\rho^a(x,x^-)\,\alpha^a(x,x^-)\right\}\,\,W[\rho]\,=\,\Sigma[S]
\end{equation}
with $S$ defined as in eq.(\ref{S}). Thus the Fourier transform of $W$ must depend only on the Wilson line $S$ and not on any other combination of the conjugate variables $\alpha$.  This of course has a very transparent physical interpretation. The functional Fourier transform of $W$ is precisely the projectile averaged scattering matrix for scattering on the "external field" $\alpha$. The restriction eq.(\ref{hilbert}) simply means that this scattering matrix must be a function of scattering matrices of individual gluons constituting the projectile, and does not depend on any other property of the external field. The normalization of the functional $W$ is determined by requiring that for $\alpha=0$ the scattering matrix is equal to unity. Thus
\begin{equation}\label{norm}
\Sigma(R=1)\,=\,1,\ \ \ \ \ \ \ \ \ \int d\rho\, W[\rho]\,=\,1
\end{equation}
Further general properties of $\Sigma$ follow from its identification as the projectile averaged scattering matrix. Let us for simplicity concentrate on the dependence of $\Sigma$ on $S$ at a particular transverse position $x$. Suppose the wave function of the projectile at this transverse position has the form (we take there to be exactly one gluon at the point $x$)
\begin{equation}\label{psi1}
|\Psi_x\rangle\,=\,\sum_{a=1}^{N^2-1}\,C_a(x)\,|a,x\rangle
\end{equation}
The scattering matrix operator when acting on $\Psi$ multiplies the gluon wave function by the matrix $S$\cite{kl1}.
We thus have
\begin{equation}
\langle\Psi_x|\hat S|\Psi_x\rangle\,=\,\sum_{a,b=1}^{N^2-1}C_a(x)\,C^*_b(x)\,\,S^{ab}(x)
\end{equation}
This generalizes easily to states with more than one gluon. Thus expanding the functional $\Sigma$ in Taylor series in $R$ we obtain the most general form compatible with its interpretation as the projectile averaged scattering matrix
\begin{equation}\label{sigmas}
\Sigma[R]\,=\,\sum_{m=0}^{\infty}\sum_{n_i=1}^{\infty}\left[C_{\{a^1\}...\{a^m\}}(x_1,...,x_m)C^*_{\{b^1\}...\{b^m\}}(x_1,...,x_m)\right]
\Pi_{i=1}^{m}\left[R^{a^i_1b^i_1}(x_i)...R^{a^i_{n_i}b^i_{n_i}}(x_i)\right]
\end{equation}
where $\{a^1\}$ denotes the set $a^1_1,a^1_2,...,a^1_{n_1}$, etc.
Simply stated this is the $S$-matrix of a state whose quantum mechanical amplitude to have $n_i$ gluons with color indices
$a^i_1,...,a^i_{n_i}$ at the transverse coordinate $x_i$ with $i=1,...,m$ is $C_{\{a^1\}...\{a^m\}}(x_1,...,x_m)$.
Thus for example the coefficients of "diagonal" terms with $a^i_j=b^i_j$ must be positive, since they have the meaning of probabilities to find the configuration with particular color indices in the projectile wave function.

Equations eq.(\ref{hilbert},\ref{norm},\ref{sigmas}) determine the Hilbert space of the
KLWMIJ Hamiltonian. We stress again that the norm in this Hilbert space as defined by eq.(\ref{norm}) is very different from the standard quantum mechanical definition. The condition of positive probability density which in standard quantum mechanics is quadratic in the wave function, in our Hilbert space is instead a linear condition eq.(\ref{sigmas}). This is a direct consequence of the fact that expectation values in our Hilbert space are defined as
\beq
\langle O[\rho]\rangle\,\equiv\,\int d\rho \,O[\rho]\,\,W[\rho]
\eeq
which is linear in the "wave functional" $W$.

The problem of calculating the $S$ matrix as a function of rapidity in the present approach is posed in the following way. First find eigenfunctions of the KLWMIJ Hamiltonian,
\begin{equation}\label{eigen1}
\chi^{KLWMIJ}[R,\delta/\delta R]\,G_q[R]\,=\,\omega_q\,G_q[R]
\end{equation}
The evolution of each eigenfunction is given by
\begin{equation}
G_q(Y)\,=\,e^{\omega_q Y}\,G_q
\end{equation}
Note that the eigenfunctions $G_q[R]$ do not have to satisfy eqs.(\ref{norm},\ref{sigmas}). Instead we can take them to satisfy the standard  Hilbert space normalization condition $\int DR G_q^*[R]G_{q}[R]=1$.
Next expand $\Sigma^P[R]$ at initial rapidity in this basis
\beq\label{bas}
\Sigma^P_0[R]\,=\,\sum_q\,\gamma_q\,G_q[R]
\eeq
The expansion coefficients are given in the standard form
\begin{equation}\label{gammaq}
\gamma_q=\int DR G_q^*[R]\Sigma^P_0[R]
\end{equation}
The expansion coefficients $\gamma_q$ are of course such that $\Sigma^P[R]$ is normalized according to eq.(\ref{sigmas}).

The total cross section at rapidity $Y$ is
\beq\label{over}
{\cal S}(Y)\,=\,\sum_q\,\gamma_q\,\,e^{\omega_q\,(Y\,-\,Y_0)}\,
\int\, DS\,\, W^T_{Y_0}[S(x)]\,\,G_q[S(x)]\,.
\eeq
Here we have assumed that the target weight functional $W^T$ depends only on $S$ and have substituted the integration over $\alpha(x,x^-)$ in eq.(\ref{ss}) by the integration over $S(x)$ (ditto in eq.(\ref{gammaq}) with respect to $R$ and $\delta/\delta\rho$).
We note that the the unitarity of the scattering amplitude requires (barring miraculous cancellations) nonpositivity of all $\omega_q$.

\subsection{Correspondence with the standard RFT language.}

Before setting out to study the approximate eigenstates and eigenvalues of the KLWMIJ Hamiltonian we want to point out to direct parallels between our setup and the approach to Reggeon field theory utilized for example in \cite{3P}.

The standard RFT approach is based on
a clear separation of $s$ - and $t$ channel
with factorization into impact factors and $t$-channel exchanges. The $t$ - channel exchanges are universal and do not depend on the nature of the projectile, while the impact factors are determined by the projectile wave function. This has direct parallels in our approach. The
projectile averaged $S$ - matrix eq.{\ref{sigmas}) is determined entirely by the wave function of the incoming projectile. As explained above, each factor $R$ in the expression eq.(\ref{sigmas}) corresponds to the scattering matrix of a single gluon in the projectile wave function. We will refer to these gluons as $s$ - channel gluons for obvious reasons. On the other hand the eigenfunctions $G_q[R]$ are completely independent of the projectile and are in this sense universal. They are determined solely by the evolution kernel $\chi^{KLWMIJ}$.

The eigenstates $G_q[R]$ are the exact analogs of, and in fact should be understood as, the $t$ - channel exchanges. To see why this is the case, we note first that as is clear from eq.(\ref{sigmas}), $\Sigma^P[R]$ carries two sets of indices. The left index labels the quantum numbers of the incoming projectile state, and the right index labels the quantum numbers of the outgoing projectile state. These sets of indices include all conserved quantum numbers of the theory, in particular color and transverse momentum. In eq.(\ref{sigmas}) we have indicated the color indices of each one of the incoming gluons $a^i_n$, the total color representation of the incoming projectile state being determined by the product of these representations and analogously for the outgoing state. We assume for simplicity that the projectile state belongs to an irreducible representation of the color  $SU_L(N)$ group and the outgoing state is in an irreducible representation of $SU_R(N)$. As for the !
 transverse momentum, since the transverse coordinates of all the gluons do not change during the interaction, $\Sigma^P$ depends only on the difference of the initial and final transverse momenta $k_\perp-p_\perp$. On the other hand the Hamiltonian $\chi^{KLWMIJ}$ has all the same symmetries as discussed earlier in this section, and so the same quantum numbers are carried by its eigenstates $G_q[R]$. More precisely $G_q[R]$ carry those quantum numbers which are not spontaneously broken by the vacuum of $\chi^{KLWMIJ}$. As we will see below the $SU_L(N)\otimes SU_R(N)$ group is broken spontaneously down to the vector subgroup $SU_V(N)$. Thus the set of quantum numbers $q$ carried by $G_q[R]$ includes the representation of $SU_V(N)$, the transverse momentum and the discrete symmetry $Z_2\otimes Z_2$ as discussed above.

Referring to the calculation for the overlap coefficients $\gamma_q$ in eq.(\ref{gammaq}) we observe, that the integration measure $DR$ is invariant under both $SU_V(N)$ and translations in the transverse space $R(x)\rightarrow R(x+a)$.
It is therefore obvious that the integral in eq.(\ref{gammaq}) gives nonzero result
only for those $G_q$ which match the change in the appropriate quantum
numbers of the projectile wave function in the scattering event. In
particular  $q_\perp=k_\perp-p_\perp$ and the $SU_V(N)$ representation of
$G_q$ is precisely that which is needed to change the $SU_L(N)$
representation of $\Sigma^P$ into the $SU_R(N)$ representation. Thus the
quantum numbers of $G_q$ that contribute to eq.(\ref{bas}) are precisely
those of the appropriate $t$ - channel exchanges that contribute to the
scattering. In this sense the eigenstates $G_q[R]$ indeed represent the
$t$ - channel exchanges with the target.

Continuing this train of thought we also see that the expansion
coefficients $\gamma_q$ are analogous to impact factors of
\cite{3P}, which describe the coupling of the $t$ - channel state with
quantum numbers $q$ to the $s$ - channel projectile state. The analogy is
indeed direct, however one has to distinguish between the impact factor of
exact eigenstate of $\chi^{KLWMIJ}$ which is $G_q$, and impact factors in the standard sense as they are used in
\cite{3P}, namely the coupling of the projectile to a fixed number of
$t$ - channel gluons. To understand this distinction, we must realize that
although an  $s$ - channel gluon is identified with the factor $R$ when it
appears in $\Sigma^P$ of eq.(\ref{sigmas}), the $t$ - channel gluon is not
identified with a factor $R$ in $G_q[R]$ but rather with a factor $\delta/ \delta\rho^a$ in its expansion. To see this we note that through eq.(\ref{over})
every factor of $\delta/\delta\rho^a(x)$ in expansion of  $\Sigma^P[R]$ becomes
a factor of the color field of the target $\alpha^a(x)$ which is indeed
naturally identified with the $t$ - channel gluon. Thus the perturbative coupling of a fixed number
$n$ of $t$ - channel gluons to a given projectile is given by
the coefficient of the $n$-th order expansion of $\Sigma^P[R]$ in powers of
$\delta/\delta\rho^a(x)$. This coefficient, which is just the $n$ gluon impact factor for the given projectile, can be written as
\begin{equation}\label{dn}
D_0^n=\int D\rho \rho^{a_1}(x_1)... \rho^{a_n}(x_n)W[\rho]
\end{equation}
The latter equality follows from the explicit form of $W[\rho]$ eq.(\ref{hilbert}).

This establishes the relation between the main elements of the standard
RFT language and the basic quantities of our approach. To recapitulate: an
$s$ - channel gluon corresponds to a single factor $R$ in $\Sigma^P[R]$; a $t$ -
channel exchange corresponds to an eigenstate $G_q$ of the Hamiltonian
$\chi^{KLWMIJ}$; a coupling of a given exchange to the projectile state is
given by $\gamma_q$; a single $t$ - channel gluon is represented by a
single factor $\delta/\delta\rho^a$ in the expansion of $\Sigma[R]$; and an impact factor of a fixed
number of $t$ - channel gluons is given by $D_0^n$, eq.(\ref{dn}).

We also note that we can define the "rapidity dependent" function
\begin{equation}\label{dn1}
D^n(Y)=\int D\rho \rho^{a_1}(x_1)... \rho^{a_n}(x_n)W_Y[\rho]
\end{equation}
which satisfies the evolution equation
\begin{equation}\label{dn2}
{d\over dY}D^n=\int D\rho \rho^{a_1}(x_1)... \rho^{a_n}(x_n)\chi^{KLWMIJ}[\rho,{\delta\over\delta\rho}]W_Y[\rho]
\end{equation}
This is the generalization of the $2$ and $4$-point functions considered in \cite{3P1,BV,Ewerz}.  The approach of \cite{3P1,BV,Ewerz} usually considers this type of functions with the projectile consisting of a single dipole and customarily splits $D^n$ into the irreducible part and the reggeized part. To make contact with this approach  we consider in Appendix A the evolution of $D^4$ for the projectile containing a single dipole in the leading order in $1/N_c$. We demonstrate there that our approach in indeed equivalent to that discussed in \cite{3P1,BV} and that the decomposition into irreducible and reggeized parts appears naturally in our approach as well.

The rest of this paper is devoted to solution of the eigenvalue problem eq.(\ref{eigen1}) in the approximation where  the KLWMIJ kernel is expanded around $R=1$. This expansion can be formulated since $R=1$ is the classical minimum of the KLWMIJ kernel. Physically this expansion is best thought of as the approximation of small target. For a small target the overlap integral in eq.(\ref{over}) is dominated by values of $S$ close to unity. Thus the leading contribution at early stages of evolution comes from the wave functions $G_q[R]$ which are large at $R$ close to unity. Those can be studied by expanding the KLWMIJ kernel around $R=1$ and keeping only the homogeneous term in the expansion in $(R-1)$. Note however that the condition that $R$ is close to unity does not necessarily mean that we have to use the standard perturbation theory in $\alpha_s$, as this closeness does not have to be parametric in $\alpha_s$. We discuss this point in Section 5. In the next section we will consider a simplified
problem of the KLWMIJ kernel reduced to its dipole limit.

\section{Reggeons in the dipole model}

A simplified version of the high energy evolution - the dipole model was introduced by Mueller in \cite{Mueller}. It describes the leading high energy behavior in the large $N_c$ limit as long as the densities in the wave functions are not too large. As shown in \cite{kl1} the dipole evolution equation is obtained as a well defined limit of the JIMWLK evolution.
We can define the projectile dipole "creation operator" $s$ and the target dipole "creation
operator" $r$:
\beq\label{pdipole}
s(x,y)\,\equiv\,{1\over N_c}Tr[S_F^\dagger(x)\,S_F(y)]\,;\ \ \ \ \ \ \ \ \ \ \ \ \ \
r(x,y)\,\equiv\,{1\over N_c}Tr[R_F^\dagger(x)\,R_F(y)]
\eeq
where $F$ indicates the fundamental representation. The term "creation operator" is not mathematically perfect - it would be more appropriate to call it the "dipole field", since it is Hermitian. We will nevertheless keep to the tradition of calling it the creation operator in this paper.
The "annihilation operators" in this parlance are  $\frac{\delta}{\delta s}$ and
$\frac{\delta}{\delta r}$ for projectile and target respectively.

If  the target weight function is a function of $s$ only, that is $W^T=W^T[s]$, the action of the JIMWLK kernel on it
in the large $N_c$ limit is equivalent to the action of the dipole kernel \cite{kl1} (see also \cite{janik}):
\beq\label{jimdipole}
\chi^{JIMWLK}\,\,W^T[s]\,=\,\chi^{dipole}_{s}\left[s,\frac{\delta}{\delta s}\right]\,\,W^T[s]
\eeq
The dipole  kernel was found in \cite{LL1,LL2} by reformulating the original Muller`s
model and was later obtained directly from eq.(\ref{jimdipole}) in \cite{kl1}
\beq\label{chidip}
\chi^{dipole}_{s}\left[s,\frac{\delta}{\delta s}\right]\,=\,\int_{x,y,z} M_{x,y,z}\,\,
\left[-\,s(x,y)\,+\,s(x,z)\,s(y,z)\right]\frac{\delta}{\delta s(x,y)}
\eeq
with $M$ - the usual dipole emission probability
\beq
M_{x,y,z}\,=\,{\bar\alpha_s\over 2\,\pi}\,{(x-y)^2\over (x-z)^2\,(y-z)^2}
\eeq

Eq.(\ref{jimdipole}) is derived under the same assumption as the JIMWLK equation - namely that the target is dense.
The dual form of this kernel describes the evolution of a hadronic state (be it a projectile or a target) which contains a small number of dipoles in its wave function. This form is derived from the KLWMIJ kernel assuming that $r$ is the only degree of freedom in the  wave function \cite{kl4} (see also \cite{MMSW,HIMS}):
\beq\label{jimdipole1}
\chi^{KLWMIJ}\,\,W[r]\,=\,\chi^{dipole}_{r}\left[r,\frac{\delta}{\delta r}\right]\,\,W[r]
\eeq
The calculation of the cross section of a projectile on a target which both are made entirely of dipoles is given in analogy with eq.(\ref{over}) by
\beq\label{overdip}
{\cal S}_D(Y)\,=\,\sum_q\,\gamma_q\,\,e^{\omega_q\,(Y\,-\,Y_0)}\,
\int\, Ds\,\, W^T_{Y_0}[s(x,y)]\,\,g_q[s(x,y)]\,.
\eeq
where the wave functions $g_q$ satisfy
\begin{equation}\label{eigend}
\chi^{dipole}_r[r,{\delta\over\delta r}]\,g_q[r]\,=\,\omega_q\,g_q[r]
\end{equation}

We start by expanding the dipole Hamiltonian around the classical solution  $r=1$.
Let us denote the dipole creation an annihilation operators by\footnote{Although we are abusing notation slightly by calling both $r$ and $1-r$ the dipole creation operator, we hope this does not cause confusion.}
$$
d^\dagger\,\equiv\,1\,-\,r\,\,; \ \ \ \ \ \ \ \ \ \ \ \ \ \ \ \ \ \ \ \ d\,\equiv \,-\,\frac{\delta}{\delta r}
$$
The dipole Hamiltonian is
\beq\label{dh}
\chi^{dipole}\,=\,H_0\,+\,H_I
\eeq
The free Hamiltonian $H_0$ is quadratic in the creation and annihilation operators
\beq
H_0\,=\,\int_{x,y,z} M_{x,y,z}\,\,
\left[-\,d^\dagger(x,y)\,+\,d^\dagger(x,z)\,+\,d^\dagger(z,y)\right]\,d(x,y)
\eeq
The Hamiltonian of the interaction $H_I$ is given by
\beq
H_I\,=\,-\,\int_{x,y,z} M_{x,y,z}\,d^\dagger(x,z)\,d^\dagger(y,z)\,d(x,y)
\eeq
We now consider the spectrum of the free Hamiltonian $H_0$.
There are of course no new results in this calculation. It only serves to establish in a very simple setting the correspondence between the Reggeon field theory and the KLWMIJ/JIMWLK approach that we want to make explicit in this paper.
This is an easy exercise, since the Hamiltonian is very similar to harmonic oscillator.
The vacuum of $H_0$ is  $|0\rangle$ such that $d|0\rangle=0$.
The wave functional of this state is simply a constant which does not depend on $r$.
The first exited state contains one dipole and is a superposition of the basis states
$d^\dagger(x,y)|0\rangle$. We take it
in the form
\beq\label{psiq}
g_q\,=\,\int_{x,y}\, {1\over (x\,-\,y)^4}\,\psi_q(x,y)\,d^\dagger(x,y)\,|0\rangle
\eeq
Acting on $g_q$ by $H_0$ we find that the wave function $\psi_q$ should satisfy
the BFKL equation
\beq\label{bfkl}
\int_{z} M_{x,y,z}\,\left[-\,\psi_q(x,y)\,+\,\psi_q(x,z)\,+\,\psi_q(y,z)\right]\,=\,\omega_q\,\psi_q(x,y)
\eeq
The eigenfunctions should vanish when two transverse coordinates coincide, since the dipole of zero size is indistinguishable from vacuum.
Solutions of the BFKL equation which satisfy this condition, the so called Moebius invariant solutions, are well known. They are
the eigenfunctions of the Casimir operators of conformal algebra \cite{Lipatov1}
\beq\label{enn}
\psi_q(x,y)\,=\,E^{n, \nu} (x-\rho, y-\rho) \, = \, \left( \frac{(x-y)}{(x-\rho) \,
(y-\rho)} \right)^{\frac{1+n}{2} + i \nu} \ \left(
\frac{(x-y)^*}{(x-\rho)^* \, (y-\rho)^*} \right)^{\frac{1-n}{2} + i
\nu}
\eeq
where the complex coordinate $x$ is defined as $x=x_1+ix_2$
The index $q$ denotes the conformal spin $n$, $\nu$, as well as the degeneracy vector $\rho$. The eigenvalues are
\beq\label{eig}
\omega_q\,=\,2 \, \bas \, \chi (n, \nu),
\eeq
where
\beq\label{chi}
\chi (n, \nu) \, = \, \psi (1) - \frac{1}{2} \, \psi \left(
\frac{1+|n|}{2} + i \nu \right)  - \frac{1}{2} \, \psi \left(
\frac{1+|n|}{2} - i \nu \right),
\eeq
These wave functions  satisfy the completeness relation \cite{Lipatov1}
\beq\label{EE}
(2 \pi)^4 \, \delta (x-\bar x) \, \delta (y-\bar y) \, = \,
\sum_{n=-\infty}^\infty \, \int_{-\infty}^\infty d \nu \, \int d^2
\rho \, \frac{16 \left(\nu^2 + \frac{n^2}{4}\right)}{|x-y|^2 \,
|\bar x-\bar y|^2} \, E^{n, \nu} (x-\rho, y-\rho) \,
E^{n, \nu \, *} (\bar x-\rho, \bar y-\rho)
\eeq
By diagonalizing $H_0$ in the one dipole Hilbert space we have found such a linear combination of $s$ - channel dipole states which couples precisely to a given "state" (fixed ($n,\ \nu)$) on the BFKL Pomeron trajectory. Note that this does not mean that the coupling of this state to the projectile and the target is via a two gluon exchange only. Since the wave functional $g_q$ depends  linearly on $r$,
and $r$ has all order expansion in powers of $\delta/\delta\rho$, the coupling of $g_q$ to the target is the full eikonal coupling which includes multiple scatterings. Nevertheless this $t$ - channel exchange with all its multiple scatterings evolves exponentially in rapidity with a fixed
exponent $2\bar\alpha_s\chi(n,\nu)$.

The exited states at the next level contain two "reggeized dipoles". Since in the dipole model all dipoles are treated as
independent degrees of freedom and $[d^\dagger(x,y),d(u,v)]=-\delta^2(x-u)\delta^2(y-v)$, we immediately infer the two dipole spectrum
\begin{equation}
H_0\,g_{q_1,q_2}\,=\,[\omega_{q_1}\,+\,\omega_{q_2}]\,g_{q_1,q_2}
\end{equation}
where
\begin{equation}
g_{q_1,q_2}\,=\,\int_{x,y,u,v}\,{1\over (x-y)^4 \,(u-v)^4}\,
\psi_{q_1}(x,y)\,\psi_{q_2}(u,v)\,d^\dagger(x,y)\,d^\dagger(u,v)|0\rangle\,.
\end{equation}

These eigenstates correspond to double Pomeron exchange.
This corresponds to the leading order in $N_c$   BKP state (which couples to a projectile consisting of two dipoles) which contains four reggeized gluons in the $t$-channel. Note however that as discussed in section 2, this exchange does not contain a fixed number of $t$-channel gluons, but rather includes all multiple scatterings. If we chose the projectile $\Sigma$ to be equal to $g_{q_1,q_2}$, such a projectile state in the present approximation would not interact via a single Pomeron exchange, but only via the double Pomeron exchange. As discussed at the end of this section however, such a projectile is not physical, as its wave function contains negative probabilities.

The generalization to the multi-dipole states is straightforward. The eigenstates and eigenvalues of $H_0$ are
\begin{equation}
H_0\,g_{q_1,...,q_n}\,=\,[\omega_{q_1}\,+...+\,\omega_{q_n}]\,g_{q_1,...,q_n}
\end{equation}
with
\begin{equation}
g_{q_1,...,q_n}\,=\,\int_{\{x_i\},\{y_i\}} {\psi_{q_1}(x_1,y_1)\over (x_1-y_1)^4}
...{\psi_{q_n}(x_n,y_n)\over (x_n-y_n)^4}\,d^\dagger(x_1,y_1)...d^\dagger(x_n,y_n)|0\rangle
\end{equation}

Thus the dipole limit of the KLWMIJ hamiltonian reproduces the leading large $N_c$ spectrum of the noninteracting multipomeron exchanges. Note that the eigenstates contain all the nonforward BFKL amplitudes and thus scattering amplitudes at arbitrary transverse momentum transfer.

This of course is simply a recasting of well known results in our framework. It is therefore also obvious that the expansion around $r=1$ violates unitarity. While the complete dipole model Hamiltonian is unitary and when solved nonperturbatively
must lead to unitary $S$ - matrix\footnote{Within additional approximation of no target correlations (see \cite{kl1} for alternatives),
finding a non-perturbative solution of the dipole Hamiltonian is equivalent to solving the non-linear BK equation. This question
was addressed in many papers \cite{BKT,Braun1,BKN,MT,MP}. The result is  consistent with the unitarity constraint.},
the Hamiltonian $H_0$ does not. One manifestation of this is the fact that the BFKL trajectory has an intercept greater than one, and therefore the scattering amplitudes grow without bound.
The same fact also manifests itself in a somewhat different fashion. In particular the eigenfunctions that we found are all (except for the vacuum) zero norm states with respect to the norm defined in eq.(\ref{norm}). All the wave functions are homogeneous polynomials of $1-r$, and thus they all, except for the constant vacuum wave function, vanish at $r=1$.

Put in other words, when interpreted as dipole states in the $s$-channel, $g_q$  contain negative probabilities. The condition eq.(\ref{sigmas}) can be adapted to the dipole limit in a straightforward manner. It states that in the expansion of $\Sigma[r]$ in powers of $r(x,y)$, individual coefficients represent probability densities of finding a particular dipole configuration, and thus all coefficients of $r(x_1,y_1)...r(x_n,y_n)$  must be positive. On the other hand consider the first excited state wave function
\begin{equation}
g_{q}\,=\,\int_{x,y}{1\over (x-y)^4}\,\psi_{q}(x,y)\,\left[r(x,y)-1\right]\,.
\end{equation}
The coefficient function $\psi_q$ as defined in eq.(\ref{enn}) is not real. Due to the degeneracy of the eigenvalues eq.(\ref{chi}), $n\rightarrow-n$ and $\nu\rightarrow-\nu$ one can construct real combinations of these functions $\psi_{n,\nu}+\psi_{-n,-\nu}$ and $i(\psi_{n,\nu}-\psi_{-n,-\nu})$. These combinations however are not strictly positive. Thus if we were to
take $\Sigma[r]=g_q[r]$, this wave function would have negative probabilities to find the dipole at some points $(x,y)$. The only eigenfunction which is positive everywhere is the one that corresponds to the forward scattering BFKL amplitude, $n=\nu=0$. Even this one however contains negative probabilities. Since $g_0$ is proportional to $r-1$ and not $r$, it has positive probability to contain one dipole (at any position $(x,y)$) but an overall negative probability to contain no dipoles at all. Similarly, the second excited state has a positive probability to contain two dipoles and no dipoles, but a negative probability to contain a single dipole.

This curious behavior is not by itself a problem, since one can construct a positive norm state with all positive probabilities at the initial rapidity by taking appropriate superposition of these zero norm states and the vacuum state.
This is also the reason why we have to keep all the eigenfunctions that we have found in the preceding analysis and not just the one with $n=\nu=0$ - we need to superpose all of them in order to construct an arbitrary positive definite scattering matrix $\Sigma[r]$.
The problem arises however at later times (higher rapidities) since the BFKL intercept is greater than one. The $S$-matrix at later times will be dominated entirely by the component in the superposition which has the largest eigenvalue, and this state by itself contains negative probabilities.
The linearized evolution therefore preserves the overall norm of the state (since $\Sigma[1]$ does not evolve with rapidity), but generates probabilities which are greater than unity as well as negative probabilities. Thus unless the unitarizing corrections due to the triple Pomeron vertex are taken into account, the evolution violates unitarity not just by driving the value of the
total scattering probability above unity, but also by making the probabilistic interpretation in the $s$ - channel impossible.

We certainly believe that this is an artifact of the perturbative approach and that the interaction will cure this problem when consistently taken into account. It is not our aim in this paper to solve nonperturbatively the KLWMIJ hamiltonian (we wish we could!) but rather to relate the perturbative approach to the standard techniques which also operate with nonunitary amplitudes. We therefore will not have much to say about the effects of the interaction. We only note that
the first perturbative correction to the single Pomeron state can be obtained by acting on the state
by the perturbation Hamiltonian $H_I$:
\beq
H_I\,g_q\,=\,\int_{x,y,z} {M_{x,y,z}\over (x-y)^4}\,\psi_q(x,y)\,d^\dagger(x,z)\,d^\dagger(y,z)\,|0\rangle\,.
\eeq
This state can be projected onto a two-particle state $g_{q_1,q_2}$.  The resulting matrix element
is
\beq\label{3P}
\langle g_{q_1,q_2}|\,H_I\,|g_q\rangle\,=\,-\,\int_{x,y,z} {M_{x,y,z}\over (x-y)^4\,(x-z)^4\,(y-z)^4}\,
\psi_q(x,y)\,\psi^*_{q_1}(x,z)\,\psi^*_{q_2}(y,z)
\eeq
\eq{3P} is the well known result  for the first unitarity correction to the BFKL Pomeron  due to the QCD
triple Pomeron vertex. In the dipole model and beyond this has been studied in detail in Refs. \cite{3P,3P1,Lotter,BV,Mueller,
NP,BLV}. It is a satisfactory feature of the present approach that this result is reproduced automatically.

We therefore conclude that the dipole model Hamiltonian considered as a second quantized 2+1 dimensional field theory
describes in the perturbative expansion the large $N_c$ BFKL Pomeron with the self-interaction given by the triple pomeron vertex. It is tempting to think about a given eigenstate of $H_0$ as a "projectile" with specific configuration of dipoles in the $s$-channel chosen so that it interacts with any target only by the specific $n$-Pomeron exchange. The wave function of such a projectile however contains negative probabilities. It is thus more correct to think about a given eigenstate $g_q$ as of a $t$ - channel exchange, as discussed in Section 2. Any physical "projectile" with all positive probabilities can be constructed at initial rapidity as a superposition of these eigenstates. The evolution generated by $H_0$ however generates negative probabilities at large enough rapidity.

In the next section we perform a similar analysis of the full KLWMIJ Hamiltonian beyond the dipole limit.

\section{Reggeons in QCD}

\subsection{The Hamiltonian}
We start with the KLWMIJ Hamiltonian written in the form eq.(\ref{KL2}). It has a classical minimum at $R=1$ and in the rest of this section we consider expansion around this configuration. Introducing
$\tilde R=1-R$ we write
\beq\label{h1}
\chi^{KLWMIJ}\,=\,H_0\,+\,H_I
\eeq
The ``free'' Hamiltonian $H_0$ is the part homogeneous in $\tilde R\,\frac{\delta}{\delta\tilde R}$, and it naturally splits into the real and virtual parts, the latter appearing due to the normal ordering in eq.(\ref{KL1}).
\beq\label{h0}
H_0\,=\,H_0^R\,+\,H_0^V
\eeq
The real part is
\begin{eqnarray}\label{h0r}
&&H_0^R\,=\,-\,\hat K_{x,y,z} \,\left\{
 - \,2\,tr\left[{\delta\over \delta \tilde R^\dagger_x}\,T^a\,(\tilde
R_x\,-\,\tilde R_z)\right]\,
tr\left[(\tilde R_y\,-\,\tilde R_z)\,T^a\,
{\delta\over \delta \tilde R^\dagger_y}\right] \right\}\,+\,
tr\left[ {\delta\over \delta \tilde R^\dagger_x}\,T^a\,(\tilde R_x\,-\,\tilde R_z)\right]\,
 \nonumber \\ &&\nonumber \\
&&\otimes\,\left.
tr\left[{\delta\over \delta \tilde R^\dagger_y}\,T^a\,(\tilde R_y\,-\,\tilde R_z)\right]\,+\,
 tr\left[(\tilde R_x\,-\,\tilde R_z)\,T^a\, {\delta\over \delta\tilde  R^\dagger_x}\right]\,
tr\left[(\tilde R_y\,-\,\tilde R_z)\,T^a\,{\delta\over \delta \tilde R^\dagger_y}\right]
\right\}
\eea
The virtual part is
\beq\label{h0v}
H_0^V\,=\,-\,2\,\int_{x,z} K_{x,x,z} \,\left\{ - \,
tr\left[T^a\,(\tilde R_x\,-\,\tilde R_z)\,T^a\,
{\delta\over \delta \tilde R^\dagger_x}\right]\,+\,N\,tr\left[(\tilde R_x\,-\,\tilde R_z)\,
{\delta\over \delta \tilde R^\dagger_x}\,\right]
\right\}
\eeq
The interaction Hamiltonian has the form
\begin{eqnarray}\label{hi}
H_{I}&=&-\,
\hat K_{x,y,z} \,tr\left[{\delta\over \delta \tilde R^\dagger_x}\,T^a\,(\tilde R_x\,-\,\tilde R_z)\right]\,\tilde R^{am}_z\,
\tilde R^{bm}_z\,
tr\left[(\tilde R_y\,-\,\tilde R_z)\,T^b\,
{\delta\over \delta \tilde R^\dagger_y}\right]  \nonumber \\ &&\nonumber \\
&+&\,\int_{x,z} K_{x,x,z} \, \left(\tilde R^{ab}_z+\tilde R^{ba}_z\right)\,
tr\left[T^a\,(\tilde R_x\,-\,\tilde R_z)\,T^b\,
{\delta\over \delta \tilde R^\dagger_x}\right] \,.
\end{eqnarray}
Here we have used the unitarity of the matrix $R$ to rewrite the first term in the form which looks like a $2\rightarrow 4$ vertex.
We have dropped the normal ordering sign in the above formulae for notational simplicity, but all the functional derivatives are understood not to act on the fields $\tilde R$ in the kernels themselves.

We view the Hamiltonian eqs.(\ref{h1},\ref{h0r},\ref{h0v},\ref{hi}) as defining the fully second quantized Reggeon Field Theory. In the following we will refer to it as either KLWMIJ or RFT Hamiltonian.

As noted in Section 2, in principle the separation of the full RFT Hamiltonian into a free term and an interaction is not unique due to the unitarity of the matrix $R$.
Our guide is to define this split in such a way that $H_0$ and $H_I$ are separately ultraviolet and infrared safe. We discuss this point further in Section 6.

Analogously to Section 3, we can think of  $\tilde R$ and $\delta/ \delta \tilde R^\dagger$ as  $s$ - channel gluon
creation and annihilation operators respectively.
Consistently with this terminology we will call the approximation based on the Hamiltonian $H_0$ in eq.(\ref{h0}) - partonic approximation to RFT. In this approximation the $s$ - channel gluons emitted throughout the evolution scatter independently and the scattering amplitude of a gluon configuration is taken to be the sum of scattering amplitudes of the individual gluons. This is encoded in the Hamiltonian $H_0$ since it is homogeneous in the gluon creation operator and thus preserves the number of $s$ - channel gluons that contribute to the scattering amplitude throughout the evolution. We will discuss the physics of this approximation as well as its relation to the standard BFKL/BKP approach in the next section.

The vacuum of the free Hamiltonian $H_0$ is clearly the state $|0\rangle$ such that
\beq
{\delta\over \delta \tilde R^\dagger}\,|0\rangle\,=\,0
\eeq
This is the state with no gluons and corresponds to a trivial "empty" projectile. The higher states are obtained from the vacuum by successive action of the creation operator $\tilde R$.
Although $H_0$ is homogeneous in the creation and annihilation operators, it is significantly more complicated than the corresponding Hamiltonian in the dipole model. It contains not only terms of the type $a^\dagger a$, but also $a^\dagger a^\dagger aa$, and its spectrum is therefore not simply that of a harmonic oscillator.

In the diagonalization of $H_0$ we will find very helpful to use its symmetries.
The expansion around $R=1$ breaks the $SU_L(N)\otimes SU_R(N)$ symmetry of the full KLWMIJ kernel down to the diagonal $SU_V(N)$:
\beq\label{u}
R(x)\rightarrow U^\dagger \,R(x)\,U\,.
\eeq
Thus the eigenstates of $H_0$ can be classified according to the representations of this $SU_V(N)$ group. As discussed above, and as we will see explicitly below, the representation of $SU_V(N)$ corresponds to the color quantum number of  the $t$ - channel exchange.

Another interesting symmetry of $H_0$ and $H_I$ is the discrete symmetry
\begin{equation}\label{signature}
R(x)\rightarrow R^\dagger(x)
\end{equation}
This transformation interchanges the row and column indices of the matrix $R$. Physically the first index of the matrix $R$ corresponds to the color index of the incoming $s$ - channel gluon, while the second index to the color of the outgoing $s$ - channel gluon. Thus the interchange of these indices interchanges the $s$ and $u$ channels. The discrete symmetry eq.(\ref{signature}) can be therefore identified with the signature discussed in the context of the Reggeon theory\footnote{We dub ``signature'' the behavior of the amplitude under the $s-u$ crossing.  This is not quite the standard definition of the signature factor
as used in the  partial wave analysis, but coincides with it for scalar exchanges. We thank Genya Levin for pointing this to us. }.
The eigenstates of $H_0$ fall into even and odd states under this transformation.

Finally another discrete symmetry of the KLWMIJ Hamiltonian is charge conjugation (C-parity). In QCD the action of the charge conjugation symmetry is conventionally defined as
\begin{equation}
A_\mu^a\tau^a\rightarrow -(A_\mu^a\tau^a)^*
\end{equation}
where $\tau^a$ are the generators of the fundamental representation.
This translates into the following transformation of the {\it fundamental} eikonal factor
\beq
R_F(x)\rightarrow R^*_F(x)
\eeq
while the adjoint eikonal factor transforms as
\beq\label{cconj}
R^{ab}(x)\,\rightarrow\, C^{ac}\,R^{cd}(x)\,C^{db}
\eeq
Here the matrix $C$ is the diagonal matrix which can be conveniently written as
\beq
C^{ab}\,=\,2\,{\rm tr}(\tau^a\,\tau^{*b})
\eeq
The transformation eq.(\ref{cconj}) flips the sign of the matrix elements of $R$ in rows and columns  whose indices correspond to the imaginary fundamental generators $\tau^a$. This symmetry is also unbroken by the configuration $R=1$.

\subsection{One-particle state: reggeized gluon and other Reggeons}

In this subsection we discuss the "one particle states" - the eigenstates of $H_0$ linear in the $s$- channel gluon creation operator $\tilde R$. We will use the terms "one particle state", "two particle state" etc. in this sense throughout the rest of the paper.

The one particle eigenstates represent $t$ - channel exchanges which in the partonic approximation couple directly to the one $s$ - channel gluon states.
We are looking for one-particle eigenstates of the form
\beq\label{ef1}
G_q^\lambda\,=\,\int_x \Psi_q(x)\eta^\lambda_{cd}\,\tilde R^{cd}(x)\,|0\rangle
\eeq
The states in this sector are multiplets of the $SU_V(N)$ symmetry, and we use cumulative  index $\lambda$ to label both the representation and the particular state vector inside the representation. The tensors $\eta_{ab}^\lambda$ essentially are projectors from the product of two adjoints to the representation $\lambda$, see below.

The eigenvalue problem we are to solve is
\beq\label{ef11}
H_0\,G_q^\lambda\,=\,\beta_q(\lambda)\,G_q^\lambda
\eeq
First we compute the action of $H_0$  on $\tilde R$
\beq\label{h0R}
H_0\,\tilde R^{cd}_x\,=\,H_0^V\,\tilde R^{cd}_x\,=\,2\,\int_{z} K_{x,x,z} \,\left\{
(T^a\,T^b)_{cd}\,(\tilde R_x^{ab}\,-\,\tilde R_z^{ab}) \,-\,N\,(\tilde R_x^{cd}\,-\,\tilde R_z^{cd})
\right\}
\eeq
The color structure can be easily diagonalized. To do that we need solutions of (no summation over $\lambda$ is implied)
\beq\label{cs}
(T^a\,T^b)_{cd}\,\eta^\lambda_{cd}\,=\,N\,k_\lambda\,\eta^\lambda_{ab}
\eeq
This is solved using the decomposition of the product of two generators into the projectors of $SU(N)$ representation (see eq.(\ref{tt}) of the Appendix C)
\beq\label{cs1}
(T^a\,T^b)_{cd}\,=\,N\,\sum_i\lambda_i\,P^{i\,ab}_{\,\,\,cd}\,=\,N\,
\left(P[1]\,+\,{1\over 2}\,P[8_A]\,+\,{1\over 2}\,P[8_S]\,-\,{1\over N}\,P[27]\,+\,{1\over N}\,P[{\cal R}_7]\right)^{ab}_{cd}\,,
\eeq
The explicit form of the projectors $P$ is given in the Appendix C. We follow in this paper the notations and nomenclature of \cite{NSZ}. Thus the different representations are labeled by the dimensionality of their counterparts in $SU(3)$, except for ${\cal R}_7$
which does not exist in the $SU(3)$ case.
Clearly choosing $\eta=P[\lambda]$ solves equation (\ref{cs}) for any representation $\lambda$ that can be constructed from the product of two adjoints.  Thus we have seven eigenfunctions corresponding
to $1$, $8_A$, $8_S$, $10$, $\overline{10}$, $27$, and ${\cal R}_7$ representations of $SU_V(N)$
with the
eigenvalues
\begin{equation}\label{lambda}\lambda_1=1,\ \ \ \ \  \lambda_{8_A}={1\over 2} , \ \ \ \ \ \ \ \lambda_{8_S}={1\over 2}, \ \ \ \ \ \ \lambda_{10}=\lambda_{\overline{10}}=0,\ \ \ \ \ \  \lambda_{27}=-{1\over N} , \ \ \ \ \ \ \ \ \lambda_{{\cal R}_7}={1\over N}.
\end{equation}

Note that each projector $P[\lambda]$ corresponds to a $t$-channel exchange in the color representation $\lambda$. This interpretation follows directly from the fact that $\tilde R$ is the $T$-matrix of the $s$-channel gluon. For $P[\lambda]$ the color indices of the incoming and outgoing gluon are projected onto the representation $\lambda$. Physically this can happen only due to an exchange with the target by an ($t$ - channel exchange) object in this representation (Fig. \ref{fig1}).

\begin{figure}[htbp]
\epsfig{file=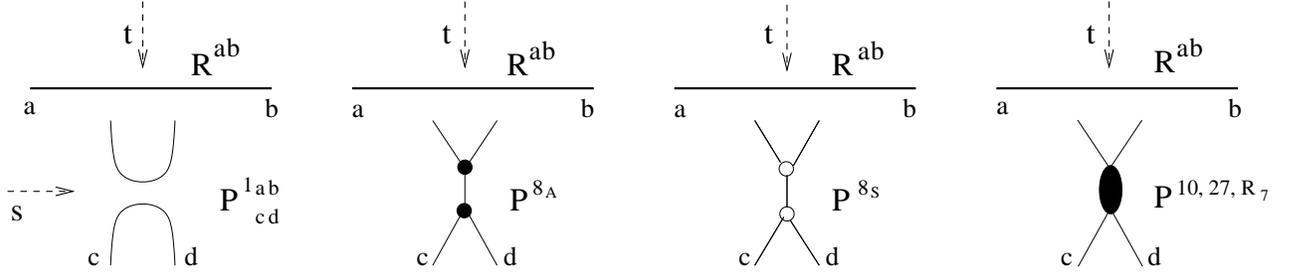,width=168mm}
\caption{\it $t$-channel projection of Wilson lines.}
\label{fig1}
\end{figure}

Once the color structure is diagonalized, the equation for $\Psi_q$ follows:
\beq\label{psi}
-\,2\,N\,(1\,-\,\lambda)\,\int_z \left[\,K_{y,y,z}\,\Psi_q(y)\,-\,
K_{z,z,y}\,\Psi_q(z)\right]\,=\,\beta_q(\lambda)\,\Psi_q(y)
\eeq
Substituting $\Psi$ in the form of a plane wave $\Psi_q(y)=e^{i\,q\,y}$ we get
\beq
-\,\frac{\alpha_s\,N}{\pi^2}\,(1\,-\,\lambda)\,
\,e^{i\,q\,y}\,\int_\mu d^2k\,{q^2\over k^2\,(q-k)^2}\,
=\,\beta_q(\lambda)\,e^{i\,q\,y}
\eeq
where $\mu$ is the infrared cutoff. The momentum $q$ is the momentum transfer in the scattering process.
For the eigenvalue $\beta_q$ we obtain ($\bar \alpha_s\equiv \alpha_s\,N/\pi$)
\beq\label{omega}
\beta_q(\lambda)\,=\,-\,{\bar\alpha_s\over \pi}\,(1\,-\,\lambda)\,\int_\mu d^2k\,{q^2\over k^2\,(q-k)^2}\,\approx\,-\,{\bar\alpha_s}\,(1\,-\,\lambda)\,
\ln{q^2\over\mu^2}
\eeq
In the last equality we have assumed $q\gg\mu$. For $q\rightarrow 0$ the eigenvalue vanishes quadratically with $q$:
\beq\beta_q(\lambda)\rightarrow_{q\rightarrow 0}\bar\alpha_s\,(1\,-\,\lambda){q^2\over \mu^2}\eeq
Note that the coefficient $1-\lambda$ is simply proportional to the second Casimir of the given representation
(tabl. \ref{table1}):
\beq
2\,N\,(1-\lambda)\,=\,C_2\,.
\eeq
\eq{omega} provides an expression for the trajectory of the Reggeon in the channel with  given color exchange\cite{3P} (see also \cite{DM}).
The ${\cal R}=8_A$ channel is special since it has the quantum number of the gluon. Indeed  we have
$$\beta_q[8_A]\,=\,-\,\frac{\alpha_s\,N}{2\,\pi}\,\ln {q^2\over \mu^2}\,;
$$
which is the standard reggeized gluon trajectory. We will refer to it as the  reggeized gluon or the $f$-Reggeon.
Its signature is negative, since the projector $P[8_A]_{ab}^{cd}$ is antisymmetric under the exchange of $a$ and $b$.
The $f$-Reggeon has a positive signature  "brother" in the ${\cal R}=8_S$ channel, which we will refer to as the $d$-Reggeon. Since $\lambda_{8_S}=\lambda_{8_A}$,
its trajectory is degenerate with that of the reggeized gluon, $\beta_q[8_S]\,=\,\beta_q[8_A]$.
 The corresponding wavefunctions are
$$
G_q^{8_A\,\,ab}\,=\,\int_x \,e^{i\,q\,x}\,f^{abk}\,f^{kcd}\,\tilde R^{cd}(x)\,;\ \ \ \ \ \ \ \ \ \ \ \
G_q^{8_S\,\,ab}\,=\,\int_x \,e^{i\,q\,x}\,d^{abk}\,d^{kcd}\,\tilde R^{cd}(x)
$$
Thus in our approach the reggeized gluon is naturally identified with an eigenstate whose eigenfunction is linear in the Wilson line $R$ projected onto the antisymmetric octet  representation in the $t$-channel. Note however that the identification is not operatorial - we do not have an operator in the Hilbert space which we could identify with Reggeized gluon. The correspondence is rather on the level of the (perturbative) eigenstate of the KLWMIJ Hamiltonian.

Note that the singlet state in the $t$-channel (${\cal R}=1$) does not contribute to the evolution of the cross section. The scattering amplitude due to the exchange of this "$s$-Reggeon" is flat with energy since $\lambda_1=1$.
Note also that the  Reggeons corresponding to $10$ and $\overline{10}$ exchanges are degenerate. Thus if the coefficient of the projector $P^{10}$ in the expansion of $\Sigma[R]$ is equal to that of $P^{\overline{10}}$ at initial rapidity, it remains so throughout the evolution. Nevertheless the $10$ and $\overline{10}$ Reggeons appear as distinct eigenstates of the RFT Hamiltonian.

The signature of the $10$ and $\overline{10}$ Reggeons
 as well as of the reggeized gluon is negative since the respective projectors are odd under the exchange of one pair of indices. The other three Reggeons - $8_S$, $27$ and ${\cal R}_7$ have positive signature.
Different types of Reggeons (including the singlet) and their relation to signature were discussed in \cite{3P,3P1,Ewerz}.

The eigenvalue at $q^2\rightarrow 0$ for all the Reggeons vanishes. For the reggeized gluon in the present approach this is the consequence of the Goldstone theorem. Our expansion breaks spontaneously the $SU_L(N)\otimes SU_R(N)$ symmetry down to $SU_V(N)$. As in any other quantum field theory, in the RFT such breaking requires the appearance of $N^2-1$ signature odd Goldstone bosons in the adjoint representation of $SU_V(N)$. These are precisely the quantum numbers of the reggeized gluons. The fact that all the other Reggeons also have vanishing eigenvalue at $q^2=0$ from this point of view is surprising. It may  signal the presence of a larger broken symmetry group which we were not able to identify.

The effect of the interaction Hamiltonian $H_I$ is to mix the one and two Reggeon states. In particular
 acting
 by the perturbation $H_I$ on the one Reggeon state sector we obtain:
\beq\label{chiGd}
H_I\,G_q^\lambda\,=\,-\,\eta^\lambda_{ab}\int_{x,z}\,
K_{x,x,z}\, \tilde R_z^{\alpha\beta}\,\left[T^\alpha\,(\tilde R_x\,-\,\tilde R_z)\,T^\beta
\right]_{ab}\,\Psi_q(x)
\eeq
Projecting the above state onto the analog of the noninteracting two-particle state using the projector
$$
{\cal P}^{\lambda_1\lambda_2}_{q_1,q_2}\,\equiv\,\int_x \eta^{\lambda_1}_{cd}\,
{\delta\over \delta \tilde R^{cd}(x)}\,\Psi^*_{q_1}
(x)\int_y \eta^{\lambda_2}_{ab}\,{\delta\over \delta\tilde R^{ab}(x)}\,\Psi^*_{q_2}(y)
$$
we obtain
\beq
{\cal P}^{\lambda_1\lambda_2}_{q_1,q_2}\,H^I\,G_q^\lambda\,=\,\eta^{\lambda_1}_{ab}\,
\eta^{\lambda_2}_{cd}\,\eta^{\lambda}_{mn}\,T^a_{mc}\,T^b_{dn}\,\,V_{1g\rightarrow 2g}(q; q_1, q_2)
\eeq
with the Reggeon "splitting vertex" given by
\bea
V_{1g\rightarrow 2g}(q; q_1, q_2)&=&-\,\int_{x,y} K_{x,x,y}\,\left\{\Psi^*_{q_1}(x)\,\Psi^*_{q_2}(y)
\left[\Psi_{q}(x)\,+\,\Psi_{q}(y)\right]\,-\,
2\,\Psi^*_{q_1}(y)\,\Psi^*_{q_2}(y)\,\Psi_{q}(x)\right\}
\nonumber \\
&=&-\,{\alpha_s\over 2\,\pi}
\,\left[\ln\frac{q_1^2}{\mu^2}\,+\,\ln\frac{q_2^2}{\mu^2}\right]\,\delta^2(q\,-\,q_1\,-\,q_2)
\eea
In accordance with the signature conservation not all of the transition are allowed. For example
the $f$-Reggeon mixes only with the $f$ and $d$ states.

\subsection{Two-particle state}
Our next task is to find eigenstate in the two particle sector. We will limit ourselves to the $SU_V(N)$ singlet states
which correspond to color singlet exchanges in the $t$-channel. It is convenient to label these states  $G_q^{S\, (CP)}$ by their eigenvalues with respect to the discrete symmetries: signature $S$, charge conjugation $C$ and space parity $P$ which for two particle states is equivalent to the symmetry of the eigenfunction with respect to transverse coordinate exchange.
The index $q$ cumulatively denotes the rest of the quantum numbers.
We will refer to the $C$-parity even eigenstates as Pomerons, while to the $C$-parity odd eigenstates as Odderons.
The most general Pomeron state has the form
\begin{equation}\label{pome}
G^{+\,(++)}_q\,=\,\sum_{i=1}^6P^{i\,bd}_{\,\,\,ac}\,\int_{u,v}\,\tilde R^{ab}(u)\,\tilde R^{cd}(v)\,\Psi_s^i(u,v)\,|0\rangle;
\end{equation}
Here the projectors
$P^i\,\equiv\,P[{\cal R}_i]$ are the same as in the previous subsection and the index $i$ runs over representations $1,\,8_A,\,8_S,\,10+
\overline{10},\,27,\,{\cal R}_7$. We have defined the projector onto the
representation $10+\overline{10}$ as $[R^{10}+R^{\overline{10}}]$. Note that the combination $P^{10}-P^{\overline {10}}$ does not appear in the Pomeron sector, since it has negative $C$-parity.

In the Odderon sector we have three possible sets of eigenstates which all differ in at least one quantum number:
\begin{eqnarray}
&G^{-\,(--)}_q&\,=\,Z^{+\,bd}_{\,\,\,ac}\,\int_{u,v}\,\tilde R^{ab}(u)\,\tilde R^{cd}(v)\,\Phi^-(u,v)\,|0\rangle\,;\nonumber\\
&G^{+\,(--)}_q&\,=\,Z^{-\,bd}_{\,\,\,ac}\,\int_{u,v}\,\tilde R^{ab}(u)\,\tilde R^{cd}(v)\,\Phi^+(u,v)\,|0\rangle\,;
\nonumber\\
&G^{-\,(-+)}_q&\,=\,[P^{10}\,-\,P^{\overline{10}}]^{bd}_{ac}\,
\int_{u,v}\,\tilde R^{ab}(u)\,\tilde R^{cd}(v)\,\Psi^-_s(u,v)\,|0\rangle\,;\label{odde}
\end{eqnarray}
\begin{figure}[htbp]
\epsfig{file=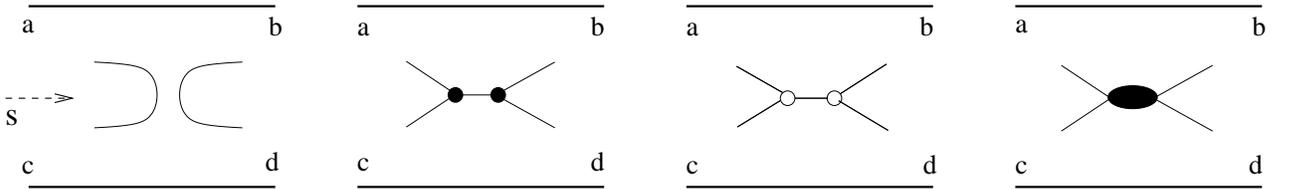,width=168mm}
\caption{\it $s$-channel projection of two Wilson lines.}
\label{fig2}
\end{figure}

The additional tensors $Z^{\pm}$ are defined as (see Appendix C).
\beq
i (Z_s^{\pm})^{ab}_{cd} =  {i \over 2} \Big( f_{bak} d_{kcd} \pm d_{bak}f_{kcd} \Big) \,
\label{zpm}
\eeq

The projectors $P^i$ are symmetric under the exchange of indices
$P^{i\,bd}_{\,\,\,ac}=P^{i\,db}_{\,\,\,ca}$
while the tensors  $Z^\pm$ are antisymmetric under the exchange of both pairs of indices.
Therefore the $\Psi_s^i(u,v)$ and $\Psi_s^-(u,v)$ are functions symmetric under the interchange of $u$ and $v$ ($P=\,+1$) while $\Phi^{\pm}(u,v)$ are antisymmetric functions ($P=\,-1$).

\eq{pome} and (\ref{odde}) have a simple interpretation from the $s$-channel point of view (Fig. \ref{fig2}). In $G^{+\,(++)}_q$,
two incoming gluons with indices $ac$ are projected onto representation ${\cal R}^i$.
The two outgoing gluons (with indices $bd$) are projected
onto the same representation ${\cal R}^i$. Thus the color representation of the two gluon state remains unchanged during the propagation through the target. Additionally the invariant tensor $P^i$ projects the product of the incoming and the outgoing representations ${\cal R}^i$ onto a color singlet.  Thus these amplitudes correspond to a color singlet exchange in the $t$-channel. It can be easily checked that these eigenstates are even ($C=+1$) under the charge conjugation transformation eq.(\ref{cconj}).
The tensors $P^i$ are symmetric under the interchange of the incoming and outgoing indices; $P^{i\,bd}_{\,\,\,ac}=P^{i\,ac}_{\,\,\,bd}$.
Hence all $G^{+\,(++)}_q$ are  positive signature eigenstates ($S=+1$).

In $G^{\pm\,(--)}$, the incoming and outgoing representations are both adjoints, but they have different charge conjugation properties. These eigenstates correspond to the change of $8_A$ into $8_S$ (or vice versa) during the interaction with the target. Again one can straightforwardly check that both $G^{\pm\,(--)}$ eigenstates are odd under the charge conjugation eq.(\ref{cconj}). They are therefore both associated with the Odderon exchange. The properties of these two sets of eigenstates under the signature transformation are however different.
The tensor $Z^+$ is odd under the interchange of the incoming and outgoing indices and therefore the set $G^{-\,(--)}$ contains negative signature eigenstates\footnote{We hope that our convention in which the tensor $Z^+$ defines the signature odd state is not hopelessly confusing. Although this convention is somewhat unfortunate, we have decided to stick literally to the notations of \cite{NSZ} rather than risk even greater confusion.}. These correspond to the standard signature odd odderon exchange. On the other hand the tensor $Z^-$ is even under the interchange of the incoming and outgoing indices. The eigenstates
$G^{+\,(--)}$ thus correspond to a signature even Odderon exchange.

The set of eigenstates $G^{-\,(-+)}$ is a peculiar one. These states have negative signature and negative charge
conjugation but are described by a symmetric wavefunction.
Because the
Hamiltonian preserves parity, charge conjugation and signature, the four sets of eigenstates of eqs. (\ref{pome},{\ref{odde}) do not mix.

The $s$-channel representation eqs.(\ref{pome},\ref{odde}) is natural for discussing the  scattering process from the $s$ channel point of view.
It emphasis the color structure of the gluon "dipole" state which propagates in the $s$ - channel.
It is however also instructive to consider an alternative $t$-channel representation (Fig. \ref{fig3}). Instead of projecting the color indices of the two incoming gluons into a definite color representation we can project the incoming and outgoing indices of the same gluon.  This corresponds to writing the wave function of the eigenstates in the form
\beq\label{G1}
G^{+\,(++)}_q\,=\,\sum_{i}P^{i\,cd}_{\,\,\,ab}\,\int_{u,v}\,\tilde R^{ab}(u)\,\tilde R^{cd}(v)\,\Psi_t^i(u,v)
\eeq
In this expression a given term in the sum eq.(\ref{G1}) represents the process whereby each one of the $s$-channel gluon exchanges color representation ${\cal R}^i$ with the target, and additionally the two ${\cal R}^i$ representations in the $t$-channel form an overall color singlet.
\begin{figure}[htbp]
\epsfig{file=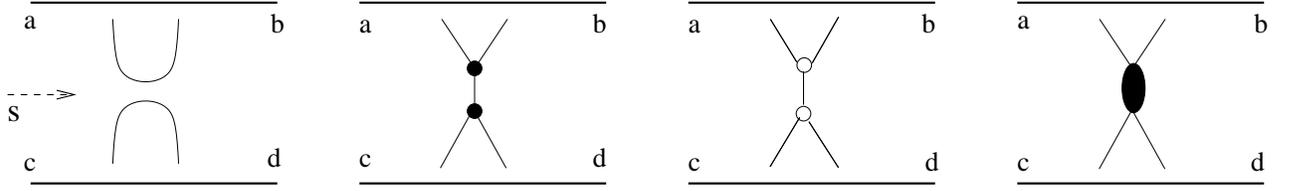,width=168mm}
\caption{\it The $t$-channel projection of two Wilson lines. Each one of the two $s$ - channel gluons exchanges a given color representation with the target. Additionally the two exchanges form an overall color singlet.}
\label{fig3}
\end{figure}
The two expression eq(\ref{pome}) and eq.(\ref{G1}) are simply related by the change of basis, since either set of projectors ($s$-channel or $t$-channel) form a complete set.
The $t$-channel coefficient functions $\Psi_t^i$ are linearly related to the functions $\Psi_s^i$:
\beq\label{basis}
\Psi_t^i\,=\,\sum_k\,C_i^k\,\Psi^k_s\,,\ \ \ \ \ \ \ \ \ \ \ \ \ \
\Psi_s^i\,=\,\sum_k\,C_i^k\,\Psi^k_t\,,
\eeq
where the crossing matrix $C$ is given in the Appendix C.

Analogous transformation can be performed for the $G_q^{-\,(--)}$, $G_q^{+\,(--)}$, $G_q^{-\,(-+)}$
 states. The $s$ channel $Z^+$ tensor remains the $Z^+$ tensor in the $t$-channel, while the $s$-channel $Z^-$ tensor becomes in the $t$-channel the difference of the projectors $[P^{10}-P^{\overline{10}}]$.
The $s$-channel $[P^{10}-P^{\overline{10}}]$ becomes $Z^-$ in the $t$ - channel.

This $t$-channel representation of the function $G_q^{S\,(CP)}$
is suggestive of the interpretation of the exchanged states in the $t$-channel as bound states of Reggeons discussed in the previous subsection. In fact this turns out to be a very convenient interpretation since as we will see the eigenstates of $H_0$ correspond to a fixed $t$-channel projector $P^i$
(that is only one $\Psi^i_t(x,y)$ is nonvanishing for a particular eigenstate).

The eigenvalue problem in the two particle sector is
\beq\label{chiG}
H_0\,G^{S\,(CP)}_q\,=\,\omega_q\,G^{S\,(CP)}_q
\eeq
Our goal now is to reformulate \eq{chiG} as  equations for $\Psi_s$ and $\Phi$. This is achieved by applying to
\eq{chiG} of the operators
$$P^{j\,\beta\delta}_{\,\,\,\alpha\gamma}\,{\delta \over \delta \tilde R^{\alpha\beta}(x)}\,
{\delta \over \delta\tilde R^{\gamma\delta}(y)}\,\ \ \ \ \ \ \ \ \,{\rm and} \ \ \  \ \ \ \ \ \ \ \
Z^{r\,\beta\delta}_{\,\,\,\alpha\gamma}\,{\delta \over \delta \tilde R^{\alpha\beta}(x)}\,
{\delta \over \delta\tilde R^{\gamma\delta}(y)}
$$
This leads to the following equations
\beq\label{chiG1}
P^{j\,\beta\delta}_{\,\,\,\alpha\gamma}\,{\delta \over \delta \tilde R^{\alpha\beta}(x)}\,
{\delta \over \delta\tilde R^{\gamma\delta}(y)}\,H_0\,G^{+\,(++)}_q\,
=\,2\,\omega_q\,tr[P^j]\,\Psi^j_s(x,y)\,=\,2\,\omega_q\,D_j\,\Psi^j_s(x,y)
\eeq
and
\beq\label{chiG2}
Z^{\pm\,\beta\delta}_{\,\,\,\,\,\,\,\alpha\gamma}\,{\delta \over \delta \tilde R^{\alpha\beta}(x)}\,
{\delta \over \delta\tilde R^{\gamma\delta}(y)}\,H_0\,G^{\mp\,(--)}_q
\,=\,\mp\,4\,\omega_q\,(N^2\,-\,4)\,D_8\,\Phi^{\mp}(x,y)
\eeq
\beq\label{chiG3}
Z^{-\,\gamma\delta}_{\,\,\,\,\,\,\,\alpha\beta}\,{\delta \over \delta \tilde R^{\alpha\beta}(x)}\,
{\delta \over \delta\tilde R^{\gamma\delta}(y)}\,H_0\,G^{-\,(-+)}_q
\,=\,4\,\omega_q\,(N^2\,-\,4)\,D_8\,\Psi_s^-(x,y)
\eeq
where $D_j\,\equiv \,dim[{\cal R}_j]$ is the dimension of the corresponding representation.
The eigenvalues $\omega_q$ and the functions $\Psi_s$, $\Phi$ are found by solving the system of homogeneous
equations   \eq{chiG1} and  \eq{chiG2}.
The details of the calculation of the left hand sides of \eq{chiG1}, \eq{chiG2} and  \eq{chiG3}
are given in the Appendix  B.
The results are summarized and discussed in the next two subsections.

\subsubsection{Pomeron and family.}
Let us consider first the equations for the $C$-parity  even sector. Equations eq.(\ref{chiG1}) can be written as (see Appendix B)
\bea\label{pom}
&&
 \left\{ \lambda_j\,+ \,\sum_i (-1)^{s_j+s_i}\,\bar C^i_j\right\} \nonumber \\  && \nonumber \\
&\times&\int_{z}\,\left[ 2\,K_{x,y,z}\,
 \,\Psi^i_s(x,y)\,-\,  K_{x,z,y}\,\Psi^i_s(x,z)
\,-\,   K_{z,y,x}\,\Psi^i_s(y,z)
\,+\,
2\,\delta(x-y)\int_u K_{z,u,x}\,\Psi^i_s(z,u)\right]
 \nonumber \\  && \nonumber \\
&+&\sum_i\,\left\{  \bar C^i_j\, -\,\delta^i_j\right\}\,
\int_{z}\,\left[2\,K_{x,x,z}
 \,\Psi^i_s(x,y)\,-\,  K_{z,z,y}\,\Psi^i_s(x,z)
\,-\,   K_{z,z,x}\,\Psi^i_s(y,z)\right] \,=\,{\omega\over 2\,N}\,\Psi^j_s(x,y) \nonumber \\
\eea
where $s_i=0$ if $i$ is a symmetric representation, $s_i=1$ if $i$ is antisymmetric, and the matrix $\bar C$ is given in eq.(\ref{barC}).
No summation over $j$ is implied.  Using the relations eq.(\ref{basis}) one can show that this system is diagonal in the $t$-channel
\bea\label{pom2}
&-& \int_{z}\,\left[ 2\,K_{x,y,z}\,
 \,\Psi^i_t(x,y)\,-\,  K_{x,z,y}\,\Psi^i_t(x,z)
\,-\,   K_{z,y,x}\,\Psi^i_t(y,z)
\,+\,
2\,\delta(x-y)\int_u K_{z,u,x}\,\Psi^i_t(z,u)\right]
 \nonumber \\  && \nonumber \\
&-&
\int_{z}\,\left[2\,K_{x,x,z}
 \,\Psi^i_t(x,y)\,-\,  K_{z,z,y}\,\Psi^i_t(x,z)
\,-\,   K_{z,z,x}\,\Psi^i_t(y,z)\right] \,=\,{\omega\over 2\,N\,(1\,-\,\lambda_i)}\,\Psi^i_t(x,y)
\eea
with $\lambda_i$ given by eq.(\ref{lambda}) and $\lambda_{10+\overline{10}}=0$.
\eq{pom2} is  the non-forward BFKL equation in the coordinate representation.
In the Fourier space
$$
\psi(k_1,k_2)\,\equiv\,\int_{x,y}\,\Psi^i_t(x,y)\,e^{i\,k_1\,x\,+\,i\,k_2\,y}
$$
the equations eq.(\ref{pom2}) read
\beq\label{BFKLm}
\left[{\omega\over 2\,(1\,-\,\lambda_i)} \,-\,\beta(k_1)\,-\,\beta(k_2)\right]\,\psi(k_1,k_2)\,=\,\int_{q_1,q_2}\,
\,K_{2\rightarrow 2}(k_1,k_2;q_1,q_2)\,\,\psi(q_1,q_2)
\eeq
The kernel $K_{2\rightarrow 2}(k_1,k_2;q_1,q_2)$ is real part of the of the standard momentum space  BFKL kernel
\beq\label{kernel22}
K_{2\rightarrow 2}(k_1,k_2;q_1,q_2)\,=\,-\,\frac{\bar\alpha_s}{2\,\pi}\,\,
\left[{(q_1\,-\,k_1)_i\over (q_1\,-\,k_1)^2}\,-\,{q_{1\,i}\over q_1^2}\right]\,\,
\left[{(q_2\,-\,k_2)_i\over (q_2\,-\,k_2)^2}\,-\,{q_{2\,i}\over q_2^2}\right]\,\,
\delta^2(k_1\,+\,k_2\,-\,q_1\,-\,q_2)\,
\eeq
The BFKL trajectory $\omega^{BFKL}$ is given by \eq{eig}.
Thus we find five independent singlet exchanges in the $t$-channel with the eigenvalues
\beq\label{eigen}
\omega^i\,=\,2\,(1\,-\,\lambda_i)\, \omega^{BFKL}
\eeq
They have a natural interpretation as "bound states" of the Reggeons discussed earlier. The compound state of two reggeized gluons is the standard BFKL Pomeron with
\begin{equation}
\omega[8_A,8_A]\,=\,\omega^{BFKL}
\end{equation}
In addition we have the following nontrivial bound states (Fig. \ref{fig6})
\begin{eqnarray}
&&\omega[8_S,8_S]\,=\,\omega^{BFKL}\,, \ \ \ \ \  \ \ \ \ \  \ \ \ \ \  \  \ \ \ \ \ \  \ \ \ \ \
\omega[10+\overline{10},10+\overline{10}]\,=\,2\, \omega^{BFKL}\,, \nonumber\\  \nonumber\\
&&\omega[{\cal R}_7,{\cal R}_7]\,=\,2\,\left(1\,-\,{1\over N}\right)\,\,\omega^{BFKL}\,,
 \ \ \ \ \ \ \ \omega[27,27]\,=\,2\,\left(1\,+\,{1\over N}\right)\,\,\omega^{BFKL}\,.
\end{eqnarray}

\begin{figure}[htbp]
\centerline{\epsfig{file=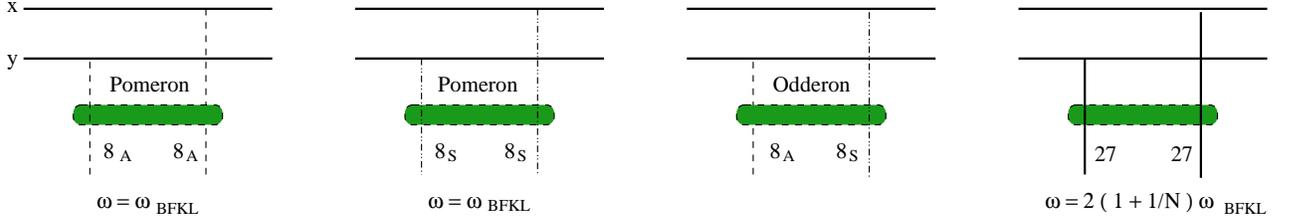,width=168mm}}
\caption{\it $t$-channel bound states of reggeized gluons.}
\label{fig6}
\end{figure}
The Pomeron that would correspond to exchange of a pair of singlet representations has zero eigenvalue just like its "constituent"  $s$ - Reggeon. Thus the contribution due to this exchange to the cross section is flat with energy.

Note that the eigenvalues in the $10\,+\,\overline{10}$, 27, and ${\cal R}_7$ channels are proportional to the second Casimir of the respective representations and are therefore greater than that of the BFKL Pomeron. The reason is that these exchanges require at least four gluon coupling in the $t$-channel as opposed to the BFKL Pomeron which in the Born approximation consists of two $t$-channel gluons. Thus the coupling of these Pomerons in perturbation theory is of order $\alpha_s^4$ and they have the same nature as the two BFKL Pomeron exchange. We will discuss this in more detail in the next subsection.

It is interesting to note that the eigenvalue of the bound state of two 27-Reggeons is higher than twice the BFKL value.
Although  four $t$ channel gluon states with intercept greater than two Pomerons have been discussed in the literature \cite{3P,LRS},
the correction we find is higher. It is of order $1/N_c$ and not $1/N_c^2$ as may be naively expected. The reason for this is that in the $N_c\rightarrow\infty$ limit the $27$ and ${\cal R}_7$ Reggeons are degenerate. In this situation the finite $N_c$ correction naturally starts with order $1/N_c$ as the $1/N_c$ perturbation theory is degenerate. This is also the reason why the $O(1/N_c)$ corrections to the
${\cal R}_7$ and $27$ states have opposite sign. A similar phenomenon is observed in the dependence of the $k$-string tensions on $N_c$ in Yang Mills theories where indeed the $1/N_c$ correction is also related with the value of the second Casimir - the so called Casimir scaling \cite{chris}.

Finally we note that the functions  $\Psi_t^i(x,y)$ are not required to vanish at $x=y$. Thus, unlike in the dipole model the solutions to the BFKL equation for $\Psi_t^i(x,y)$ are not limited to the Moebius invariant ones but include a larger set of solutions which are nonvanishing for the configuration where the transverse coordinates of two $s$ - channel gluons coincide. This will be important for the discussion of Section 5.

\subsubsection{Odderon and friends.}
In the $C$-parity odd sector \eq{chiG2} and
\eq{chiG3} lead to three decoupled equations for the functions $\Phi^+$, $\Phi^-$ and $\Psi^{-}_s$.
For the $\Phi^-$ component we have
\bea\label{od1}
&-&\int_u \left
[K_{u,x,y}\,\Phi^-(u,x)\,+\,K_{u,y,x}\,\Phi^-(y,u)\right]
\,+\,\int_v \left[K_{v,v,y}\, \Phi^-(x,v)\,
\,+\,K_{v,v,x}\,\Phi^-(v,y)\right ]
\nonumber \\ \nonumber && \\
&-&2\,\int_{z} \left[K_{x,y,z}\,-\,  K_{x,x,z}\right]
\Phi^-(x,y)\,=\,{\omega\over \,N}\,\Phi^-(y,x)
\eea
This is  the BFKL equation for the Odderon. The solutions are  antisymmetric BFKL eigenfunctions.
This solution is known as the BLV Odderon and was originally found in \cite{BLVod}.
This Odderon is a "bound state" of the reggeized gluon and the $d$-Reggeon.

For the signature even $\Phi^+$ component we obtain the very same equation but with the eigenvalues twice as big
\bea\label{od2}
&-&\int_u \left
[K_{u,x,y}\,\Phi^+(u,x)\,+\,K_{u,y,x}\,\Phi^+(y,u)\right]
\,+\,\int_v \left[K_{v,v,y}\, \Phi^+(x,v)\,
\,+\,K_{v,v,x}\,\Phi^+(v,y)\right ]
\nonumber \\ \nonumber && \\
&-&2\,\int_{z} \left[K_{x,y,z}\,-\,  K_{x,x,z}\right]
\Phi^+(x,y)\,=\,{\omega\over 2\,N}\,\Phi^+(y,x)
\eea
In the $t$-channel the color tensor structure of this eigenstate is that of $[P^{10}-P^{\overline{10}}]$. Its interpretation therefore is that of the bound state of the (degenerate) $10$ and $\overline{10}$ Reggeons which is antisymmetric both in color and coordinate spaces.
The eigenvalues for eigenstates on this Odderon trajectory have twice the value of the corresponding BLV eigenstates. Since for the BLV Odderon  $\omega\le 0$, the exchange due to this second Odderon is always suppressed. As we will show in the next subsection its coupling in the perturbative regime is also suppressed by $\alpha_s^3$ relative to the BLV Odderon, since this solution corresponds to the $t$-channel state containing at least six gluons.

Surprisingly we find that the $P$ even Odderon solution, $\Psi_s^-$, obeys the ordinary BFKL equation:
\bea\label{od3}
&-& \int_{z}\left[ 2\,K_{x,y,z}\,\Psi_s^-(x,y)\,-\,  K_{x,z,y}\,\Psi_s^-(x,z)
\,-\,   K_{z,y,x}\Psi_s^-(y,z) \right. \nonumber \\  && \nonumber \\
 && \left. \ \ \ \ \ \ \ \ \ \ + \ \
2\,\delta(x-y)\,\,\int_u K_{z,u,x}\,\Psi_s^-(z,u)\right]
 \\  && \nonumber \\
&-&
\int_{z}\,\left[2\,K_{x,x,z}
 \,\Psi_s^-(x,y)\,-\,  K_{z,z,y}\,\Psi_s^-(x,z)
\,-\,   K_{z,z,x}\,\Psi_s^-(y,z)\right] \,=\,{\omega\over N}\,\Psi_s^-(x,y) \nonumber
\eea
Since wave function is symmetric under the interchange of the coordinates the eigenvalues corresponding to this solution lie on the very same trajectory as the BFKL Pomeron $\omega^{BFKL}$. This Odderon grows with energy and dominates over all previously known $C$ odd exchanges.
We will comment on its couplings in the next subsection.

\subsection{Scattering of a gluonic dipole}
In this subsection we comment on the relation of the  eigenstates discussed above to the cross section of the scattering of an $s$ - channel gluonic dipole.
The basic relation is very similar to that of the dipole model. Consider the evolution of the scattering amplitude of a single gluonic dipole.
The projectile wavefunction at initial rapidity eq.(\ref{psi1}) consists of two gluons at transverse coordinates
$x,\,y$ in a
color singlet state. The projectile averaged $s$-matrix $\Sigma$ eq.(\ref{sigmas}) has the form
\beq
\Sigma[R]\,=\,{1\over N^2-1}tr[R(x)\,R^\dagger(y)]\,=\,P^{1\,\alpha\beta}_{\,\,\,\,\gamma\delta}\,R^{\alpha\gamma}(x)\,R^{\beta\delta}(y)
\,=\,{1\over N^2\,-\,1} \,\sum_i\,P^{i\,\alpha\gamma}_{\,\,\,\,\beta\delta}\,R^{\alpha\gamma}(x)\,R^{\beta\delta}(y)
\eeq
The latter form allows us to express this wave function in terms of the two particle eigenstates of $H_0$.
We use the completeness relation \eq{EE} to write:
\begin{eqnarray}\label{RRE}
P^{i\,\alpha\gamma}_{\,\,\,\,\beta\delta}\, R^{\alpha\gamma}(x)\, R^{\beta\delta}(y)\,&=&\,
\sum_{n=-\infty}^\infty \, \int_{-\infty}^\infty d \nu \, \int d^2
\rho \, \frac{ \left(\nu^2 + \frac{n^2}{4}\right)}{\pi^4\,|x-y|^4 } \, G_{n,\nu,\rho}^{+\,i}\, \,
E^{n, \nu \, *} ( x-\rho,  y-\rho)\nonumber\\ && \nonumber\\
&+&\delta^{i1}\,\left[{\rm tr}R(x)\,+\,{\rm tr}R(y)\,-\,(N^2-1)\right]
\end{eqnarray}
\eq{RRE} has the structure of \eq{bas}. Let us define  $\gamma_q$ as
$$\gamma_q(x,y)\,=\,{1\over N^2\,-\,1} \,\frac{ \left(\nu^2 + \frac{n^2}{4}\right)}{\pi^4\,|x-y|^4 } \,E^{n, \nu \, *} ( x-\rho,  y-\rho)
$$
The evolved amplitude of the scattering  of the dipole off a target specified by a wave function $W^T$
at rapidity $Y$ reads
\begin{eqnarray}\label{gd}
{\cal S}(Y)\,&=&\,\sum_i  \sum_q \gamma_q \,e^{\omega_i\,(Y\,-\,Y_0)}\,
\int\, DS\,\, W^T_{Y_0}[S]\,\,G_{\lambda}^{+\,i}[S]\nonumber\\
&+&\int\, DS\,\, W^T_{Y_0}[S]{1\over N^2-1}\left[{\rm tr}S(x)\,+\,{\rm tr}S(y)\,-\,(N^2-1)\right]
\end{eqnarray}
where the sum over $q$ denotes summation over $n$ and integrations over $\nu$ and $\rho$.
Just like in the dipole model this expression violates unitarity. Even if we arrange $\Sigma_{Y_0}[S]$ so that all coefficients satisfy eq.(\ref{sigmas}) at the initial rapidity $Y_0$, this property is violated at higher rapidities. Some "probabilities" in eq.(\ref{gd}) grow beyond unity while some become negative. The overall norm of the state remains normalized, $\Sigma_Y[1]=1$.

The situation is in fact even more complicated, since eq.(\ref{RRE}) notwithstanding, gluonic dipole couples also to higher multi Reggeon states. Thus for example the gluonic dipole will also couple to "Pomeron" containing three reggeized gluons. We discuss this in detail in the next section.
These problems are endemic to any approximation that treats the interaction term in the RFT Hamiltonian perturbatively and can only be solved beyond this perturbative expansion.

Setting these problems aside for the moment we would like to ask a very specific question about the minimal $t$-channel gluon content of the Pomerons we have found in this section. To this end, we expand the expression for $\cal S$
in perturbative series. In our approach, the coupling constant enters explicitly only in the phase of the matrix $S$:
\beq\label{sexp}
S^{\alpha\gamma}(x)\,=\,1\,+\,i\,T^a_{\alpha\gamma}\,\alpha^a(x)\,-\,{1\over 2}\,(T^a\,T^b)_{\alpha\gamma}\,
\alpha^a(x)\,\alpha^b(x)\,+\,...
\eeq
In perturbation theory (for small targets) the magnitude of the field $\alpha$ is proportional to $\alpha_s$.  Thus to find the leading perturbative content of each Pomeron we have to expand the appropriate eigenstate $G^{+\,(++)}[S]$ to the lowest order in $\alpha$. To order $\alpha_s^2$ only the BFKL Pomeron eigenstate contributes
\beq
P^{8_A\,\alpha\gamma}_{\,\,\,\,\,\,\,\beta\delta}\,S^{\alpha\gamma}(x)\,S^{\beta\delta}(y)\,=\,N\,\alpha^a(x)\,\alpha^a(y)\,.
\eeq
To this order the scattering amplitude (we omit the terms which do not evolve with energy) is:
\beq\label{gd1}
{{\cal S}_{\alpha_s^2}}(Y)\,=\,N^2\,  \,e^{\omega^{BFKL}\,(Y\,-\,Y_0)}\,\sum_q \gamma_q(x,y)\,
\int_{u,v} \langle \alpha^a(u)\,\alpha^a(v)\rangle_T \,E^{n, \nu } ( u-\rho,  v-\rho)
\eeq
where the target average is defined as
$$
\langle {\cal O}\rangle_T\,\equiv\,\int\, DS\,\, W^T_{Y_0}[S]\,\cal O[S]
$$
In order $\alpha_s^4$ we get contributions also from all symmetric representations
\beq\label{alpha4}
P^{i\,\alpha\gamma}_{\,\,\,\,\beta\delta}\,S^{\alpha\gamma}(x)\,S^{\beta\delta}(y)\,=\,
{N^2\,\lambda_i^2\over 4}P^{i\,ab}_{\,\,\,\,cd}  \,\alpha^a(x)\,\alpha^b(x)\,\alpha^c(y)\,\alpha^d(y),\ \ \ \ \ \ \ \ \ \ \ i\,=\,8_S, \ 27, \ {\cal R}_7
\eeq
Those come from expanding each factor of $S$ to second order in $\alpha$. Note that there is also $O(\alpha_s^4)$ correction to the nonlinear relation between $\alpha(x)$ and  $\rho(x)$ according to eq.(\ref{alpha}). To this order this contributes only to the reggeized gluon exchange and therefore only to the BFKL Pomeron. The $O(\alpha_s^4)$ contribution to the $S$-matrix of the gluonic dipole is
\bea\label{gd2}
{{\cal S}_{\alpha_s^4}}(Y)&=&N^2\, \sum_i{\lambda_i^2\over 4}
 \,e^{2\,(1-\lambda_i)\,\omega^{BFKL}\,(Y\,-\,Y_0)}\,\sum_q
\gamma_q(x,y)\nonumber \\
&\times&P^{i\,ab}_{\,\,\,\,cd}\,\int_{u,v}\,\langle \alpha^a(u)\,\alpha^b(u)\,\alpha^c(v)\,\alpha^d(v)\rangle_T
\,\,E^{n, \nu } ( u-\rho,  v-\rho)
\eea
The dominant exponent at the four $t$ - channel gluon level comes from $i=27$ unless the target has a very particular structure which suppresses its contribution through the $S$-integral in eq.(\ref{gd2}).
Note that for a projectile containing four gluons in the $s$-channel at order $O(\alpha^4_S)$ there is also a contribution  due to the double Pomeron exchange.
Compared to the double Pomeron exchange, the residue of the $[27,\ 27]$ exchange is suppressed by $1/N^2$.

It is also interesting that the $[10+\overline{10},10+\overline{10}]$ Pomeron does not couple to the gluon dipole at $O(\alpha_s^4)$. This state contains at least six $t$ channel gluons and has a further suppression in its coupling strength to dilute targets.

Let us now consider a projectile which couples to a negative charge conjugation exchange such as the Odderon.
Note that a  charge conjugation odd exchange does not contribute directly to the forward scattering of any projectile which is an eigenstate of $C$-parity. The physical states in pure Yang-Mills theory are indeed eigenstates of the $C$-parity since they are color singlets due to the confinement of color. In perturbation theory however the color is not confined and one is in principle free to consider projectiles which are not color singlets. Such projectile wave functions also do not have to be $C$-parity eigenstates. In particular one can consider a gluonic octet dipole which is a superposition of the $8_A$ and $8_S$ representations\footnote{Note that within perturbation theory it is perfectly legitimate to consider projectiles which are not color singlets, even though the cross section of scattering of two nonsinglet objects is infrared divergent. The divergence appears as the result of the integration over the impact parameter, while the scattering probability at fixed imp!
 act parameter, which is the focus of our discussion, is finite for colored as well as colorless objects.}. The Odderon does contribute to the forward scattering amplitude of such a state. Another example of such a state is the quark dipole. Although it is a color singlet, the dipole with a fixed transverse position of a quark and an antiquark is not an eigenstate of $C$-parity, and the Odderon was found to contribute to its forward scattering amplitude in\cite{KSW},\cite{oderon}.
>From our perspective considering such states is merely a convenient trick to make explicit what is the $t$-channel gluon content of Odderon states.

Consider therefore the "state"
\beq
\Sigma[R]\,=\,Z^{+\,\alpha\beta}_{\,\,\,\,\gamma\delta}\,R^{\alpha\gamma}(x)\,R^{\beta\delta}(y)
\eeq
The lowest order perturbative expansion starts with the term of order $\alpha^3$ where two $t$ channel gluons couple to one leg of the dipole and the third one to the second leg. We find
\beq\label{zz+}
Z^{+\,\alpha\beta}_{\,\,\,\,\gamma\delta}\,S^{\alpha\gamma}(x)\,S^{\beta\delta}(y)=
{N^2\over 4}\,d^{abc}\,[\alpha^a(x)\,\alpha^b(x)\,\alpha^c(y)\,-\,
\alpha^a(y)\,\alpha^b(y)\,\alpha^c(x)]
\eeq
The $S$-matrix to this order is
\beq\label{gd3}
{{\cal S}_{\alpha_s^3}}(Y)\,=\,{N^2\over 4}\sum_{q} \gamma_{q}(x,y)
e^{\omega^{odd}\,(Y\,-\,Y_0)}
d^{abc}\int_{u,v}\langle\alpha^a(u)\alpha^b(u)\alpha^c(v)-
\alpha^a(v)\alpha^b(v)\alpha^c(u)\rangle_T\,\Phi^{n, \nu } ( u-\rho,  v-\rho)
\eeq
Where $\Phi$ are antisymmetric solutions of the BFKL equation, see \cite{BLVod}. Thus the  $Z^+$ exchange indeed behaves as the usual (BLV) Odderon with intercept
$1+\omega^{odd}(q=0)$.

For the $Z^-$ state we can simply refer to the derivation in the Pomeron sector, since as mentioned above it has the color structure of $[P^{10}-P^{\overline{10}}]$ in the $t$-channel. Therefore just like the $[10+\overline{10}, 10+\overline{10}]$ Pomeron it contains at least six $t$-channel gluons.
We have checked explicitly that the coupling of this state at the six gluon level does not vanish.

Finally, the $P$ even Odderon solution which arose in our calculation couples to the "state"
\beq
\Sigma[R]\,=\,[P^{10}-P^{\overline{10}}]^{\alpha\beta}_{\gamma\delta}\,R^{\alpha\gamma}(x)\,R^{\beta\delta}(y)\,=\,
Z^{-\,\alpha\gamma}_{\,\,\,\,\beta\delta}\,R^{\alpha\gamma}(x)\,R^{\beta\delta}(y)
\eeq
Expanding to the lowest order in $\alpha$ we find
\beq\label{zz-}
Z^{-\,\alpha\gamma}_{\,\,\,\,\beta\delta}\,S^{\alpha\gamma}(x)\,S^{\beta\delta}(y)\,=\,
{N^2\over 4}\,d^{abc}\,[\alpha^a(x)\,\alpha^b(x)\,\alpha^c(y)\,+\,
\alpha^a(y)\,\alpha^b(y)\,\alpha^c(x)]
\eeq
The $S$-matrix to this gluon order is
\bea\label{gd4}
{{\cal S}_{\alpha_s^3}}(Y)&=&{N^2\over 4}\sum_{q} \gamma_{q}(x,y)\,
e^{\omega^{BFKL}\,(Y\,-\,Y_0)} \nonumber  \\
&&\times\,d^{abc}\,\int_{u,v}\langle\alpha^a(u)\,\alpha^b(u)\,\alpha^c(v)\,+\,
\alpha^a(v)\,\alpha^b(v)\,\alpha^c(u)\rangle_T\,E^{n, \nu } ( u-\rho,  v-\rho)
\eea
Thus the minimal content of this Odderon solution is also three gluons and it's coupling is not suppressed by powers of $1/N^2$ relative to the BLV Odderon.
It may seem odd that this solution as opposed to the BLV Odderon has not been discussed in the framework of the scattering matrix of a fundamental dipole.
The reason is that the fundamental dipole does not couple to this solution at any order. The fundamental dipole state is
\beq
D_F(x,y)={1\over N}{\rm tr} [R^\dagger_F(x)R_F(y)]
\eeq
The $C$-parity transforms it into
\beq
C^\dagger D_F(x,y)C={1\over N}{\rm tr} [R^\dagger_F(y)R_F(x)]
\eeq
Thus in order to couple to this state  a $C$-odd exchange must also be $P$-odd. This excludes the $G^{-\{-+\}}$ Odderon. This is also the reason why this Odderon does not appear among the dipole model eigenstates discussed in Section 3 even though it is formally not suppressed by $1/N$.
It would be interesting to perform a similar analysis for more general physical states in order to understand whether there are any physical states that can couple to this Odderon.

\subsection{The perturbation.}

The first unitarity corrections in perturbative RFT originate from applying
the perturbation $H_I$ to a generic $t$ - channel exchange in the two particle sector described by the wavefunction $\Omega^\lambda_{gg}$:
\beq
|gg\rangle\,=\,\int_{u,v}\tilde R_u^{\alpha\beta}\,R_v^{\gamma\delta}\,\Omega^\lambda_{gg}(u,v)\,|0\rangle
\eeq
For color singlet $t$-channel exchanges the state $|gg\rangle$ can be decomposed into components along the  Pomeron and Odderon directions.
More generally an arbitrary $t$ channel exchange $|gg\rangle$ does not have to be a color singlet and therefore has projections onto non singlet $t$-channel exchanges which we have not studied in this paper.
Quite generally however,  the perturbation $H_I$ when acting on the state in a two gluon sector $|gg\rangle$ creates a state in the four gluon sector $|gggg\rangle $:
\bea\label{RT}
|gggg\rangle\,\equiv\,H_I\,|gg\rangle&=&\hat K_{x,y,z} \,\left[
T^a\,(\tilde R_x\,-\,\tilde R_z)\right]_{\alpha\beta}\,\tilde R^{am}_z\,
\tilde R^{bm}_z\,
\left[(\tilde R_y\,-\,\tilde R_z)\,T^b\,\right]_{\gamma\delta}  \,\Omega_{gg}(x,y) |0\rangle\nonumber \\ &&\nonumber \\
&+&\hat K_{x,y,z} \,\left[
T^a\,(\tilde R_x\,-\,\tilde R_z)\right]_{\gamma\delta}\,\tilde R^{am}_z\,
\tilde R^{bm}_z\,
\left[(\tilde R_y\,-\,\tilde R_z)\,T^b\,\right]_{\alpha\delta}  \,\Omega_{gg}(y,x) |0\rangle \nonumber \\ &&\nonumber \\
&-&\int_{x,y,z} K_{x,x,z} \,\tilde R^{\alpha\beta}_y\,\tilde R^{ab}_z\,
\left[T^a\,(\tilde R_x\,-\,\tilde R_z)\,T^b\,\right]_{\gamma\delta} \,\Omega_{gg}(y,x) |0\rangle \nonumber \\ &&\nonumber \\
&-&\int_{x,y,z} K_{y,y,z} \,\tilde R^{\gamma\delta}_x\,\tilde R^{ab}_z\,
\left[T^a\,(\tilde R_y\,-\,\tilde R_z)\,T^b\,\right]_{\alpha\beta} \,\Omega_{gg}(y,x) |0\rangle
\eea
\eq{RT} is the most general form of the transition vertex from two Reggeons into four. To fully analyze the vertex in terms of the four Reggeon states we would need first to find spectrum of $H_0$ in the four particle sector.
This analysis is much more complicated and must involve solution of the BKP equation in the four gluon sector. It is beyond the scope of the present paper.
We expect however that the proper analysis of the transition vertex should contain, among other things the following elements. The  "matrix element" of $H_I$ between the eigenstate $|gg\rangle$ projected onto a single Pomeron (bound state of two reggeized gluons discussed above) and an eigenstate $|gggg\rangle$ projected onto two Pomerons, should reduce to the
triple Pomeron vertex of Refs. \cite{3P,3P1}. A generic singlet eigenstate $|gggg\rangle$ is expected to have projections onto two Odderons
as well as two Pomerons. Thus the vertex $P\rightarrow OO$ of Ref. \cite{BE} is also expected to appear in this calculation. If the initial state $|gg\rangle$ is projected onto Odderon, we expect to find the vertex
$O\rightarrow OP$\cite{KSW}. This study is postponed until later date.

\section{Partonic approximation versus BFKL/BKP  }

Only a small number of the multiple "Reggeons", "Pomerons" and "Odderons" that we have found in the previous section appear in the standard BFKL/BKP approach. It is therefore important to understand the relation of our approximation to BFKL and what additional physics, if any it contains.

To make this connection, first let us formulate the BFKL approximation in terms of the approximation to the KLWMIJ Hamiltonian. In the BFKL approximation every gluon which is emitted in the process of the evolution is assumed to scatter on the target only once, and the scattering probability is assumed to be a small parameter. This corresponds to approximating the unitary matrix $R$ and its functional derivative by
\begin{eqnarray}\label{expan}
R^{ab}(x)&\rightarrow& \delta^{ab}+T^{ab}_c\int dx^-{\delta\over\delta \rho^c(x, x^-)}\\
{\delta\over \delta R^{ab}(x)}&\rightarrow & T^{ab}_c\int dx^-{ \rho^c(x,x^-)} \nonumber
\end{eqnarray}
Denoting $\int dx^-\rho^a(x,x^-)$ by $\rho^a_x$, and dropping terms of order $[{\delta\over\delta \rho^c(x)}]^3$ in eq.(\ref{KL2}), the KLWMIJ Hamiltonian reduces to
\begin{equation}\label{hbfkl}
H^{BFKL}\,=\,\hat K_{x,y,z}\,(T^aT^b)_{cd}\,
\rho^a_x\,\left[{\delta\over \delta \rho^c_x}-{\delta\over \delta \rho^c_z}\right]\,
\left[{\delta\over \delta \rho^d_y}-{\delta\over \delta \rho^d_z}\right]\,\rho^b_y
\end{equation}

Note that the approximation eq.(\ref{expan}) by itself as an approximation to $H^{KLWMIJ}$ does not have an expansion parameter. It becomes an expansion in $\alpha_s$ only once we agree to consider the scattering of the projectile whose evolution we describe on dilute targets with colour fields of order $\alpha_T=O(\alpha_s)$. As discussed above, after averaging over the projectile wave function, factors of $\delta/\delta\rho$ turn into factors of the target field $\alpha$, which has to be averaged over the target wave function. For dilute targets therefore, every additional factor of $\delta/\delta\rho$ brings an extra power of $\alpha_s$.

We can now search for the eigenstates and the eigenvalues of this functional BFKL Hamiltonian
\footnote{We emphasize the word ``functional'' to distinguish this Hamiltonian from the commonly used BFKL Hamiltonian
which acts on space of functions of two transverse coordinates.  The  Hamiltonian eq.(\ref{hbfkl}) generates the
standard BFKL equation for the Pomeron in the two gluon sector, but also the  whole hierarchy of the BKP equations for the higher "Fock space" states - see below.}.
The calculation follows the same line as in the previous section except it is much simpler.
The eigenstates have the general form
\beq\label{eigenbfkl}
G^{BFKL}_{n,\, \lambda,\,  q}\,=\,
P^{ab...c}_\lambda\,\,\int_{x_1...,x_n}\,\Psi_{n,\, \lambda,\, q}(x_1,x_2,...,x_n)\,\,
{\delta\over \rho^a_{x_1}}\,{\delta\over \rho^b_{x_2}}...{\delta\over \rho^c_{x_n}}\,|0\rangle
\eeq
where $P^{ab...c}_\lambda$ projects the product of $n$ adjoint representations onto the representation $\lambda$ and the functions
$\Psi_{n,\, \lambda,\, q}(x_1,x_2,...,x_n)$
satisfying the $n$-gluon BKP equations with eigenvalues $\omega_{n,\, \lambda,\, q}$
\footnote{In the rest of this section all multi-gluon BKP states will be referred to as the eigenstates of the functional
BFKL Hamiltonian.
}.
The resulting eigenstates are those and only those which are present in the standard BFKL/BKP approximation.
In particular the only state in the one gluon sector ($n=1$) is the reggeized gluon. In the two gluon sector the only $C$-even singlet exchange is the BFKL Pomeron, and the only $C$-odd one is the BLV Odderon. The nonsinglet states again are only those that can be constructed from two reggeized gluons. There is no trace of the "Reggeon" states corresponding to other color representations we have discussed in the previous section, as well as of the additional Pomeron and Odderon states constructed from these Reggeons.

The next question one should address is, given the eigenfunctions of the functional BFKL Hamiltonian (BKP eigenstates)
how do we calculate the scattering amplitude of a physical state  $\Phi$. To do this, in principle one has to expand this state in the basis of eigenstates and evolve each eigenstate with its own exponential factor
\beq\label{evol}
\Phi_Y\,=\,\sum_{n,\, \lambda,\, q}e^{\omega_{n,\, \lambda,\,q}Y}\,\,\gamma_{n,\, \lambda,\, q}\,\,G^{BFKL}_{n,\, \lambda,\,  q}
\eeq
with
\beq\label{evol1}
\gamma_{n,\, \lambda,\,  q}\,=\,\langle G^{BFKL}_{n,\, \lambda,\, q}|\Phi_{Y=0}\rangle
\eeq
This however presupposes that the eigenstates of the BFKL Hamiltonian form a complete basis of normalizable states. This assumption is problematic to say the least, as the states $G^{BFKL}_{n,\lambda,q}$ are not normalizable. The reason for this is easy to understand. The expansion of $R$ to first order in Taylor series in eq.(\ref{expan}) is equivalent to Taylor expansion of the eigenfunctions to the lowest order in $\delta/\delta\rho$\footnote{We do not mean to imply that this is expansion of {\it exact} eigenfunctions which correspond to {\it exact} eigenvalues of $H^{KLWMIJ}$. In particular the spectrum of $H^{BFKL}$ is certainly not the same as that of $H^{KLWMIJ}$, as the latter operator is positive definite, while some of the BFKL eigenvalues are negative. Thus the effect of approximation eq.(\ref{expan}) is not only to expand the eigenfunctions, but also to distort the spectrum of $H^{KLWMIJ}$.}. Thus the eigenfunctions in eq.(\ref{eigenbfkl}) being homogeneous polynomials are 
clearly not normalizable.

One possible strategy to avoid this problem would be to expand the wavefunction of a physical state at initial rapidity in Taylor series in $\delta/\delta\rho$ and algebraically match the coefficients to the BKP solutions.
For example for a single $s$ - channel gluon state we have
\footnote{As discussed in \cite{kl4} when the operator $R$ acts on a symmetric function of $\rho^a(x,x^-)$, the path ordering in its definition is unimportant. We use this fact in writing eq.(\ref{physex}).}
\beq \label{R1}
R^{ab}(x)|0\rangle\,=\,\left(\delta^{ab}\,+\,T^{ab}_e{\delta\over \delta\rho^e(x)}\,+\,{1\over 2}(T_eT_f)^{ab}{\delta\over \delta\rho^e(x)}
{\delta\over \delta\rho^f(x)}\,+\,\cdots \right)
\eeq
Then the coefficient of each power of $\delta\over \delta\rho$ is expanded in the BKP eigenfunctions
$\Psi_{n,\lambda,q}$.
For example the coefficient of the linear term is given in term of the eigenfunctions of the reggeized gluon,
 which are simply plane waves,
\beq
{\delta\over \delta\rho^a(x)}\,\,=\,\,\int d^2q \,e^{i\,q\,x}\,\,G_{1,\,a,\,q}\, \, .
\eeq
The coefficient of the quadratic term can be expanded in the BFKL  eigenstates of eq.(\ref{enn}) analogously to eq.
(\ref{RRE}).  Clearly only Moebius noninvariant  solutions of the BFKL equation  contribute to this expansion. The third order terms in $\delta/\delta\rho$ can be "projected" onto three gluon BKP states and so on.

 Similarly to the section 4.2, we can use  $\eta^\lambda_{ab}$ in order to project
the state eq.(\ref{R1}) onto a definite color representation in the $t$-channel $\eta^\lambda_{ab}R^{ab}$. Only  $\lambda=8_A$ contributes to the expansion of the linear term while all the other representations
contribute to the quadratic or higher terms only. The quadratic term has the structure of a "bound state" of two reggeized gluons at the same point in transverse plane projected onto a given representation $\lambda$. Note that the BFKL equation for the channels  $\lambda=1$ and $\lambda=27\,({\cal R}_7)$
leads to a cut and not to a simple pole in the Mellin representation.
This is different from the contribution of the `exotic'' colored Reggeons found in section 4.2 which lead to a pole structure. We will discuss the nature of the difference between these contributions later.

Analogously to \eq{R1}, for the two $s$ - channel gluon state we would write
\begin{eqnarray}\label{physex}
&&R^{ab}(x)\,R^{cd}(y)|0\rangle\,=\,
\left(\delta^{ab}\delta^{cd}+T^{ab}_e{\delta\over \delta\rho^e(x)}\delta^{cd}+
T^{cd}_e{\delta\over \delta\rho^e(y)}\delta^{ab}+\right.\\
&&\ \ \ \ \left.
+{1\over 2}(T_eT_f)^{ab}{\delta\over \delta\rho^e(x)}{\delta\over \delta\rho^f(x)}\delta^{cd}+{1\over 2}(T_eT_f)^{cd}{\delta\over \delta\rho^e(y)}{\delta\over \delta\rho^f(y)}\delta^{ab}+T^{ab}_e{\delta\over \delta\rho^e(x)}T^{cd}_f{\delta\over \delta\rho^f(y)}+\dots\right )|0\rangle\nonumber
\end{eqnarray}
Identifying each term in the expansion with a linear combination of BKP states one obtains an expansion of the form
\beq\label{who}
R^{ab}(x)\,R^{cd}(y)|0\rangle\,=\,\sum_{n,\lambda,q}\tilde\gamma_{n,\,\lambda,\,q}\,\,G^{BFKL}_{n,\,\lambda,\,q}
\eeq
whose evolution in rapidity is then given by eq.(\ref{evol}). Note that even though some of the individual eigenvalues of the BKP hierarchy are positive, the scattering amplitude calculated in this way is expected to be unitary due to eikonalization of the multigluon exchanges inherent in the exponential form of $R$.

Strictly speaking keeping higher order terms in the expansions eq.(\ref{R1},\ref{physex}) is beyond the accuracy of the approximation
eq.(\ref{expan}). Within this approximation only the lowest term in the expansion of the projectile wave function can be kept.
More precisely one should keep the term which becomes leading after averaging over the target wave function. Thus for example, if the target is assumed to have only color singlet correlators (as any physical hadron should), the leading term is quadratic  in $\delta/\delta\rho$ for {\it any} projectile wave function. This term describes single BFKL Pomeron exchange. Contributions of all higher BKP states after averaging over the wave function of a dilute target are suppressed by powers of $\alpha_s$.

Even if one decides to keep the higher order terms it is important to realize that the procedure sketched in eqs.(\ref{physex},\ref{who}) is by no means unique. The reason is that the BKP states (\ref{eigenbfkl})
are only determined to lowest order in Taylor expansion. Thus for example adding higher order terms to the eigenfunctions is no less accurate than keeping higher order terms in the expansion eq.(\ref{physex}). On the positive side, such a modification can make eigenfunctions normalizable. In particular consider the following natural modification: in all the eigenstates of eq.(\ref{eigenbfkl}) make the substitution
\beq\label{sub}
{\delta\over \delta\rho^a}\,\rightarrow \,-\,{1\over N}\,{\rm Tr}(T^a\,R)
\eeq
This modification does not change the leading order terms, but now since the matrix $R$ is unitary, the modified states are normalizable.
With these modified states one could go back to the original prescription and calculate the evolution according to eqs.(\ref{evol},\ref{evol1}). This unfortunately is not quite kosher either, since although the modified states are normalizable, they are not necessarily orthogonal to each other even if they correspond to different eigenvalues. In particular a state
$G_{n,\lambda,q}^{BFKL}$ has a nonvanishing overlap with a state $G_{n+1,\lambda,q}^{BFKL}$ if the wave function $\Psi_{n+1}(x_1,...,x_{n+1})$ does not vanish when two of the coordinates coincide, $x_i=x_j$ and the overall color representations of the two states are the same. This is again clearly the result of the ambiguity of the higher order terms in $\delta/\delta\rho$ which is not parametrically suppressed within the KLWMIJ framework proper.

This discussion is of course not original. We presented it to demonstrate that just like in the standard Feynman diagram approach, in the framework of the BFKL truncation of the KLWMIJ Hamiltonian only the one Pomeron exchange constitutes the proper weak coupling expansion, but also to stress that the powers of the coupling constant enter only through the procedure of averaging over the target. As long as we do not commit to scatter the projectile on a dilute target, that is within the KLWMIJ Hamiltonian proper there is no parameter which makes the BFKL truncation leading in any sense. The one Pomeron exchange approximation of course violates unitarity. The unitarity can be restored by summing contributions of higher BKP states, but this summation is ambiguous. In order to properly unitarize BFKL approximation truncation of the scattering matrices $R$ in the Hamiltonian itself is not allowed.

The truncation eq.(\ref{expan}) of the KLWMIJ Hamiltonian has two consequences, which are physically quite distinct. The first one is that the number of gluons in the wave function does not change throughout the evolution. This is obvious since each factor of $\delta/\delta\rho$ is associated with one gluon, and the BFKL Hamiltonian is homogeneous in $\delta/\delta\rho$. As a result the eigenstates eq.(\ref{eigenbfkl}) have a fixed number of gluons. Second, each gluon in the wave function is only allowed to scatter on the target once. Thus the single gluon scattering matrix $1-R$
in this approximation is always in the adjoint representation in the $t$ - channel, since only single gluon exchange between any projectile gluon and the target is allowed.

The complete treatment of KLWMIJ Hamiltonian will therefore bring two types of corrections to the BFKL approximation. The first one is due to the possibility of the gluons of the projectile to multiply scatter on the target. This in particular allows processes whereby a single gluon of the projectile exchanges color representations other than just the adjoint with the target. The second type of correction involves what one can call the generalized triple Pomeron vertex. It takes properly into account the fact that the number of gluons in the projectile wave function does not stay fixed through the evolution.

>From this perspective the partonic approximation discussed in the previous section goes beyond the BFKL approximation in that it allows multiple scattering of the projectile gluons.
It can be checked directly that if we restrict the matrix $\tilde R=1-R$ within the Hamiltonian eqs.(\ref{h0r},\ref{h0v}) to have only adjoint component:
\beq
\tilde R^{ab}\,\rightarrow \,T_c^{ab}\,{\delta\over \delta \bar\rho^c}
\eeq
the partonic approximation Hamiltonian eq.(\ref{h0}) reduces directly to the BFKL Hamiltonian eq.(\ref{hbfkl}).
Thus clearly the partonic approximation contains all the BKP eigenstates with the same eigenvalues as the standard calculation. Their interpretation is just like in the BFKL approximation - "bound states" of reggeized gluons. The only difference is that now the amplitude
$\delta\over \delta \bar\rho^c$ contains in principle not only single gluon exchange but also multiple exchanges with net color in
the adjoint representation.
The other Reggeon states in the parton approximation all correspond to $t$-channel exchanges in higher color representation and their "bound states" which all can occur only if each projectile gluon is allowed to multiply scatter on the target. As explained above these exchanges are not accounted for in the BFKL approximation, but instead are additional contributions which still preserve the total number of $s$ - channel gluons in the projectile.

To put things into perspective, the accuracy of including the extra Reggeon contributions into the calculation of the scattering amplitude is no better or worse than that of the higher BKP states. To begin with we have treated the variable $R$ in the previous section as a linear variable rather than a unitary one, and so the states in eqs.(\ref{pome},\ref{odde})
are strictly speaking not normalizable. However in complete analogy with the treatment of the higher BKP states in eq.(\ref{sub}) we can treat $R$ as unitary when calculating the norm of the eigenstates. This procedure then makes the multi-Reggeon states normalizable,
 but non-orthogonal. The non-orthogonality is of very same nature as for the modified BFKL/BKP states discussed above. If a wavefunction $\Psi(x_1,...,x_{n+m})$ does not vanish when any $m$ transverse coordinates in a state $G_{m+n}$ coincide, then this state is not orthogonal to a state $G_n$ as long as the two states have the same color quantum numbers.
In this way the states which formally have $n$ powers of $R$ contribute to scattering amplitude of any state with $m\le n$ gluons. For example the Moebius noninvariant Pomeron state contributes to the scattering amplitude of a single $s$-channel gluon via a singlet $t$-channel exchange. The $t$ - channel exchange amplitude in the representation 27 will have a pole contribution from the 27 Reggeon state as well as a cut contribution from the two reggeized gluon state with total color representation 27 as well as from higher multi Reggeon states.

The question whether the extra Reggeons we discussed here and their "composites" give important contributions to scattering amplitudes can only be decided with the knowledge of the structure of the target. As we have noted above, the partonic approximation is identical to the BFKL approximation if the target is dilute so that only single scattering is allowed. However one can imagine a situation in which the target is not so dilute that multiple scatterings are important but still not dense so that the scattering amplitude of a single gluon is significantly smaller than unity. The smallness in this case will not be parametric in $\alpha_s$, but for example simply numerical. In that case the other Reggeons are a priory as important as the reggeized gluon and their "composites" are no less important than the BKP states.

To summarize this discussion, the Reggeon states discussed in the previous section are complementary to the higher BKP states. Both take into account multiple scattering corrections, but the corrections are of different kinds. The BKP states encode corrections due to scattering of multiple gluons of the projectile where each gluons scatters on the target only once. The extra Reggeons and their composites on the other hand encode unitarization corrections due to processes where each projectile gluon scatters on the target many times.
In order to treat unitarization consistently in addition to those two types of corrections one also has to take into account the non-conservation of the $s$ - channel gluon number due to the interaction term in the KLWMIJ Hamiltonian. We note an interesting point that some of the additional Reggeon composites we have discussed above lead to faster growth of the scattering amplitude with energy than the corresponding BKP states that contribute nominally in the same  order in $\alpha_s$. For example as discussed above the $[27,27]$ exchange grows significantly faster than the known four gluon BKP states (its overall amplitude is however suppressed by the factor $1/N^2$).
This may be an indication that the extra Reggeons and their composites are no less important for unitarization than the standard BKP states.

\section{Discussion}

Our goal in this paper was to relate the language of the
KLWMIJ/JIMWLK evolution to that of the Reggeon field theory. We have
shown that the KLWMIJ (or equivalently JIMWLK) Hamiltonian should be
understood directly as the second quantized Hamiltonian of the
Reggeon field theory. We have described precise correspondence
between many elements of the standard QCD Reggeon theory approach
and the calculation of QCD scattering amplitudes using JIMWLK
evolution. We note that our discussion is not in terms of cut
amplitudes - the objects of standard use in much of Reggeon field
theory literature. Filling this missing link is an interesting
question which should be addressed in future work.

 The quantum field degree of freedom
in the Reggeon field theory discussed here is the $SU(N)$ matrix
valued field which directly corresponds to the eikonal scattering
matrix of a single gluon. The fact that the basic degree of freedom
is a local field in the transverse space makes the theory similar to
the original pre-QCD Reggeon field theory of Gribov analyzed in
\cite{Amati} (and also recently in \cite{BF}). In this respect it is
simpler than the Pomeron Hamiltonian considered in Section 3 (see
also \cite{braunlast}) based on the dipole model, whose basic degree
of freedom is the bilocal dipole field. It is however much more
complicated than the Gribov theory in two aspects. First, the basic
field is a matrix, and thus the number of degrees of freedom is
significantly greater. And second, the interaction is nonlocal but
rather  decreases in the transverse plane only as $1/x^2$ which
leads to the violation of the Froisart bound \cite{urs}. Both these
features are related to the specific nature of the QCD interaction.
The proliferation of degrees of freedom is simply due to the fact
that the basic partons of any QCD state are gluons,
 and the nonlocality of the interaction is the direct consequence of the perturbative masslessness of the gluon.
One would hope that proper account of confinement can lead to an effective theory, valid at large transverse distances $|x|>1/\Lambda_{QCD}$ which operates directly only with scattering amplitudes of gauge invariant states and is local on these transverse distance scales. Such a theory should follow from the fundamental QCD RFT by operator product expansion. It is however unclear to us whether the number of degrees of freedom in such a theory should remain finite.
At any rate this question is far beyond the scope of the present research.

In this work we have considered perturbative approach in the framework of the QCD RFT and have found several eigenstates of the free Hamiltonian. In particular we have found all eigenstates in the one particle channel and color singlet states in the two particle states.

In the one particle sector we find the reggeized gluon as well as a set of additional Reggeons corresponding to the exchanges in the color representations $8_S, \ 10, \ \overline{10}, \ 27$ and ${\cal R}_7$. In the color singlet two particle sector we find states which are naturally interpreted as bound states of these Reggeons. In particular we find the standard BFKL Pomeron which is the "bound state" of two reggeized gluons, and also Pomerons that correspond to symmetric bound states of two $8_S$, two
$27$, two $10+\overline{10}$ and two ${\cal R}_7$ Reggeons. All these correspond to charge conjugation even exchanges in the $t$-channel and have positive signature.

An interesting observation here is that the $[27,27]$ Pomeron exchange grows faster with energy than two BFKL Pomerons.
Also the large $N_c$ correction to its intercept is $O(1/N_c)$ and not $O(1/N_c^2)$ as could be naively expected.
The reason for this large correction is that in the $N_c\rightarrow\infty$ limit the $[27,27]$ and
$[{\cal R}_7,{\cal R}_7]$ Pomerons are degenerate and so the $1/N_c$ expansion is actually a degenerate perturbation theory.

We also analyzed the $C$ parity odd sector. Here we find three sets of eigenstates. One is the signature odd bound state of the reggeized gluon and the $8_S$ Reggeon, which has all the properties of the BLV Odderon.
Another set of eigenstates can be thought of as an antisymmetric bound state of $10$ and $\overline{10}$ Reggeons. These eigenstates are charge conjugation odd and parity odd. They are quite distinct from the standard Odderon however in that they have even signature. The intercept of this trajectory is unity and so these exchanges lead to cross section that does not rise with energy.
Finally the third set of states in the Odderon sector is somewhat mysterious. The eigenstates of this set have negative signature but positive parity. Their wave functions are symmetric under the exchange of the transverse coordinates of constituents (the reggeized gluon and the $8_S$ Reggeon). The intercept of this trajectory is equal to that of the BFKL Pomeron.

We have analyzed the $t$ - channel gluon content of these eigenstates and have found that the BFKL Pomeron (unsurprisingly) contains at least two gluons, the $[8_S,8_S] \ [27,27]$ and $[R_7,R_7]$ states contain at least four gluons, the two signature odd Odderons contains at least $3$ gluons while both $[10,\overline{10}]$ bound states have at least $6$ gluons. Thus perturbatively the couplings of $[8_S,8_S], \ [27,27], \ [{\cal R}_7,{\cal R}_7]$
exchanges are suppressed by $\alpha_s^2$ and of the $10$ exchanges by $\alpha_s^4$ relative to the BFKL Pomeron.

>From the $s$-channel scattering point of view this means that these higher Pomeron exchanges are due to multiple scatterings. Thus for example in order to exchange the $[27,27]$ Pomeron each projectile gluon has to scatter at least twice from the target as is clear from the  derivation in eq.(\ref{alpha4}).
>From the point of view of the target this means that the target field must be dense enough so that at a given point in the transverse plane there is significant probability to find a multigluon state in the representation $27$ of the color group.
Such scattering is suppressed if the target is dilute, that is if the probability to find an extra gluon at a given point in transverse plane is $O(\alpha_s)$.
However in the high density regime this suppression disappears, since the probabilities to find an octet or a $27$ are equal when the average color charge density is large.

We have analyzed the relation of the spectrum we have found in partonic approximation to KLWMIJ with the standard BFKL/BKP approach. We have shown that all the BKP states do appear as eigenstates on the partonic Hamiltonian. The extra states we found are outside the BFKL/BKP framework, as multiple scatterings of a single projectile gluon are {\it a fortifiory} excluded from the BKP approximation. These are therefore physically distinct unitarization corrections. Since they are related to multiple scatterings of a single projectile gluon, it is natural to expect that they are suppressed by powers of $1/N$ relative to the corresponding BKP states. This is indeed what we find. For example the contribution of $[27,27]$ Pomeron which has an intercept significantly greater than any of the four gluon BKP states, has a $1/N^2$ suppression factor relative to the leading BKP state.

Next we discuss some issues related to the structure of the symmetries of the RFT.
As we have already remarked, the theory has a $SU_L(N)\otimes SU_R(N)$ symmetry as well as the discrete charge conjugation symmetry $R\rightarrow CRC$ and the signature symmetry $R\rightarrow R^\dagger$. Expansion around $R=1$ breaks spontaneously $SU_L(N)\otimes SU_L(N)$ down to $SU_V(N)$ while both discrete symmetries remain unbroken. The breaking of the symmetry leads to the appearance of the Goldstone boson, which is the reggeized gluon.
Since the intercept of the other Reggeons is also unity it suggests that the Hamiltonian may have higher symmetry. The most general transformation that one can define on the space of adjoint unitary matrices is $GL(N^2-1,R)$. This is obvious since $R$ can be written in terms of $N^2-1$ real parameters as
$R^{ab}=[\exp\{\xi^cT^c\}]^{ab}$ and the parameters $\xi^c$ can be transformed by an arbitrary real linear transformation. The $SU_L(N)\otimes SU_R(N)$ transformation is a small subgroup of this group. It would be interesting to check whether or not the KLWMIJ Hamiltonian is invariant under a larger subgroup of $GL(N^2-1,R)$ which is also broken by $R=1$ configuration.
There is also a possibility that the appearance of many massless particles has a similar origin as in supersymmetric theories \cite{susy}, where additional symmetry appears if one artificially extends the Hilbert space of the theory. We note that it has been previously suggested in Ref. \cite{BLVod} that the degeneracy between $f$- and $d$- trajectories may be related to a variant of supersymmetry \cite{LipatovDual}.

Another issue related to symmetries is the status of the conformal symmetry in the framework of the second quantized RFT. It is well known that the BFKL equation is conformally invariant on the Moebius invariant sector, that is on the set of functions $F(x,y)$ which vanish when $x=y$\cite{LipatovCI,BLW}. The nonlinear Kovchegov equation \cite{Kovchegov} is also conformally invariant.
In fact it is easy to see that the full KLWMIJ Hamiltonian is conformally invariant if we were to substitute the Weiszacker-Williams kernel in eq.(\ref{kernel}) by the dipole kernel
\begin{equation}\label{dif}
K_{x,y,z}\rightarrow M(x,y,z)\,=\,-\,\frac{\alpha_s}{4\pi^2}\, {(x-y)^2
\over (z-x)^2(z-y)^2}=K_{x,y,z}-\frac{\alpha_s}{4\pi^2}\, \left[{1
\over (z-x)^2}+{1
\over (z-y)^2}\right]
\end{equation}
Under the inversion transformation (with $x\equiv x_1+ix_2$ etc.)
\beq\label{rx}
R(x)\rightarrow R(1/x), \ \ \ {\delta\over \delta R(x)}\rightarrow (x^*x)^{-2} {\delta\over \delta R(1/x)}
\eeq
The dipole kernel can be written
as
\begin{equation}\label{rx1}
M(x,y,z)\,=\, (z^*z)^{-2}\,M(1/x,1/y,1/z)
\end{equation}
Changing the integration variables from $x,y,z$ to $1/x,1/y,1/z$ we see that the Hamiltonian eq.(\ref{KL}) with the dipole kernel is invariant under this transformation. The full conformal invariance follows from the dilatational and inversion invariance.

The KLWMIJ Hamiltonian eq.(\ref{KL}) is not invariant under inversion, as it can be easily checked that the property eq.(\ref{rx1}) is not shared by the Weiszacker-Williams kernel $K(x,y,z)$. However as has been extensively discussed in the literature (see \cite{BLV} or \cite{oderon}) if one limits oneself to consideration of color singlet impact factors and color singlet exchanges, the kernels $K$ and $M$ are interchangeable. Formally speaking, taking the difference between the corresponding KLWMIJ Hamiltonians one finds terms proportional to either
$$Q^a_L=\int d^2x tr\left[ {\delta\over \delta R^\dagger_x}\,T^a\,R_x\right]$$ or
$$Q^a_R=\int d^2x tr\left[R_x\,T^a\, {\delta\over \delta R^\dagger_x}\right]\ .$$ These are the generators of the $SU_L(N)$ and $SU_R(N)$ transformations respectively.
Thus when acting on states which are $SU_L(N)\otimes SU_R(N)$ invariant the two Hamiltonians are identical.

We therefore conclude that although the KLWMIJ Hamiltonian is not conformally invariant, part of its spectrum is.
Note however that the perturbative approach described in the present paper breaks spontaneously the  $SU_L(N)\otimes SU_R(N)$ symmetry. This type of perturbation theory would give different results for the two Hamiltonians. In particular the dipole variant of the KLWMIJ Hamiltonian has no virtual terms of the type eq.(\ref{h0v}). Its one particle spectrum is therefore trivial - any one particle state is its eigenstate with zero eigenvalue. On the other hand the spectrum of the Pomerons and Odderons discussed above remains unaffected. This may have to do with the fact that even though the perturbative expansion is not $SU_L(N)\otimes SU_R(N)$ invariant, some eigenstates nevertheless do have the full symmetry of the Hamiltonian.

We note that we have not made a serious attempt to formalize the approximation scheme employed in this paper. In particular it has two important elements. First we have split the RFT Hamiltonian into the free and the interaction parts according to eqs.(\ref{h0}) and (\ref{hi}). The idea behind it is to separate $H$ into a homogeneous and
an inhomogeneous part. The homogeneous part preserves the number of $s$-channel gluons throughout the evolution, while the inhomogeneous term increases this number.

However such a separation is not unique due to the fact that $R$ is a unitary matrix.
Our choice of the split was guided by two criteria: we required that the free and the interaction parts are separately ultraviolet and infrared finite. The infrared finiteness is straightforward, it simply asks not to emit new gluons far away from the existing gluons with probability greater than the square of the Weiszacker-Williams field. The origin of the requirement of the ultraviolet finiteness in physical terms is the following.
Calculating the cross sections using the homogeneous part of the Hamiltonian corresponds to the partonic approximation, which treats the scattering of the projectile gluons as completely independent. However physically it is clear that if two gluons in the wave function are separated by a distance which is smaller than the correlation length of the fields in the target, such partonic approximation is not valid. In this case the two gluon system scatters like a single object in the color representation that combines the two gluons and the color correlations can not be neglected. In particular if this two gluon state emerges from a single gluon through a step in the evolution, such state should scatter as a single octet, and not as an independent product of two octets.  For separations greater than the correlation scale the partonic approximation makes perfect sense. However our approach is designed for an arbitrary target and the split into $H_0$ and $H_I$ can not depend  on !
 the characteristic of the target. The physical alternative then is to choose such a split which suppresses the emission of a gluon at a neighboring point without introducing a scale, that is in a dilatationally invariant way. Technically this is ensured in eqs.(\ref{h0r},\ref{h0v}) by the presence of the factors $\tilde R(x)-\tilde R(z)$. This is the only possibility we found to satisfy the condition of ultraviolet finiteness, but we have not proven that it is unique.

The second important element of our approach was the diagonalization of $H_0$. To do this we had to differentiate functions of $R$. Since $R$ is a unitary matrix, this has to be done with care. The complete RFT Hamiltonian  preserves the unitarity of $R$. This means that when acting on a wave function which vanishes for nonunitary $R$, the Hamiltonian transforms it into another wave function with the same property. In this situation one can substitute the constraint differentiation with respect to unitary $R$ by an unconstrained one as it does not change the action of $H$ on physical states.
However this property is not satisfied by $H_0$ and $H_I$ separately. In our calculation we have disregarded this subtlety. A more careful approach would be to introduce the Lagrange multiplier that imposes the unitarity of the matrix $R$. Although we do not believe that this will affect our results, it is a very important question worth detailed study. In particular there is in principle a possibility that some of the states we found may be pushed into the unphysical part of the Hilbert space once the constraint on the $R$ matrix is imposed. The prime suspect here could be the parity even Odderon, since it does not couple to simple color singlet states.

As we have discussed in Section 4, the perturbative treatment of $H$ violates unitarity. This leads to appearance of negative probabilities as well as to the unbounded growth of the cross section. In this context the Pomeron states are tachyonic, since they have negative eigenvalues at low momentum transfers. In fact the Hamiltonian $H_0$ is not bounded from below and it is clear that such tachyons exist in sectors with arbitrary number of $s$ - channel gluons with ever increasing intercepts.
The cure for this can only come by including the interaction $H_I$ nonperturbatively. We certainly believe that the
JIMWLK/KLWMIJ Hamiltonian does define a unitary theory. Numerical studies of both the
BK equation and of the full JIMWLK equation lead to this conclusion \cite{BKN}.
It would be very interesting to devise a nonperturbative approach to RFT, perhaps along the lines of a mean field approximation with $\langle R\rangle=0$.

A state with $\langle R\rangle=0$ corresponds to the black disk limit.
The $SU(N)\otimes SU(N)$ symmetry is restored in such a state. In physical terms the restoration of the symmetry means that the color index of an incoming gluons is not correlated with the color index of the outgoing one. Thus the color is completely randomized during the interaction with the target.
It is usual in the quantum field theory that the restoration of symmetry is associated with condensation of the excitations of the ordered phase. In our case those excitations are Pomerons. We can think therefore of the black disk limit as condensation of Pomerons similar to the old ideas of \cite{Amati} (see also \cite{BF} for the renewed interest in this problem).

This analogy is not perfect. There is no phase transition in RFT, since there really are no two distinct phases as a function of some external parameter like temperature or coupling constant. The only parameter in the RFT Hamiltonian eq.(\ref{KL}) is $\alpha_s$, but it enters as an overall multiplicative constant and thus its value can not affect the phase structure. It only affects the overall scale of the energy and therefore the speed with which a given state evolves towards the vacuum. So one and the same phase must be realized at all values of $\alpha_s$. The Pomeron condensation  therefore occurs always. This is consistent with the fact that as excitations they have tachyonic nature already in the perturbative expansion. In principle the situation might change when the additional terms responsible for Pomeron loops are included (see discussion below), although we would still expect the black disc limit to be
the ground state for all values of $\alpha_s$ at least as long as nonperturbative confining effects are not accounted for.

A similar situation is encountered in the Gribov effective Reggeon field theory\cite{Amati}, where the nonperturbative account of the Pomeron interactions leads to a gray disc limiting behavior with the growth of the total cross section which satisfies the Froissart bound.
There are however many differences between the two cases. Apart from the ones already mentioned above, one very important difference is the fact that the KLWMIJ Hamiltonian which has been the basis of the present work violates $t$-channel unitarity as it does not include Pomeron loops.

There has been a lot of activity recently in an attempt to incorporate the Pomeron loops
into  the JIMWLK/KLWMIJ formalism \cite{MSW,IT,LL3,kl1,kl,something,SMITH,Balitsky05,blaizot,Kovchegov05}.
In spite of some progress in this direction the complete evolution kernel is still unknown. We therefore
 still do not have a completely unitary RFT.
One could try naively to restore the $t$-channel unitarity by symmetrizing the Hamiltonian
with respect to $R$ and $\delta\over \delta R$. This would be analogous to the Gribov theory which is symmetric under the interchange of the Pomeron field and its conjugate momentum. This idea would lead us to add terms of the type $RR{\delta\over
\delta R}{ \delta\over \delta R}{ \delta\over \delta R}$ to the Hamiltonian. However these terms correspond to the process of recombination of three $s$ - channel gluons into two as a result of the boost of the projectile wave function.
The $s$-channel gluons however do not recombine, but rather
are created by the evolution (see \cite{kl5} for discussion). The nonlinearities in the JIMWLK evolution for example correspond to the decrease in the emission probability of an extra gluon, but not to the appearance of recombination in $s$ - channel. In Refs. \cite{kl1,kl,something} we have derived an extended version
of the JIMWLK/KLWMIJ equations (JIMWLK+/KLWMIJ+), which  account  partially for the effects of Pomeron loops. Similar results were obtained in the path integral approach in  \cite{SMITH,Balitsky05}.  This work suggests that the $t$-channel unitarity corrections should occur as higher powers
of $R{\delta\over \delta R}$. This should have an effect of generating $n \rightarrow m$ ($n\ge m$) Reggeon transitions alongside the
$m\rightarrow n$ transitions present in KLWMIJ. It remains unclear whether the effects accounted for in the JIMWLK+/KLWMIJ+ equations are
indeed sufficient to restore the $t$-channel unitarity of the RFT. It would be interesting to understand in general  whether the selfduality
condition of Ref. \cite{something} ensures the $t$-channel unitarity.
The $t$-channel interpretation which emerged naturally in the present paper should be helpful in addressing these questions.

Finally we note that in this paper we considered RFT in gluonic sector only. Although in the leading eikonal approximation only gluons are created in the evolution, it is not difficult to consider initial states which contain quarks.  To this end essentially all one has to do is to rewrite the left and right color rotation generators in eq.(\ref{KL}) in terms of the fundamental rather than adjoint matrices $R_F$. One can then repeat the perturbative analysis of the present paper without any additional complications. It may be interesting to consider
the quark dipole scattering as it is more directly related to physical process (such as DIS) than the gluon dipole.

\section*{Acknowledgments}

We are grateful to  Ian  Balitsky, Jochen Bartels, Yuri Dokshitzer, Genya Levin, Lev Lipatov, Larry McLerran, Edward Shuryak, Anna  Stasto,  George Sterman and Urs Wiedemann for useful discussions and correspondence related to the subject of this work. We also thank anonymous referee for valuable remarks which prompted us to think more thoroughly about the relation of our results with the standard GLLA and EGLLA approximations.

\appendix
\section{Appendix: Evolution of $D^4$.}\label{sec:A}
In this appendix we consider the evolution of the four gluon $t$ - channel state coupled to a single dipole ($s$ - channel) projectile. The discussion is in the framework of the leading large $N$ approximation.
The state with $n$ gluons in $t$-channel is defined by eq.(\ref{dn1}). We consider two such states - with two and four gluons. We also project the gluons pairwise on color singlet states as appropriate for the large $N$ limit. We thus consider
\begin{equation}
D^2(Y)(x_1,x_2)={1\over N}\int D\rho \rho^{a}(x_1)\rho^{a}(x_2)W_Y[\rho]
\end{equation}
and
\begin{equation}\label{d4def}
D^4(Y)(x_1,x_2,x_3,x_4)={1\over N^2}\int D\rho \rho^{a}(x_1)\rho^{a}(x_2)\rho^{b}(x_3)\rho^{b}(x_4)W_Y[\rho]
\end{equation}
There is one subtle point in this definition which turns out to be important for the calculation of the initial condition. Namely when any two transverse coordinates are equal ($x_1=x_3,\,\, etc.$) the appropriate charge density operators do not commute with each other. In this case one has to remember that in the integral in eq.(\ref{d4def}) the charge density variable is also endowed with the longitudinal coordinate $x^-$, which keeps track of the ordering of the operators $\rho^a(x_1)\rightarrow \rho^a(x_1,x^-_1)$. This has been discussed in great detail in \cite{kl4} where we have also developed a technique to calculate this type of averages taking the noncommutativity into account. We do not give details here but interested reader can consult \cite{kl4}.

The function $D^4$ is the one considered in \cite{BV}. We choose to concentrate on it rather than on a function completely symmetric with respect to all four gluons \cite{3P1} because the color algebra here is marginally simpler.
The evolution with rapidity is given by
\begin{eqnarray}\label{eqd}
{d\over dY}D^2&=&{1\over N}\int D\rho \rho^{a}(x_1)\rho^{a}(x_2)\chi^{KLWMIJ}[\rho,{\delta\over\delta\rho}]W_Y[\rho]\\
{d\over dY}D^4&=&{1\over N^2}\int D\rho \rho^{a}(x_1)\rho^{a}(x_2)\rho^{b}(x_3)\rho^{b}(x_4)\chi^{KLWMIJ}[\rho,{\delta\over\delta\rho}]W_Y[\rho]\nonumber
\end{eqnarray}.
For a "virtual photon" projectile state with wave function $P^\gamma(u,v)$ the initial projectile weight functional is
\begin{equation}
W[\rho]=\int_{u,v}P^\gamma(u,v)r(u,v)\delta[\rho]
\end{equation}
with $r(,v)$ defined in eq.(\ref{pdipole}).).
For such a projectile the initial conditions for eq.(\ref{eqd}) are
\begin{eqnarray}\label{init}
D^2(Y=0)&=&D^2_0(x_1,x_2)={1\over N}\int_{u,v}P^\gamma(u,v)\int D\rho \rho^{a}(x_1)\rho^{a}(x_2)r(u,v)\delta[\rho]\nonumber \\&=&\int_vP^\gamma(x_1,v)\delta_{x_1,x_2}-   P^\gamma(x_1,x_2)
\end{eqnarray}
\begin{eqnarray}
&&D^4(Y=0)=D^4_0(x_1,x_2,x_3,x_4)={1\over N^2}\int_{u,v}P^\gamma(u,v)\int D\rho \rho^{a}(x_1)\rho^{a}(x_2)\rho^{b}(x_3)\rho^{b}(x_4)r(u,v)\delta[\rho]\nonumber\\
&&=\,-\,\frac{1}{2}\,\Big[D^2_0(x_1,x_3)\,
\delta_{x_1,x_2}\,\delta_{x_3,x_4}\,+\,D^2_0(x_1,x_2)\,
\delta_{x_1,x_3}\,\delta_{x_2,x_4}\,+\,D^2_0(x_1,x_2)\,
\delta_{x_1,x_4}\,\delta_{x_2,x_3} \\
&&-\,
 D^2_0(x_1,x_2)\,\delta_{x_2,x_3}\,\delta_{x_2,x_4}\,-\,
D^2_0(x_1,x_2)\,\delta_{x_1,x_3}\,\delta_{x_1,x_4}\,-\,D^2_0(x_1,x_3)\,\delta_{x_1,x_2}\,\delta_{x_1,x_4}\,-\,
D^2_0(x_1,x_4)\,\delta_{x_1,x_2}\,\delta_{x_1,x_3}
\Big]\nonumber
\end{eqnarray}
It is straightforward to verify explicitly by expanding $r(u,v)$ in powers of $\delta/\delta\rho$ that the functions $D^2_0$ and $D^4_0$ are the standard impact factors for coupling of two and four gluons to a dipole projectile. The explicit relation between $D^4_0$ and $D^2_0$ is the same as that of \cite{BV}. We stress again that in the derivation of eq.(\ref{init}) it is important to account correctly for noncommutativity of operators $\rho^a$ at the same transverse coordinate \cite{kl4}.

To write explicitly the evolution equations  for $D^2$ and $D^4$ we use $\chi^{KLWMIJ}$ in the dipole limit eq.(\ref{chidip}) and act with the kernel to the left in eq.(\ref{eqd})\footnote{Note that the full KLWMIJ Hamiltonian beyond the dipole limit contains additional terms in the expansion to fourth order in derivatives. The additional terms however do not contribute when acting on $D^2$ and $D^4$ as defined above, see \cite{kl4},\cite{MMSW}.} Since $\chi^{KLWMIJ}$ acts at most on fourth power of $\rho$, the nonvanishing contributions come only from expanding operators $r$ ($d-1$) in the interaction part of the kernel to order $(\delta/\delta\rho)^2$. The leading large $N$ contribution of the free part comes from the same order of expansion of $r$. Thus for the purpose of the present calculation the dipole creation and annihilation operators are
\begin{equation}
d^\dagger(x,y)={1\over 4N}\left[{\delta\over\delta\rho^a(x)}-{\delta\over\delta\rho^a(x)}\right]^2;\ \ \ \ \ \ \ \ d(x,y)={1\over N}\rho^a(x)\rho^a(y)
\end{equation}
and the kernel can be written as
\begin{eqnarray}
\chi&=&{1\over 4N^2}\int_{x,y,z}M_{x,y,z}\rho^a(x)\rho^a(y)\times\\
&&\left[2\left({\delta\over\delta\rho^b(x)}-{\delta\over\delta\rho^b(z)}\right)\left({\delta\over\delta\rho^b(y)}-{\delta\over\delta\rho^b(z)}\right)-{1\over 4N}\left({\delta\over\delta\rho^b(x)}-{\delta\over\delta\rho^b(z)}\right)^2\left({\delta\over\delta\rho^c(y)}-{\delta\over\delta\rho^c(z)}\right)^2\right]\nonumber
\end{eqnarray}
Now acting with this kernel in eq.(\ref{eqd}) we obtain
\begin{equation}\label{eqn2}
{d\over dY}D^2(x_1,x_2)=\int_{y_1,y_2}K(x_1,x_2;y_1,y_2)D^2(y_1,y_2)
\end{equation}
and
\begin{eqnarray}\label{eqn4}
{d\over dY}D^4(x_1,x_2,x_3,x_4)=
\int_{y_1,y_2}&&\left[K(x_1,x_2;y_1,y_2)D^4(y_1,y_2,x_3,x_4)+K(x_3,x_4;y_1,y_2)D^4(x_1,x_2,y_1,y_2)\right.\nonumber\\
&-&\left.V(x_1,x_2,x_3,x_4,y_1,y_2)D^2(y_1,y_2)\right]
\end{eqnarray}
where $K(x_1,x_2;y_1,y_2)$ is the nonforward BFKL kernel and $V$ is the so called three Pomeron vertex
\begin{eqnarray}\label{vert}
V(x_1,x_2,x_3,x_4,y_1,y_2)&&={1\over 2}\Big[M(x_1,x_3,x_2)\delta_{y_1,x_1}\delta_{y_2,x_3}\delta_{x_2,x_4}+M(x_1,x_4,x_2)\delta_{y_1,x_1}\delta_{y_2,x_4}\delta_{x_2,x_3}\nonumber\\
&&+M(x_2,x_3,x_1)\delta_{y_1,x_2}\delta_{y_2,x_3}\delta_{x_1,x_4}
+M(x_2,x_4,x_1)\delta_{y_1,x_2}\delta_{y_2,x_4}\delta_{x_1,x_3}\Big]
\end{eqnarray}
In eq.(\ref{vert}) we have not written explicitly terms proportional to $\delta_{x_1,x_2}$ and/or $\delta_{x_3,x_4}$ since they vanish when multiplied by the Pomeron eigenfunctions of the dipole model and thus do not contribute to solution of eq.(\ref{eqn4}), see \cite{BV} for discussion.

Eq.(\ref{eqn2}) is clearly the BFKL equation. Solution of the nonhomogeneous  eq.(\ref{eqn4}) can be written in the form
\begin{equation}
D^4(Y)=D^4_I(Y)+D^4_R(Y)
\end{equation}
Here the irreducible piece $D^4_I$ is defined as the solution of eq.(\ref{eqn4}) with initial condition $D^4_I(Y=0)=0$, while the reggeized piece $D^4_R$ is defined as the solution of the homogeneous part of the equation (eq.(\ref{eqn4}) excluding the last vertex piece) with the initial condition $D^4_R(Y=0)=D_0^4$.
The piece $D_R^4$ is precisely the reggeized piece in the standard terminology. Its dependence on energy is that of the single Pomeron exchange rather than the double Pomeron exchange. The reason is that the initial condition $D^4_0$ of eq.(\ref{init}) is given by a linear combination of $D^2_0$ due to the fact that the dipole projectile has only two transverse coordinates. On the other hand the homogeneous part of the evolution equation eq.(\ref{eqn4}) has the property that it propagates such an initial condition with the BFKL kernel.

The reason why the reggeized piece appears in the present analysis but does not appear in the analysis of the dipole model presented in \cite{BV} is the following. The authors of \cite{BV} analyzed the equation for the inclusive double dipole density  rather than for the four gluon correlator. The two quantities although similar, are not completely
equivalent. In particular it is obvious that the inclusive double dipole density vanishes in the initial single dipole state. Thus even though in the large $N$ limit it satisfies the same evolution equation as the four gluon correlator, the solutions of the two equations differ precisely by the reggeized piece.

We thus see that our approach reproduces the standard splitting of the gluon four point function into the irreducible part and the reggeized part at least in the large $N$ limit. Comparison of this aspect of the two approaches in more general setting is beyond the scope of this work.

\section{Appendix: Derivation of the eigenvalue equations} \label{sec:B}

In this Appendix we give details of computation of the left hand side of eqs. (\ref{chiG1}, \ref{chiG2}). We start by acting
with the Hamiltonian $H_0$ on the functions $G^{S\,(CP)}$. We begin with $G^{+\,(++)}$ and then repeat the analysis for the function $G^{\pm\,(--)}$.
Acting with the "real" part of the Hamiltonian $H_0^R$ we obtain
\bea\label{chiGR}
&&H_0^R\,G^{+\,(++)}\,=\,-\,2\,
\hat K_{u,v,z} \,\left\{  T^{\bar a}_{a\gamma}\,[\tilde R_u\,-\,\tilde R_z]^{\gamma b}\,T^{\bar a}_{ck}\,
[\tilde R_v\,-\,\tilde R_z]^{kd}\,+\,[\tilde R_u\,-\,\tilde R_z]^{a\beta}\,T^{\bar a}_{\beta b}\,
[\tilde R_v\,-\,\tilde R_z]^{cn}\,T^{\bar a}_{nd}\right.\nonumber \\ \nonumber && \\
&&-\left.
\,T^{\bar a}_{a \gamma}\,[\tilde R_u\,-\,\tilde R_z]^{\gamma b}\,[\tilde R_v\,-\,\tilde R_z]^{cn}\,T^{\bar a}_{nd}\,-\,
T^{\bar a}_{c \gamma}\,[\tilde R_v\,-\,\tilde R_z]^{\gamma d}\,[\tilde R_u\,-\,\tilde R_z]^{an}\,T^{\bar a}_{nb} \right\}
\,P^{i\,bd}_{\,\,ac}\, \Psi^i_s(u,v)
\eea
Similarly applying the ``virtual'' Hamiltonian $H_0^V$ gives
\bea\label{chiGV}
H_0^V\,G^{+\,(++)}\,=\,4\,\int_{u,v,z} K_{v,v,z}\,P^{i\,bd}_{\,\,ac}\, \Psi^i_s(u,v)
  \,\left\{
(T^{\bar a}\,T^{\bar b})_{cd}\,[\tilde R_v\,-\,\tilde R_z]^{\bar a \bar b} \,-\,N\,[\tilde R_v\,-\,\tilde R_z]^{cd}
\right\}\,\tilde R^{ab}(u)
  \nonumber  \\
\eea
The next step is to differentiate \eq{chiGR} and \eq{chiGV}
 with respect to ${\delta \over \delta \tilde R^{\alpha\beta}(x)}\,
{\delta \over \delta\tilde R^{\gamma\delta}(y)}$:
\bea\label{chiGR1}
&&{\delta \over \delta \tilde R^{\alpha\beta}_x}\,
{\delta \over \delta\tilde R^{\gamma\delta}_y}\,\,
H_0^R\,G^{+\,(++)}\,
=\,-\,\left\{
T^{\bar a}_{a\alpha}\,T^{\bar a}_{c\gamma} \,\,P^{i\,\beta \delta}_{\,\,ac}\, +\,
T^{\bar a}_{\beta b}\,T^{\bar a}_{\delta d}\,\, P^{i\,bd}_{\,\,\alpha\gamma}\,-\,
T^{\bar a}_{a\alpha}\,T^{\bar a}_{\delta d}\,\,P^{i\,\beta d}_{\,\,a\gamma}
 \,-\,
T^{\bar a}_{\beta d}\,T^{\bar a}_{a\gamma}\,\,P^{i\,d\delta}_{\,\,\alpha a}\right\}
\nonumber \\ \nonumber && \\
&&\otimes\,2\,\int_{z}\,\left[ 2\,K_{x,y,z}\,
 \,\Psi^i_s(x,y)\,-\,  K_{x,z,y}\,\Psi^i_s(x,z)
\,-\,   K_{z,y,x}\,\Psi^i_s(y,z)\,+\,2\,\delta(x-y)\int_u K_{z,u,x}\,\Psi^i_s(z,u)\right]\nonumber \\ \nonumber && \\
&&
=
\,-\, 4\,N\,
 \left\{ -\,\lambda_i\,P^{i\,\beta\delta}_{\,\,\,\alpha\gamma}\,
\,-\, (-1)^{s_i+s_k}\,\bar C^i_k\,P^{k\,\beta\delta}_{\,\,\,\alpha \gamma}\right\}\,\nonumber \\ \nonumber && \\
&&\otimes
\int_{z}\,\left[ 2\,K_{x,y,z}\,
 \,\Psi^i_s(x,y)\,-\,  K_{x,z,y}\,\Psi^i_s(x,z)
\,-\,   K_{z,y,x}\,\Psi^i_s(y,z)+\,2\,\delta(x-y)\int_u K_{z,u,x}\,\Psi^i_s(z,u)\right]
\eea
The last equality follows by using the crossing properties of the projectors \eq{eq:C.8}-\eq{PTT}.
The contribution of the virtual part is
\bea\label{chiGV1}
&&{\delta \over \delta \tilde R^{\alpha\beta}(x)}\,
{\delta \over \delta\tilde R^{\gamma\delta}(y)}\,\,
H_0^V\,G^{+\,(++)}\,=\,4
 \,\left\{
(T^{\gamma}\,T^{\delta})_{cd}\, P^{i\,\beta d}_{\,\,\alpha c}\,
\,-\,N\,P^{i\,\beta \delta}_{\,\,\alpha \gamma}\right\}\nonumber  \\
&&\otimes\,\int_{z}\,\left[ (K_{y,y,z}\,+\,K_{x,x,z})
 \,\Psi^i_s(x,y)\,-\,  K_{z,z,y}\,\Psi^i_s(x,z)
\,-\,   K_{z,z,x}\,\Psi^i_s(y,z)\right] \nonumber
\\
&&=\,4N \left[ \bar C^i_k \,P^k\,-\,P^i\right]^{\beta\delta}_{\alpha \gamma}
\,\int_{z}\,\left[2\,K_{x,x,z}
 \,\Psi^i_s(x,y)\,-\,  K_{z,z,y}\,\Psi^i_s(x,z)
\,-\,   K_{z,z,x}\,\Psi^i_s(y,z)\right]
\eea
where again \eq{eq:C.8}-\eq{PTT} were used to get the last equality.

We repeat the same algebra but now applied to the function $G^{\pm\,(--)}$. The ``real'' and ``virtual'' terms read
\bea\label{chiGR-}
&&H_0^R\,G^{\mp\,(--)}\,=\,-\,2\,
\hat K_{u,v,z} \,\left\{  T^{\bar a}_{a\gamma}\,[\tilde R_u\,-\,\tilde R_z]^{\gamma b}\,T^{\bar a}_{ck}\,
[\tilde R_v\,-\,\tilde R_z]^{kd}\,+\,[\tilde R_u\,-\,\tilde R_z]^{a\beta}\,T^{\bar a}_{\beta b}\,
[\tilde R_v\,-\,\tilde R_z]^{cn}\,T^{\bar a}_{nd}\right.\nonumber \\ \nonumber && \\
&&-\left.
\,T^{\bar a}_{a \gamma}\,[\tilde R_u\,-\,\tilde R_z]^{\gamma b}\,[\tilde R_v\,-\,\tilde R_z]^{cn}\,T^{\bar a}_{nd}\,-\,
T^{\bar a}_{c \gamma}\,[\tilde R_v\,-\,\tilde R_z]^{\gamma d}\,[\tilde R_u\,-\,\tilde R_z]^{an}\,T^{\bar a}_{nb} \right\}
\,Z^{\pm\,bd}_{\,\,ac}\, \Phi^{\mp}(u,v)
\eea
\bea\label{chiGV-}
H_0^V\,G^{\mp\,(--)}\,=\,4\,\int_{u,v,z} K_{v,v,z}\,Z^{\pm\,bd}_{\,\,ac}\, \Phi^{\mp}(u,v)
  \,\left\{
(T^{\bar a}\,T^{\bar b})_{cd}\,[\tilde R_v\,-\,\tilde R_z]^{\bar a \bar b} \,-\,N\,[\tilde R_v\,-\,\tilde R_z]^{cd}
\right\}\,\tilde R^{ab}(u)
  \nonumber  \\
\eea
Differentiating \eq{chiGR-} and \eq{chiGV-}
 with respect to ${\delta \over \delta \tilde R^{\alpha\beta}(x)}\,
{\delta \over \delta\tilde R^{\gamma\delta}(y)}$ we obtain
\bea\label{chiGR1-}
{\delta \over \delta \tilde R^{\alpha\beta}_x}\,
{\delta \over \delta\tilde R^{\gamma\delta}_y}\,\,
H_0^R\,G^{\mp\,(--)}\,
&= &-\,\left\{
T^{\bar a}_{a\alpha}\,T^{\bar a}_{c\gamma} \,\,Z^{\pm\,\beta \delta}_{\,\,ac}\, +\,
T^{\bar a}_{\beta b}\,T^{\bar a}_{\delta d}\,\, Z^{\pm\,bd}_{\,\,\alpha\gamma}\,-\,
T^{\bar a}_{a\alpha}\,T^{\bar a}_{\delta d}\,\,Z^{\pm\,\beta d}_{\,\,a\gamma}
 \,-\,
T^{\bar a}_{\beta d}\,T^{\bar a}_{a\gamma}\,\,Z^{\pm\,d\delta}_{\,\,\alpha a}\right\}
\nonumber \\ \nonumber && \\
&\otimes&2\,\int_{z}\left[ 2\,K_{x,y,z}
 \,\Phi^{\mp}(x,y)\,-\,  K_{x,z,y}\,\Phi^{\mp}(x,z)
\,-\,   K_{z,y,x}\,\Phi^{\mp}(z,y)\right]\nonumber \\ \nonumber && \\
&
=
& 2^{(3\mp 1)/2}\,N\,Z^{\pm\,\beta\delta}_{\,\,\,\alpha\gamma}
\int_{z}\left[ 2\,K_{x,y,z}
 \Phi^{\mp}(x,y)-  K_{x,z,y}\,\Phi^{\mp}(x,z)
-   K_{z,y,x}\Phi^{\mp}(y,z)\right] \nonumber \\
\eea
We again rely on the crossing properties of the projectors \eq{eq:C.8}-\eq{last}.
Finally the contribution of the virtual part is
\bea\label{chiGV1-}
&&{\delta \over \delta \tilde R^{\alpha\beta}_x}\,
{\delta \over \delta\tilde R^{\gamma\delta}_y}\,\,
H_0^V\,G^{\mp\,(--)}\,=\,4
 \,\left\{
(T^{\gamma}\,T^{\delta})_{cd}\, Z^{\pm\,\beta d}_{\,\,\alpha c}\,
\,-\,N\,Z^{\pm\,\beta \delta}_{\,\,\alpha \gamma}\right\}\nonumber  \\
&&\ \ \ \ \ \ \ \ \otimes\,\int_{z}\,\left[ (K_{y,y,z}\,+\,K_{x,x,z})
 \,\Phi^{\mp}(x,y)\,-\,  K_{z,z,y}\,\Phi^{\mp}(x,z)
\,-\,   K_{z,z,x}\,\Phi^{\mp}(z,y)\right] \nonumber
\\
&&\ \ \ \ \ \ \ \ =\,-\,2^{(3\mp1)/2}\,N\, Z^{\pm\,\beta\delta}_{\,\,\,\alpha\gamma}
\,\int_{z}\,\left[2\,K_{x,x,z}
 \,\Phi^{\mp}(x,y)\,-\,  K_{z,z,y}\,\Phi^{\mp}(x,z)
\,-\,   K_{z,z,x}\,\Phi^{\mp}(z,y)\right]
\eea

The same algebra  applied to the function $G^{-\,(-+)}$ gives (we note that $[P^{10}-P^{\overline{10}}]^{bd}_{ac}\,=\,Z^{-\,cd}_{\,\,\,\,\,ab}$)
\bea\label{chiGR10}
&&H_0^R\,G^{-\,(-+)}\,=\,-\,2\,
\hat K_{u,v,z} \,\left\{  T^{\bar a}_{a\gamma}\,[\tilde R_u\,-\,\tilde R_z]^{\gamma b}\,T^{\bar a}_{ck}\,
[\tilde R_v\,-\,\tilde R_z]^{kd}\,+\,[\tilde R_u\,-\,\tilde R_z]^{a\beta}\,T^{\bar a}_{\beta b}\,
[\tilde R_v\,-\,\tilde R_z]^{cn}\,T^{\bar a}_{nd}\right.\nonumber \\ \nonumber && \\
&&-\left.
\,T^{\bar a}_{a \gamma}\,[\tilde R_u\,-\,\tilde R_z]^{\gamma b}\,[\tilde R_v\,-\,\tilde R_z]^{cn}\,T^{\bar a}_{nd}\,-\,
T^{\bar a}_{c \gamma}\,[\tilde R_v\,-\,\tilde R_z]^{\gamma d}\,[\tilde R_u\,-\,\tilde R_z]^{an}\,T^{\bar a}_{nb} \right\}
\,Z^{-\,cd}_{\,\,\,\,\,ab}\, \Psi_s^-(u,v)
\eea
\bea\label{chiGV10}
H_0^V\,G^{-\,(-+)}\,=\,4\,\int_{u,v,z} K_{v,v,z}\,Z^{-\,cd}_{\,\,\,\,\,ab}\, \Psi_s^-(u,v)
  \,\left\{
(T^{\bar a}\,T^{\bar b})_{cd}\,[\tilde R_v\,-\,\tilde R_z]^{\bar a \bar b} \,-\,N\,[\tilde R_v\,-\,\tilde R_z]^{cd}
\right\}\,\tilde R^{ab}(u)
  \nonumber  \\
\eea
Differentiating \eq{chiGR10} and \eq{chiGV10}
 with respect to ${\delta \over \delta \tilde R^{\alpha\beta}(x)}\,
{\delta \over \delta\tilde R^{\gamma\delta}(y)}$ we obtain
\bea\label{chiGR110}
{\delta \over \delta \tilde R^{\alpha\beta}_x}\,
{\delta \over \delta\tilde R^{\gamma\delta}_y}\,\,
H_0^R\,G^{-\,(-+)}\,
&= &-\,\left\{
T^{\bar a}_{a\alpha}\,T^{\bar a}_{c\gamma} \,\,Z^{-\,c \delta}_{\,\,a\beta}\, +\,
T^{\bar a}_{\beta b}\,T^{\bar a}_{\delta d}\,\, Z^{-\,\gamma d}_{\,\,\alpha b}\,-\,
T^{\bar a}_{a\alpha}\,T^{\bar a}_{\delta d}\,\,Z^{-\,\gamma d}_{\,\,a\beta}
 \,-\,
T^{\bar a}_{\beta d}\,T^{\bar a}_{a\gamma}\,\,Z^{-\,a\delta}_{\,\,\alpha d}\right\}
\nonumber \\ \nonumber && \\
&\otimes&2\,\int_{z}\left[ 2\,K_{x,y,z}
 \,\Psi_s^-(x,y)\,-\,  K_{x,z,y}\,\Psi_s^-(x,z)
\,-\,   K_{z,y,x}\,\Psi_s^-(z,y)\right]\nonumber \\ \nonumber && \\
&
=
& 2\,N\,Z^{-\,\gamma\delta}_{\,\,\,\alpha\beta}
\int_{z}\left[ 2\,K_{x,y,z}
 \Psi_s^-(x,y)-  K_{x,z,y}\,\Psi_s^-(x,z)
-   K_{z,y,x}\Psi_s^-(y,z) \right. \nonumber \\
&+&\left. 2\,\delta(x-y)\int_u K_{z,u,x}\,\Psi_s^-(z,u)\right]
\eea
The contribution of the virtual part is
\bea\label{chiGV110}
&&{\delta \over \delta \tilde R^{\alpha\beta}_x}\,
{\delta \over \delta\tilde R^{\gamma\delta}_y}\,\,
H_0^V\,G^{-\,(-+)}\,=\,4
 \,\left\{
(T^{\gamma}\,T^{\delta})_{cd}\, Z^{-\,c d}_{\,\,\alpha \beta}\,
\,-\,N\,Z^{-\,\gamma \delta}_{\,\,\alpha \beta}\right\}\nonumber  \\
&&\ \ \ \ \ \ \ \ \otimes\,\int_{z}\,\left[ (K_{y,y,z}\,+\,K_{x,x,z})
 \,\Psi_s^-(x,y)\,-\,  K_{z,z,y}\,\Psi_s^-(x,z)
\,-\,   K_{z,z,x}\,\Psi_s^-(z,y)\right] \nonumber
\\
&&=\,-\,2\,N\, Z^{-\,\gamma\delta}_{\,\,\,\alpha\beta}
\,\int_{z}\,\left[2\,K_{x,x,z}
 \,\Psi_s^-(x,y)\,-\,  K_{z,z,y}\,\Psi_s^-(x,z)
\,-\,   K_{z,z,x}\,\Psi_s^-(z,y)\right]
\eea

\section{Appendix: Projectors in $SU(N)$} \label{sec:C}

In this Appendix for convenience of reference we collected some standard formulae for the $SU(N)$ group and its representations.
In particular throughout our calculations we have used extensively the decomposition of the product of two adjoint representations into a direct sum of irreducible representations.
We give definitions of the relevant projectors and summarize their properties. This is taken almost entirely from
Ref. \cite{NSZ} and we use notations of that paper. The same algebra has also been worked out in \cite{DM}. For more details we refer the
reader to Refs. \cite{NSZ,DM,KOS}. The  $SU(3)$ results can also be found in \cite{3P,LRS}.

The $SU(N)$-generators in the fundamental representation, $t^a, a = 1 \dots N^2-1$
define
the familiar $f$-- and $d$--tensors through
\bea
t^a t^b = {1 \over 2 N} \delta_{ab} \,
+ { 1\over 2} \Big( d_{abc} + i f_{abc} \Big) t^c \, ,
\label{eq:4.1.1}
\eea
It follows that
\bea
i f_{abc} = 2 \, \Tr\Big([t^a,t^b]t^c\Big) \, , \, d_{abc} = 2 \, \Tr\Big(\{t^a,t^b\}t^c\Big) \, .
\label{eq:A.2}
\eea
The $f$ and $d$ tensors obey  the Jacobi identities
\bea
if_{kam}if_{mbl} -if_{kbm}if_{mal} = i f_{abm} if_{kml} \, ,
\label{eq:A.4}
\eea
\bea
f_{kam} d_{mbl} - d_{kbm} f_{mal} =  f_{abm} d_{kml} \, ,
\label{eq:A.5}
\eea
and  summation rules
\bea
f_{aij}f_{bij} &&= N \, \delta_{ab} \, , \nonumber \\
 d_{aij}d_{bij} &&= {N^2 - 4 \over N} \delta_{ab} \, , \\
 f_{aij}d_{bij} &&= 0 \, . \nonumber
\label{eq:A.8}
\eea
\bea
f_{iaj}f_{jbk}f_{kci} &&= -{N \over 2} f_{abc} \, , \nonumber \\
f_{iaj}f_{jbk}d_{kci} &&= -{N \over 2} d_{abc} \nonumber \\
f_{iaj}d_{jbk}d_{kci} &&= {N^2 - 4  \over 2N} f_{abc} \\
d_{iaj}d_{jbk}d_{kci} &&= {N^2 - 12 \over 2N} d_{abc} \, . \nonumber
\label{eq:A.9}
\eea

The $SU(N)$ generators in the adjoint representation
$ T^a_{bc} = -i f_{abc} $ satisfy standard commutation relations $[T^a,T^b] = i f_{abc} T^c$.

We now consider  the product of two adjoint representations of the $SU(N)$ group and its decomposition
into a direct sum of irreducible representations.
\bea
(N^2-1) \times (N^2-1)&=& 1 + (N^2-1)_A + (N^2-1)_S \nonumber  \\
&& + {(N^2-4)(N^2-1) \over 4} +
\Big[{(N^2-4)(N^2-1) \over 4}\Big]^* \nonumber \\
&& + {N^2 (N+3)(N-1) \over 4} + {N^2 (N-3)(N+1) \over 4} \nonumber \\
&=& 1 + 8_A + 8_S + 10 + \overline{10} + 27 + {\cal R}_7 \, .
\label{eq:4.0.1}
\eea
We follow the nomenclature of \cite{NSZ} in denoting the representations by their dimensions in the $SU(3)$ case.
The symmetric representation ${\cal R}_7$ does not exist in $SU(3)$.
In Ref. \cite{DM} this representation is denoted as $0$.

The $SU(N)$-projectors into the singlet as well as the
two adjoint multiplets have
manifestly the same form as
their well-known $N=3$ counterparts:
\bea
P[1]^{ab}_{cd} &=& {1 \over N^2-1} \delta_{ab} \delta_{cd} \\
P[8_A]^{ab}_{cd} &=& {1 \over N} f_{abk}f_{kcd} = {1 \over N} i f_{abk} i f_{kdc} \\
P[8_S]^{ab}_{cd} &=& {N \over N^2-4}  d_{abk}d_{kcd} =  {N \over N^2-4}\, [D_s]^{ab}_{cd}
\label{eq:4.1.7}
\eea
The projectors onto the antisymmetric $10$ and $\overline{10}$ are
\bea
P[10]^{ab}_{cd}  &=& {1 \over 2} \Big( {\cal{A}}^{ab}_{cd} - P[8_A]^{ab}_{cd} + i Y^{ab}_{cd} \Big)
\label{eq:4.1.12}\nonumber\\
P[\overline{10}]^{ab}_{cd}  &=& {1 \over 2} \Big( {\cal{A}}^{ab}_{cd}
- P[8_A]^{ab}_{cd} - i Y^{ab}_{cd} \Big) \, .
\label{eq:4.1.13}
\eea
The remaining representations are symmetric and  the corresponding projectors are
\bea
P[27]^{ab}_{cd} &=& {1 \over 2N} \Big( (N+2) {\cal{S}}^{ab}_{cd} - (N+2)(N-1) P[1]^{ab}_{cd}
\nonumber \\ &-& {1 \over 2} (N-2)(N+4) P[8_S]^{ab}_{cd}
 + {N\over 2} ([D_t]^{ab}_{cd}+ [D_u]^{ab}_{cd})
\Big) \label{eq:4.1.14}\\
P[{\cal R}_7]^{ab}_{cd} & =& {1 \over 2N} \Big( (N-2) {\cal{S}}^{ab}_{cd} + (N-2)(N+1) P[1]^{ab}_{cd}
\nonumber \\ &+&  {1 \over 2} (N+2)(N-4) P[8_S]^{ab}_{cd}
 - {N\over 2} ([D_t]^{ab}_{cd}+ [D_u]^{ab}_{cd})\Big) \, .
\label{eq:4.1.15}
\eea
The tensors $A$, $Y$, $S$, and $D$ are defined as follows
\bea
{\cal{S}}^{ab}_{cd}  \equiv {1 \over 2}\Big( \delta_{ac} \delta_{bd} +
\delta_{ad} \delta_{bc} \Big)\,, \,\, \,\,\,\,\,\,
{\cal{A}}^{ab}_{cd}  \equiv {1 \over 2}\Big( \delta_{ac} \delta_{bd} -
\delta_{ad} \delta_{bc} \Big) \, .
\label{eq:4.1.2}
\eea
\bea
i Y^{ab}_{cd} && \equiv {i \over 2} \Big(  d_{adk}f_{kbc} +  f_{adk} d_{kbc} \Big)
\label{eq:4.1.4}
\eea
\bea
[D_t]^{ab}_{cd} \equiv d_{ack}d_{kbd}\, , \,\,\,\,\, [D_u]^{ab}_{cd} \equiv d_{adk}d_{kbc} \, ,\,\,\,\,\,\,
[D_s]^{ab}_{cd} \equiv d_{abk}d_{kcd}
\label{eq:4.1.5}
\eea

\begin{table}
\begin{tabular}{lccccccc}
& \multicolumn{4}{|c|} {symmetric} &  \multicolumn{3}{c} {antisymmetric} \\
\hline
Name of rep.  &  $1$  &   $8_S$  &  $27$ &  ${\cal R}_7$  &  $8_A$  &  $10$  &  $\overline{10}$ \\
 Dimension  &  1  & $N^2-1$  & ${N^2 (N+3)(N-1) \over 4}$ &
${N^2 (N-3)(N+1) \over 4}$ & $ N^2-1$ & $ {(N^2-4)(N^2-1) \over 4} $ &
$ {(N^2-4)(N^2-1) \over 4} $ \\
Casimir $C_2[{\cal R}]$ & 0 & $N$ &  $2(N+1)$ &  $2(N-1)$ &  $N$ &  $2N$ &  $2N$ \\
\hline
\end{tabular}
\caption{Properties of Multiplets}
\label{table1}
\end{table}

Apart from the complex, but hermitian structure
$i Y_s = P_s[10] - P_s[\overline{10}],$
which already appeared in the decuplet projectors, the full set of color-singlet
four-gluon states includes two more complex hermitian tensor structures
\bea
&&i (Z_s^{+})^{ab}_{cd} =  {i \over 2} \Big( f_{bak} d_{kcd} + d_{bak}f_{kcd} \Big) \, , \nonumber \\
&&i (Z_s^{-})^{ab}_{cd} =  {i \over 2} \Big( f_{bak} d_{kcd} - d_{bak}f_{kcd} \Big) \, .
\label{eq:C.12}
\eea
These tensors $i Z_s^{(\pm)}$ "project" onto mixed $\ket{8_A 8_S}$--states.

The projectors satisfy the Fierz--type identities:
 \bea
 P_t[{\cal R}_j] = \sum_{i=1}^9 C^j_i \, P_s[{\cal R}_i] \, ,
\label{eq:C.8}\eea
where the $t$--channel projectors are defined as:
\bea
 P_t[{\cal R}]^{ab}_{cd} \equiv  P_s[{\cal R}]^{ac}_{bd} \, .
\eea
The Fierz relations are therefore
\bea
 P[{\cal R}_j]^{ac}_{bd}    = \sum_{i=1}^9 C^j_i \, P[{\cal R}_i]^{ab}_{cd} \, ,
\label{eq:C.9}\eea
with the crossing matrix $C^j_i$ (Fig. \ref{tC}) is now obtained as
\bea
 C^j_i =  {P[{\cal R}_j]^{ac}_{bd}\cdot P[{\cal R}_i]^{cd}_{ab} \over  P[{\cal R}_i]^{ab}_{cd}P[{\cal R}_i]^{cd}_{ab}}
= {P[{\cal R}_j]^{ac}_{bd}\cdot P[{\cal R}_i]^{cd}_{ab} \over \dim[{\cal R}_i]} \, .
\label{eq:C.10}\eea
Note that eq.(\ref{eq:C.9}) includes all nine tensors (i.e. $P$ and  $Z^\pm$).

\begin{figure}[!ht]
\begin{center}
\includegraphics[width = 15 cm, height = 22 cm, angle=180]{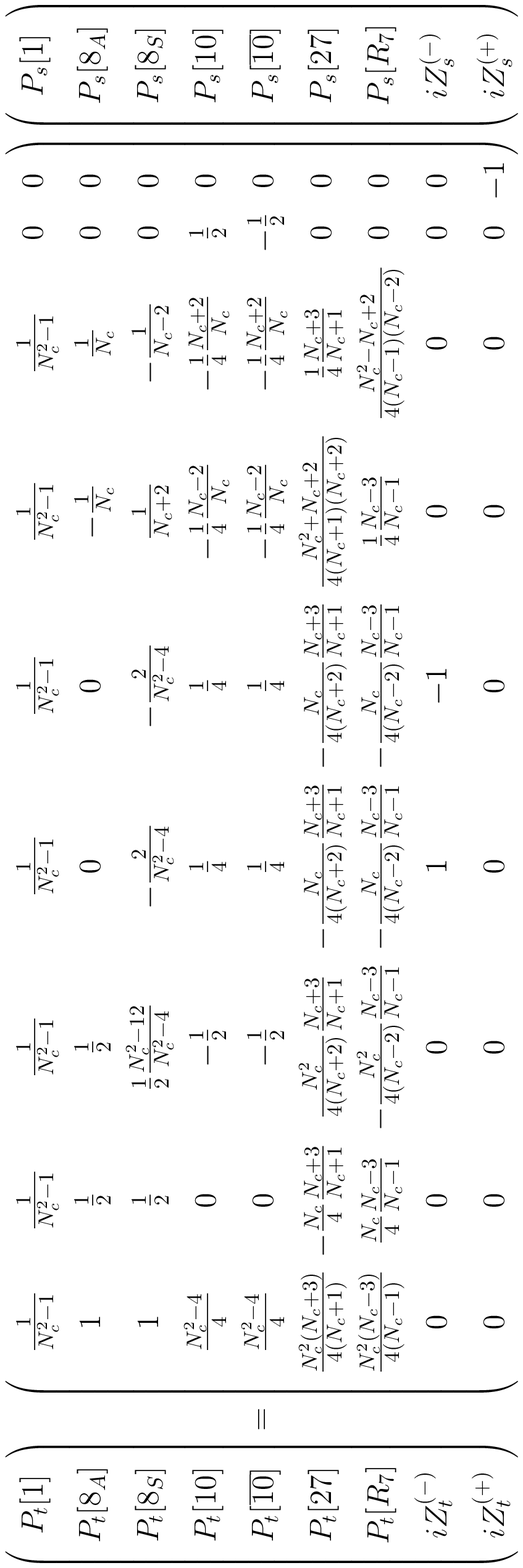}
\caption{The crossing matrix $C$}
\label{tC}
\end{center}
\end{figure}

Below we present several identities which we found useful.
\beq\label{tt}
(T^a\,T^c)_{bd}\,=\,N\,\lambda_j\,P[{\cal R}_j]^{ac}_{bd}
\eeq
where $\lambda_j$ is defined as the second line of the matrix $C$:
\beq\label{lambdaC}
\lambda_j\,\equiv\,C_j^{8_A}
\eeq
If $D_i$ as the dimension of $i$-th representation  we have
\beq
C^i_{8_A}\,=\,\lambda_i\,{D_i\over  D_8}
\eeq
It is useful to introduce the matrix $\bar C$ (table \ref{t3})
\beq\label{barC}
\bar C^i_k\,\equiv \sum_{j=1}^7 \lambda_j\,C^i_j\,C^j_k
\eeq
\begin{table}
\begin{tabular}{c|c|c|c|c|c|c|c|}
Name of rep.  &  $1$  &   $8_A$  &  $8_S$ &  $10$  &  $\overline{10}$  &  $27$  &  ${\cal R}_7$ \\
\hline
 $1$  &  0  & ${1\over N^2-1}$  & $0$ & $0$ & $ 0$ & $ 0$ & $ 0 $ \\
$8_A$ & 1 & ${1\over 4}$ &  $1\over 4$ &  $0$ &  $0$ &  ${1\over N^2}$ & ${ 1\over N^2}$ \\
$8_S$ & 0 & ${1\over 4}$ &  $1\over 4$ &  $1\over N^2\,-\,4$ &  $1\over N^2\,-\,4$ &  $0$ & $0$ \\
$10$ & 0 & $0$ &  $1\over 4$ &  $1\over 4$ &  $1\over 4$ &  $(N\,+\,1)\,(N\,-\,2)\over 4\,N^2$ &
$(N\,-\,1)\,(N\,+\,2)\over 4\,N^2$ \\
$\overline{10}$ & 0 & $0$ &  $1\over 4$ &  $1\over 4$ &  $1\over 4$ &  $(N\,+\,1)\,(N\,-\,2)\over 4\,N^2$ &
$(N\,-\,1)\,(N\,+\,2)\over 4\,N^2$ \\
$27$ & 0 & $N\,+\,3\over 4\,(N\,+\,1)$ &  $ 0 $ &  $N\,+\,3\over 4\,(N\,+\,2)$ &
 $N\,+\,3\over 4\,(N\,+\,2)$ &  $(N\,+\,1)\over 2\,N $ &
$0$ \\
${\cal R}_7$ & 0 & $N\,-\,3\over 4\,(N\,-\,1)$ &  $ 0 $ &  $N\,-\,3\over 4\,(N\,-\,2)$ &
 $N\,-\,3\over 4\,(N\,-\,2)$ & $0$ &  $(N\,-\,1)\over 2\,N $  \\
\hline
\end{tabular}
\caption{Matrix $\bar C$}
\label{t3}
\end{table}
which obeys
\beq
\sum_j\,(-1)^{s_j}\,\lambda_j \,D_j\,\bar C^i_j\,=\,(-1)^{s_i}\,(-\,\lambda_i^2\,+\,\lambda_i/2) \,D_i
\eeq
were we introduced the symmetry index $s_i=s[{\cal R}_i]$ via
\beq
P^{i\,ab}_{\,\,\,cd}\,=\,(-1)^{s_i}\,P^{i\,ba}_{\,\,\,cd}\,=\,(-1)^{s_i}\,P^{i\,ab}_{\,\,\,dc}
\eeq
For symmetric representations $s_i=0$ while for antisymmetric ones  $s_i=1$ (table \ref{table1}).
The matrix $\bar C$ arises when a projector multiplies a product of two matrices T:
\beq\label{PTT}
P^{i\,\beta d}_{\,\,\,a\gamma}\,T^{\bar a}_{\alpha a}\,T^{\bar a}_{d \delta}\,=\,
P^{i\,\beta d}_{\,\,\,a\gamma}\,N\,\lambda_j\,P^{j\,a d}_{\,\,\,\alpha\delta}\,=\,
N\,(-1)^{s_i}\,P^{i\,\beta d}_{\,\,\,\gamma a}\,N\,\lambda_j\,P^{j\,a d}_{\,\,\,\alpha\delta}\,=\,
N\,(-1)^{s_i+s_k}\,\bar C^i_k\,P^{k\,\beta \delta}_{\,\,\,\alpha\gamma}\,.
\eeq
We also have
\beq
P^{k\,\beta \delta}_{\,\,\,\alpha\gamma}\,P^{i\,b d}_{\,\,\,\,a\alpha}\,
T^{\delta}_{\beta d}\,T^{\gamma}_{ab}\,=\,N\,\lambda_i\,\delta^{k, 8_A}\,D_i\,.
\eeq

We now list some useful properties obeyed by the tensors $Z$. Under the exchange of indices they behave as:
\beq
Z^{+\,ab}_{\,\,\,\,\,cd}\,=\,-\,Z^{+\,cd}_{\,\,\,\,\,\,ab}\,;
\,\,\,\,\,\,\,\,\,\,\,\,\,\,\,\,\,\,
Z^{-\,ab}_{\,\,\,\,\,cd}\,=\,Z^{-\,cd}_{\,\,\,\,\,\,ab}\,;
\,\,\,\,\,\,\,\,\,\,\,\,\,\,\,\,\,\,
Z^{+\,ab}_{\,\,\,\,\,cd}\,=\,Z^{-\,ab}_{\,\,\,\,\,\,dc}\,;
\,\,\,\,\,\,\,\,\,\,\,\,\,\,\,\,\,\,
Z^{+\,ab}_{\,\,\,\,\,cd}\,=\,-\,Z^{-\,ba}_{\,\,\,\,\,\,cd}\,.
\eeq
Multiplication properties of $Z$  are
\bea\label{last}
&&Z^{\pm\,ab}_{\,\,\,\,\,cd}\,P[8_A]^{cd}_{kl}\,=\,\pm\,{1\over 2}\,(Z^+\,-\,Z^-)^{ab}_{kl}\,;\nonumber\\
&&
Z^{\pm\,ab}_{\,\,\,\,\,cd}\,P[8_S]^{cd}_{kl}\,=\,{1\over 2}\,(Z^+\,+\,Z^-)^{ab}_{kl}\nonumber\\
&&
Z^{\pm\,ab}_{\,\,\,\,\,cd}\,Z^{\pm\,cd}_{\,\,\,\,\,\,kl}\,=\,\mp\,(N^2\,-\,4)\,(P[8_A]\,+
\,P[8_S])^{ab}_{kl}\,;\nonumber\\
&&Z^{+\,ab}_{\,\,\,\,\,cd}\,Z^{-\,cd}_{\,\,\,\,\,\,kl}\,=\,(N^2\,-\,4)\,(P[8_A]\,-
\,P[8_S])^{ab}_{kl}\nonumber\\
&&
Z^{+\, d\delta}_{\,\,\,\,\alpha a}\,T^{\bar a}_{a\gamma}\,T^{\bar a}_{ \beta d}\,=\,
Z^{-\, \delta d}_{\,\,\,\,\alpha a}\,N\,\lambda_j\,P^{j\,a d}_{\,\,\,\gamma\beta}\,=\,
{N\over 2}\,(P^{10}\,-\,P^{\overline{10}})^{ \delta \alpha}_{d a}\,\lambda_j\,P^{j\,a d}_{\,\,\,\gamma\beta}
\,=\,0\nonumber\\
&&
Z^{-\, d\delta}_{\,\,\,\,\alpha a}\,T^{\bar a}_{a\gamma}\,T^{\bar a}_{ \beta d}\,=\,
Z^{+\, \delta d}_{\,\,\,\,\alpha a}\,N\,\lambda_j\,P^{j\,a d}_{\,\,\,\gamma\beta}\,=\,-\,
N\,Z^{+\, \delta \alpha}_{\,\,\,\,\,d a}\,\lambda_j\,P^{j\,a d}_{\,\,\,\gamma\beta}\,=\,
-\,N\,Z^{+\, \delta \alpha}_{\,\,\,\,\,\gamma\beta}\,\lambda_8\,=\,
-\,\frac{N}{2}\,Z^{-\, \beta\delta}_{\,\,\,\,\,\alpha \gamma}\nonumber\\
&&
Z^{\pm\,\beta \delta}_{\,\,\,\,\,\alpha\gamma}\,P^{i\,b d}_{\,\,\,\,a\alpha}\,
T^{d}_{\delta \beta }\,T^{\gamma}_{ab}\,=\,\frac{N}{4}\,D_{10}\,[\delta^{i, 10}\,-\,\delta^{i, \overline{10}}]
\,;\nonumber\\ &&
Z^{\pm\,\beta \delta}_{\,\,\,\,\,\alpha\gamma}\,P^{i\,d\beta}_{\,\,\,\,ca}\,
T^{a}_{\alpha \gamma }\,T^{\delta}_{cd}\,=\,\pm\,\frac{N}{4}\,D_{10}\,[\delta^{i, 10}\,-\,\delta^{i, \overline{10}}]
\eea


\begin{thebibliography}{99}

\bibitem{Gribov}
  V.~N.~Gribov,
  Sov.\ Phys.\ JETP {\bf 26}, 414 (1968)
  [Zh.\ Eksp.\ Teor.\ Fiz.\  {\bf 53}, 654 (1967)].

\bibitem{Amati}
  D.~Amati, M.~Le Bellac, G.~Marchesini and M.~Ciafaloni,
  Nucl.\ Phys.\ B {\bf 112}, 107 (1976).
\\
V.~Alessandrini, D.~Amati and M.~Ciafaloni,
  Nucl.\ Phys.\ B {\bf 130}, 429 (1977).
\\
  D.~Amati, G.~Marchesini, M.~Ciafaloni and G.~Parisi,
  Nucl.\ Phys.\ B {\bf 114}, 483 (1976).
\\
  M.~Ciafaloni and G.~Marchesini,
  Nucl.\ Phys.\ B {\bf 109}, 261 (1976).


\bibitem{Baker}
  M.~Baker and K.~A.~Ter-Martirosian,
  Phys.\ Rept.\  {\bf 28}, 1 (1976).



\bibitem{BFKL}
 E. A. Kuraev, L. N. Lipatov, and F. S. Fadin,  Sov. Phys. JETP
                {\bf 45} (1977) 199 ; \\
Ya. Ya. Balitsky and L. N. Lipatov,
               {  Sov. J. Nucl. Phys.}\, {\bf 28} (1978) 22 .



\bibitem{LipatovRgluon}
  L.~N.~Lipatov,
  Sov.\ J.\ Nucl.\ Phys.\  {\bf 23}, 338 (1976)
  [Yad.\ Fiz.\  {\bf 23}, 642 (1976)].

\bibitem{FS} L.~L.~Frankfurt and V.~E.~Sherman,
  Sov.\ J.\ Nucl.\ Phys.\  {\bf 23} (1976) 581.


\bibitem{Fadin}
V.~S.~Fadin, M.~I.~Kotsky and R.~Fiore,
  Phys.\ Lett.\ B {\bf 359} (1995) 181.


\bibitem{ELLA}
  J.~Bartels, preprint DESY-91-074; J. Bartels,
  Nucl.\ Phys.\ B {\bf 151}, 293 (1979).


\bibitem{BKP}
J.~Bartels,
%
Nucl.\ Phys.\  {\bf B175}, 365 (1980);\,\,\,\,
J.~Kwiecinski and M.~Praszalowicz,
%
Phys.\ Lett.\  {\bf B94}, 413 (1980).

\bibitem{GLR}  L.V.~Gribov, E.~Levin and M.~Ryskin, Phys. Rep.
100:1,1983.

\bibitem{3P} J.~Bartels, Z.Phys. C60, 471, 1993.

\bibitem{3P1} J.~Bartels and M.~Wusthoff, Z. Phys. {\bf C 66}, 157, 1995.


\bibitem{LipatovInteg}
  L.~N.~Lipatov,
  JETP Lett.\  {\bf 59}, 596 (1994)
  [Pisma Zh.\ Eksp.\ Teor.\ Fiz.\  {\bf 59}, 571 (1994)].

\bibitem{FK}
  L.~D.~Faddeev and G.~P.~Korchemsky,
  Phys.\ Lett.\ B {\bf 342}, 311 (1995)
  [arXiv:hep-th/9404173].

\bibitem{Korchemsky}
  G.~P.~Korchemsky,
  Nucl.\ Phys.\ B {\bf 462}, 333 (1996)
  [arXiv:hep-th/9508025];
  Nucl.\ Phys.\ B {\bf 443}, 255 (1995)
  [arXiv:hep-ph/9501232];
  Nucl.\ Phys.\ B {\bf 498}, 68 (1997)
  [arXiv:hep-th/9609123];
G.~P.~Korchemsky, J.~Kotanski and A.~N.~Manashov,
Phys.\ Rev.\ Lett.\  {\bf 88} (2002) 122002
[arXiv:hep-ph/0111185].

\bibitem{devega} H.J. de Vega and L.N. Lipatov, Phys.Rev.D64:114019,2001, [arXive: hep-ph/0107225]; Phys.Rev.D66:074013,2002.
[arXive: hep-ph/0204245]


\bibitem{BLW}
  J.~Bartels, L.~N.~Lipatov and M.~Wusthoff,
  Nucl.\ Phys.\ B {\bf 464}, 298 (1996)
  [arXiv:hep-ph/9509303].

\bibitem{Lotter}

  H.~Lotter,
  arXiv:hep-ph/9705288.

\bibitem{BV}
  M.~A.~Braun and G.~P.~Vacca,
  Eur.\ Phys.\ J.\ C {\bf 6}, 147 (1999)
  [arXiv:hep-ph/9711486].


\bibitem{BE} J.~Bartels and C.~Ewerz, JHEP  9909, 026, 1999
[e-Print Archive:hep-ph/9908454].

\bibitem{ES}
  C.~Ewerz and V.~Schatz,
  Nucl.\ Phys.\ A {\bf 736}, 371 (2004)
  [arXiv:hep-ph/0308056].



\bibitem{BBV}
  J.~Bartels, M.~Braun and G.~P.~Vacca,
  Eur.\ Phys.\ J.\ C {\bf 40}, 419 (2005)
  [arXiv:hep-ph/0412218].

\bibitem{BLV} J.~Bartels, L.~N.~Lipatov and G.~P.~Vacca,
  Nucl.\ Phys.\ B {\bf 706}, 391 (2005)
  [arXiv:hep-ph/0404110].



\bibitem{Braun1} M. Braun, {\it Eur. Phys. J.} {\bf C 16} (2000) 337.

\bibitem{Braun2}
M. Braun, Phys. Lett. B483, 115, 2000 [e-Print Archive: hep-ph/0003004].

\bibitem{braunlast} M. Braun, e-Print Archive: hep-ph/0512057.

\bibitem{MUQI}
A. H. Mueller and J. Qiu,  Nucl. Phys. {\bf B 268} (1986) 427.


\bibitem{Mueller} A. Mueller, {\it
Nucl. Phys.} {\bf B335} 115 (1990); {\it ibid} {\bf B 415};
373 (1994); {\it ibid} {\bf B437} 107 (1995).

\bibitem{Reggeon}A.~H.~Mueller and B.~Patel, Nucl. Phys. B 425, 471, 1994.

\bibitem{NP}
H.~Navelet and R.~Peschanski,
 Nucl.\ Phys.\,  {\bf B634} (2002) 291
[arXiv:hep-ph/0201285]; \,\,\,
 Phys.\ Rev.\ Lett.\,  {\bf 82} (1999) 137,,
[arXiv:hep-ph/9809474];\,\,\,
Nucl.\ Phys.\,  {\bf B507} (1997)  353,
[arXiv:hep-ph/9703238]\,\,\,\,\,\,\,A.~Bialas, H.~Navelet
and R.~Peschanski,
 Phys.\ Rev.\, {\bf D57} (1998) 6585;\,\,\,\,\,R.~Peschanski,
Phys.\ Lett. \, {\bf B409} (1997) 491.


\bibitem{mv} L.~McLerran and R.~Venugopalan, Phys. Rev. D49:2233-2241,1994
; Phys.Rev.D49:3352-3355,1994.


\bibitem{balitsky} I. Balitsky, {\it Nucl. Phys.}  {\bf B463} 99 (1996);
{\it Phys. Rev. Lett.} {\bf 81} 2024 (1998); 
{\it Phys. Rev.}{\bf D60} 014020 (1999).


\bibitem{Kovchegov}
  Y.~V.~Kovchegov,
  Phys.\ Rev.\ D {\bf 61}, 074018 (2000)
  [arXiv:hep-ph/9905214].

\bibitem{JIMWLK} J. Jalilian Marian, A. Kovner, A.Leonidov and H.
Weigert,
{\it Nucl. Phys.}{\bf  B504} 415 (1997); 
{\it Phys. Rev.} {\bf D59} 014014 (1999); 
J. Jalilian Marian, A. Kovner and H. Weigert, {\it Phys. Rev.}{\bf D59}
014015 (1999);
A. Kovner and J.G. Milhano, {\it Phys. Rev.} {\bf D61} 014012 (2000) .
 A. Kovner, J.G. Milhano and H. Weigert,
{\it Phys.Rev.} {\bf D62} 114005 (2000);
 H. Weigert, {\it Nucl.Phys.} {\bf A 703} (2002) 823.


\bibitem{cgc}  E.Iancu, A. Leonidov and L. McLerran, {\it Nucl. Phys.}
{\bf A 692} (2001) 583; {Phys. Lett.} {\bf B
510} (2001) 133;
E. Ferreiro, E. Iancu, A. Leonidov, L. McLerran;
{\it Nucl. Phys.}{\bf A703} (2002) 489.

\bibitem{IM}
  E.~Iancu and A.~H.~Mueller,
  Nucl.\ Phys.\ A {\bf 730}, 494 (2004)
  [arXiv:hep-ph/0309276].

\bibitem{shoshi1} A. Mueller and A.I. Shoshi, Nucl. Phys. B {\bf 692}, 175 (2004).

\bibitem{IMM} E. Iancu, A.H. Mueller and S. Munier, Phys.\ Lett.\ B {\bf 606}, 342 (2005).

\bibitem{IT}  E. Iancu and  D. N. Triantafyllopoulos,  Phys.\ Lett.\ B {\bf 610}, 253 (2005);
Nucl.\ Phys.\ A {\bf 756}, 419 (2005); E.~Iancu, G.~Soyez and D.~N.~Triantafyllopoulos,
  arXiv:hep-ph/0510094.

\bibitem{something} A. Kovner and M. Lublinsky, Phys.\ Rev.\ Lett.\  {\bf 94}, 181603 (2005).

\bibitem{kl4} A.~Kovner and M.~Lublinsky,
Phys.\ Rev.\ D {\bf 72}, 074023 (2005)
  [arXiv:hep-ph/0503155].

\bibitem{balitsky1} I. Balitsky, in *Shifman, M. (ed.): At the frontier of particle physics, vol. 2* 1237-1342;
e-print arxive hep-ph/0101042;
  Phys.\ Rev.\ D {\bf 70}, 114030 (2004)
  [arXiv:hep-ph/0409314].



\bibitem{kozlov} M. Kozlov and E. Levin,
{\it Nucl.Phys.} {\bf A739} 291 (2004).




\bibitem{kl}  A. Kovner and M. Lublinsky, Phys.\ Rev.\ D {\bf 71}, 085004 (2005).

\bibitem{MSW}A. Mueller, A. Shoshi and S. Wong, Nucl.\ Phys.\ B {\bf 715}, 440 (2005).

\bibitem{LL3} E. Levin and M. Lublinsky; Nucl. Phys. A {\bf 763}, 172 (2005)
[arxive hep-ph/0501173].

\bibitem{kl1} A. Kovner and M. Lublinsky, JHEP {\bf 0503}, 001 (2005).

\bibitem{kl5}
  A.~Kovner and M.~Lublinsky,
  arXiv:hep-ph/0510047.

\bibitem{SMITH}
  Y.~Hatta, E.~Iancu, L.~McLerran, A.~Stasto and D.~N.~Triantafyllopoulos,
  arXiv:hep-ph/0504182.


\bibitem{genya} E. Levin, e-print arxive hep-ph/0502243;
  arXiv:hep-ph/0511074.


\bibitem{Balitsky05}
  I.~Balitsky,
  arXiv:hep-ph/0507237.

\bibitem{zakopane} A. Kovner, Lectures at the Zakopane summer school, e-print arxive hep-ph/0508232.



\bibitem{Verlinde}
  H.~L.~Verlinde and E.~P.~Verlinde,
  arXiv:hep-th/9302104.


\bibitem{LipatovFT}
  L.~N.~Lipatov,
  Nucl.\ Phys.\ B {\bf 365}, 614 (1991).
  Nucl.\ Phys.\ B {\bf 452}, 369 (1995)
  [arXiv:hep-ph/9502308]; \\
R.~Kirschner, L.~N.~Lipatov and L.~Szymanowski,
  Nucl.\ Phys.\ B {\bf 425}, 579 (1994)
  [arXiv:hep-th/9402010].
  Phys.\ Rev.\ D {\bf 51}, 838 (1995)
  [arXiv:hep-th/9403082]; \\
L.~N.~Lipatov,
  Phys.\ Rept.\  {\bf 286}, 131 (1997)
  [arXiv:hep-ph/9610276].





\bibitem{KSW}
  Y.~V.~Kovchegov, L.~Szymanowski and S.~Wallon,
  Phys.\ Lett.\ B {\bf 586}, 267 (2004)
  [arXiv:hep-ph/0309281].

\bibitem{oderon} Y.~Hatta, E.~Iancu, K.~Itakura and L.~McLerran,
  Nucl.\ Phys.\ A {\bf 760}, 172 (2005)
  [arXiv:hep-ph/0501171].
 .


\bibitem{NSZ}
  N.~N.~Nikolaev, W.~Schafer and B.~G.~Zakharov,
  arXiv:hep-ph/0508310.

\bibitem{Ewerzod}
C.~Ewerz,
arXiv:hep-ph/0306137.

\bibitem{BLVod}
J.~Bartels, L.~N.~Lipatov and G.~P.~Vacca,
Phys.\ Lett.\ B {\bf 477} (2000) 178
[arXiv:hep-ph/9912423].



\bibitem{first} J. Jalilian-Marian, A. Kovner, L. D. McLerran and H. Weigert, Phys.\ Rev.\ {\bf D55} (1997), 5414;
e-Print Archive: hep-ph/9606337.


\bibitem{Hatta}
  Y.~Hatta,
  arXiv:hep-ph/0511287.

\bibitem{Fukushima}
  K.~Fukushima,
  arXiv:hep-ph/0512138.

\bibitem{janik} R. Janik, Phys.Lett.B604:192-198,2004, hep-ph/0409256.

\bibitem{LL1}
E.~Levin and M.~Lublinsky,
Nucl.\ Phys.\, {\bf A730} (2004) 191.
\bibitem{LL2}
E.~Levin and M.~Lublinsky, Phys.\ Lett.\ {\bf B607}, (2005) 131.

\bibitem{MMSW}
  C.~Marquet, A.~H.~Mueller, A.~I.~Shoshi and S.~M.~H.~Wong,
  arXiv:hep-ph/0505229.

\bibitem{HIMS}
  Y.~Hatta, E.~Iancu, L.~McLerran and A.~Stasto,
  arXiv:hep-ph/0505235.


\bibitem{Lipatov1}
L.~N.~Lipatov,
Sov.\ Phys.\ JETP {\bf 63}, 904 (1986)
[Zh.\ Eksp.\ Teor.\ Fiz.\  {\bf 90}, 1536 (1986)].


\bibitem{Braun05r}
  M.~A.~Braun,
  arXiv:hep-ph/0503041.


\bibitem{BKT}
Yu. Kovchegov,  { \it Phys. Rev.} {\bf D 61} (2000) 074018; \\
 E. Levin and K. Tuchin, {\it Nucl. Phys.} {\bf B 573} (2000) 833;
 {\it Nucl.\ Phys.} {\bf A 691} (2001) 779;\\
E.~Iancu, K.~Itakura and L.~McLerran,
{\it Nucl.\ Phys.}  {\bf A  708} (2002) 327.


\bibitem{BKN}
 N. Armesto and M. Braun,  {\it Eur. Phys. J.} {\bf
    C 20} (2001) 517;\\
M.~Lublinsky,
{\it Eur.\ Phys.\ J.}  {\bf C 21} (2001) 513;
E.~Levin and M.~Lublinsky,
{\it Nucl.\ Phys.}  {\bf A 712} (2002) 95;
  Nucl.\ Phys.\ A {\bf 712}, 95 (2002);
  Nucl.\ Phys.\ A {\bf 696}, 833 (2001)
  [arXiv:hep-ph/0104108];
  Eur.\ Phys.\ J.\ C {\bf 22}, 647 (2002); \\
M.~Lublinsky, E.~Gotsman, E.~Levin and U.~Maor,
{\it Nucl.\ Phys.} {\bf A 696} (2001) 851;
E.~Gotsman, E.~Levin, M.~Lublinsky and U.~Maor,
{\it Eur. Phys. J} {\bf C 27} (2003) 411;\\
  K. Golec-Biernat, L. Motyka, A. Stasto, {\it Phys. Rev.}
  {\bf D 65} (2002) 074037;\\
K. Rummukainen and H. Weigert, {\it Nucl. Phys.} {\bf A 739} (2004) 183;\\
K. Golec-Biernat and  A.M. Stasto,
{\it Nucl. Phys.} {\bf B 668} (2003) 345; \\
       E. Gotsman, M. Kozlov, E. Levin, U. Maor, and E. Naftali, Nucl.\ Phys.\ A {\bf 742}, 55 (2004); \\
 K. Kutak and A.M. Stasto, Eur.\ Phys.\ J.\ C {\bf 41}, 343 (2005);\\
G. Chachamis, M. Lublinsky and A. Sabio-Vera, Nucl.\ Phys.\ A {\bf 748}, 649 (2005);\\
J.~L.~Albacete, N.~Armesto, J.~G.~Milhano, C.~A.~Salgado and U.~A.~Wiedemann, Phys.\ Rev.\ D {\bf 71}, 014003 (2005).

\bibitem{MT}
  A.~H.~Mueller and D.~N.~Triantafyllopoulos,
  Nucl.\ Phys.\ B {\bf 640}, 331 (2002)
  [arXiv:hep-ph/0205167].

\bibitem{MP}
S. Munier and R. Peschanski, {\it Phys. Rev.} {\bf D 69}
(2004) 034008; {\it Phys. Rev. Lett.} {\bf 91} (2003) 232001; hep-ph/0401215.
C.~Marquet, R.~Peschanski and G.~Soyez,
  Nucl.\ Phys.\ A {\bf 756}, 399 (2005)
  [arXiv:hep-ph/0502020];
  arXiv:hep-ph/0509074.
R.~Peschanski,
  Phys.\ Lett.\ B {\bf 622}, 178 (2005)
  [arXiv:hep-ph/0505237].





\bibitem{DM}
  Y.~L.~Dokshitzer and G.~Marchesini,
  arXiv:hep-ph/0509078;
  arXiv:hep-ph/0508130.

 \bibitem{Ewerz}
  C.~Ewerz,
  JHEP {\bf 0104}, 031 (2001)
  [arXiv:hep-ph/0103260].

\bibitem{LRS}
E.~M.~Levin, M.~G.~Ryskin and A.~G.~Shuvaev,
in
  Nucl.\ Phys.\ B {\bf 387} (1992) 589.


\bibitem{chris} C. Korthals Altes and H. Meyer,  hep-ph/0509018.


\bibitem{urs} A. Kovner and U. Wiedemann,  Phys.Rev.D66:051502,2002
e-Print Archive: hep-ph/0112140;  Phys.Rev.D66:034031,2002
e-Print Archive: hep-ph/0204277; Phys.Lett.B551:311-316,2003
e-Print Archive: hep-ph/0207335.

\bibitem{JW}
  R.~A.~Janik and J.~Wosiek,
  Phys.\ Rev.\ Lett.\  {\bf 82}, 1092 (1999)
  [arXiv:hep-th/9802100].

\bibitem{susy} D. Amati, K. Konishi, Y. Meurice, G.C. Rossi and G. Veneziano (CERN),  Phys.Rept.162:169-248,1988.

\bibitem{LipatovDual}
  L.~N.~Lipatov,
  Nucl.\ Phys.\ B {\bf 548}, 328 (1999)
  [arXiv:hep-ph/9812336].


\bibitem{LipatovCI}
  L.~N.~Lipatov,
  Sov.\ Phys.\ JETP {\bf 63}, 904 (1986)
  [Zh.\ Eksp.\ Teor.\ Fiz.\  {\bf 90}, 1536 (1986)].


\bibitem{BF}
  B.~Blok and L.~Frankfurt,
  arXiv:hep-ph/0508218;  hep-ph/0512225.

\bibitem{blaizot} J.P. Blaizot et. al., Phys.\ Lett.\ B {\bf 615}, 221 (2005).


\bibitem{Kovchegov05}
  Y.~V.~Kovchegov,
  arXiv:hep-ph/0508276.





\bibitem{KOS}
  N.~Kidonakis, G.~Oderda and G.~Sterman,
  Nucl.\ Phys.\ B {\bf 531}, 365 (1998)
  [arXiv:hep-ph/9803241].

\end{thebibliography}
\end{document}